\documentclass[final]{linearFeng}
\newcommand{\ws}[1]{{}}
\newcommand{\arxiv}[1]{#1}

\renewcommand{\cite}[1]{[\refcite{#1}]}

\usepackage{graphicx}

\usepackage{makeidx}

\makeindex

\begin{document}

%%%%%%%%%%%%%%%%%%%%%%%%%please do not remove these lines %%%%%%%%%%%%%%%%%%%%%%%
%\titlepages 
%please leave this as it is. Pages 1-4 are reserved for publisher.
%%%%%%%%%%%%%%%%%%%%%%%%%%%%%%%%%%%%%%%%%%%%%%%%%%%%%%%%%%%%%%%%%%%%%%%%%%%%%%%%%

%JLF\begin{dedication}
%%%%% Please input your dedication data here.%%%%%%%%%%%%%%%%%%%%%

%JLFto World Scientific Publishing Company

%JLF\end{dedication}

%JLF\begin{preface}
%% please have your text for preface in a seperate file 
%% (only the contents of the preface).

%JLF\input preface.tex

%JLF\end{preface}

%%%%%%%%%%%%%%%%%%%%%%%%%%%%%%%% Need not the change anything %%%%%%%%%%%%%%%%%
%\begin{tblofcontents}
%%%%% Need not to input any data here. 
%%%%% System will input automatically.
%\end{tblofcontents}
%%%%%%%%%%%%%%%%%%%%%%%%%%%%%%%%%%%%%%%%%%%%%%%%%%%%%%%%%%%%%%%%%%%%%%%%%%%%%%%

\setcounter{chapter}{0}
\setcounter{page}{1}

%JLF MACROS:
\newcommand{\rem}[1]{{\bf #1}}
\newcommand{\postscript}[2]{\setlength{\epsfxsize}{#2\hsize}
   \centerline{\epsfbox{#1}}}
\newcommand{\ifb}{\text{fb}^{-1}}
\newcommand{\ev}{\text{eV}}
\newcommand{\kev}{\text{keV}}
\newcommand{\mev}{\text{MeV}}
\newcommand{\gev}{\text{GeV}}
\newcommand{\tev}{\text{TeV}}
\newcommand{\pb}{\text{pb}}
\newcommand{\mb}{\text{mb}}
\newcommand{\cm}{\text{cm}}
\newcommand{\km}{\text{km}}
\newcommand{\g}{\text{g}}
\newcommand{\s}{\text{s}}
\newcommand{\yr}{\text{yr}}
\newcommand{\sr}{\text{sr}}
\newcommand{\etal}{{\em et al.}}
\newcommand{\eg}{{\em e.g.}}
\newcommand{\ie}{{\em i.e.}}
\newcommand{\ibid}{{\em ibid.}}
\renewcommand{\eqref}[1]{Eq.~(\ref{#1})}
\newcommand{\eqsref}[2]{Eqs.~(\ref{#1}) and (\ref{#2})}
\newcommand{\secref}[1]{Sec.~(\ref{#1})}
\newcommand{\secsref}[2]{Secs.~(\ref{#1}) and (\ref{#2})}
\renewcommand{\text}[1]{{\rm #1}}
\newcommand{\agt}{ \mathop{}_{\textstyle \sim}^{\textstyle >} }
\newcommand{\alt}{ \mathop{}_{\textstyle \sim}^{\textstyle <} }

\newcommand{\chc}{\tilde{\chi}^{\pm}}
\newcommand{\chcflip}{\tilde{\chi}^{\mp}}
\newcommand{\chargino}{\tilde{\chi}^{\pm}_1}
\newcommand{\charginotwo}{\tilde{\chi}^{\pm}_2}
\newcommand{\mchargino}{m_{\tilde{\chi}^{\pm}_1}}
\newcommand{\mcharginotwo}{m_{\tilde{\chi}^{\pm}_2}}
\newcommand{\chcp}{\tilde{\chi}^+}
\newcommand{\chcm}{\tilde{\chi}^-}
\newcommand{\chn}{\tilde{\chi}^0}
\newcommand{\tb}{\tan\beta}
\newcommand{\mplanck}{M_{\text{Pl}}}
\newcommand{\mgaugino}{M_{1/2}}
\newcommand{\sign}{\text{sign}}
\newcommand{\mmess}{M_{\text{mess}}}
\newcommand{\epem}{e^+e^-}
\newcommand{\emem}{e^-e^-}
\newcommand{\ebeam}{E_{\text{beam}}}
\newcommand{\fdsb}{F_{\text{DSB}}}
\newcommand{\Arg}{\text{Arg}}

%MMN MACROS:
\newcommand{\cosb}{c_\beta}
\newcommand{\sinb}{s_\beta}
\newcommand{\cw}{c_w}
\newcommand{\sw}{s_w}
\newcommand{\tw}{t_w}
\newcommand{\sws}{s_w^2}
\newcommand{\cws}{c_w^2}
\newcommand{\tws}{t_w^2}
\newcommand{\aiiR}{a_{1i}^R}
\newcommand{\aiiL}{a_{1i}^L}
\newcommand{\ptau}{P_{\tau}(\tilde{\tau}\to\tau\tilde{\chi})}
\newcommand{\dbarchi}{\Delta\bar{\chi}^2}
\newcommand{\se}{\tilde{e}}
\newcommand{\mserl}{m_{\tilde{e}_{LR}}}
\newcommand{\smu}{\tilde{\mu}}
\newcommand{\stau}{\tilde{\tau}}
\newcommand{\stl}{\tilde{\tau}_L}
\newcommand{\str}{\tilde{\tau}_R}
\newcommand{\sti}{\tilde{\tau}_1}
\newcommand{\stii}{\tilde{\tau}_2}
\newcommand{\cost}{\cos\theta_{\tilde{\tau}}}
\newcommand{\sint}{\sin\theta_{\tilde{\tau}}}
\newcommand{\mstau}{m_{\tilde{\tau}}}
\newcommand{\mstl}{m_{\tilde{\tau}_L}}
\newcommand{\mstr} {m_{\tilde{\tau}_R}}
\newcommand{\msti}{m_{\tilde{\tau}_1}}
\newcommand{\mstii}{m_{\tilde{\tau}_2}}
\newcommand{\mstrl}{m_{\tilde{\tau}_{LR}}}
\newcommand{\mchi}{m_{\tilde{\chi}^0_1}}
\newcommand{\mweak}{M_{\rm weak}}
\newcommand{\mgut}{M_{\rm GUT}}
\newcommand{\be}{\begin{equation}}
\newcommand{\ee}{\end{equation}}
\newcommand{\beq}{\begin{eqnarray}}
\newcommand{\eeq}{\end{eqnarray}}
\newcommand{\ben}{\begin{mathletters}}
\newcommand{\een}{\end{mathletters}}
\newcommand{\sla}[1]{\! \not{\! {#1}}}

\arxiv{\chapter*{Supersymmetry and the Linear Collider}}
\ws{\chapter{Supersymmetry and the Linear Collider}}

\label{Schap:supersymmetry}

\markboth{J.~L.~Feng and M.~M.~Nojiri}{Supersymmetry and the Linear
Collider}

\arxiv{
\vspace*{-1.8in}
{\hfill UCI--TR--02--37}

{\hfill YITP--02--64}

{\hfill hep-ph/0210390}

\vspace*{1.5in}
}

\author{
\arxiv{\vspace*{-.2in}}
Jonathan L.~Feng$^a$ and Mihoko M.~Nojiri$^b$}

\address{
\vspace*{.1in}
$^a$ Department of Physics and Astronomy \\
University of California, Irvine, CA 92697, USA \\ 
\vspace*{.1in}
$^b$ Yukawa Institute of Theoretical Physics \\
Kyoto University, Kyoto 606-8502, Japan}

\arxiv{\vspace*{.2in}}

\begin{abstract}
We present a pedagogical introduction to supersymmetry and
supersymmetric models and give an overview of the potential of the
linear collider for studying them.  If supersymmetry is found, its
discovery will bring with it many more questions than answers.  These
include: \\
\hspace*{0.1in} Are the newly discovered particles really superpartners? \\
\hspace*{0.1in} If not all superpartners are discovered, where are the rest
of them? \\
\hspace*{0.1in} Do the electromagnetic, weak, and strong forces unify? \\
\hspace*{0.1in} Is a supersymmetric particle the dark matter? \\
\hspace*{0.1in} How are the supersymmetric flavor and CP problems solved? \\
\hspace*{0.1in} What is the scale of supersymmetry breaking? \\
\hspace*{0.1in} What are the fundamental interactions at the Planck
scale? \\ 
We review how the linear collider will provide definitive
answers to some of these and may shed light on the rest, even if only
one or a few superpartners are kinematically accessible.

\arxiv{
\begin{center}
\vspace*{.3in}
To appear as a chapter in\\ 
\vspace*{.1in}
{\em Linear Collider Physics in the New Millennium}\\ 
\vspace*{.1in}
published by World Scientific\\
editors David Miller, Keisuke Fujii and Amarjit Soni
\end{center}
}

\end{abstract}

\arxiv{\newpage}

\tableofcontents

\section{Introduction} 
\label{Ssec:intro}

Supersymmetry relates fermions to bosons and postulates the existence
of a partner particle for every known particle. Its motivations range
from the gauge hierarchy problem and the central role of supersymmetry
in quantum theories of gravity to the unification of gauge couplings
and the existence of dark matter.  The search for supersymmetry in
particle physics has been carried out on numerous fronts, and its
discovery would mark the culmination of an intense research effort
spanning several decades.

At the same time, the discovery of supersymmetry will bring with it
many more questions than answers.  These include:
\begin{itemize}
\item Are the newly discovered particles really superpartners?
\item If not all superpartners are discovered, where are the rest
of them?
\item Do the electromagnetic, weak, and strong forces unify?
\item Is a supersymmetric particle the dark matter?
\item How are the supersymmetric flavor and CP problems solved?
\item What is the scale of supersymmetry breaking?
\item What are the fundamental interactions at the Planck scale? 
\end{itemize}
\noindent Insights into these and other fundamental questions will
have far-reaching implications, transforming not only particle
physics, but also astrophysics and cosmology.

Here we present a pedagogical overview of what the linear collider may
contribute to answering these questions.  The large and growing
world-wide interest in linear colliders has motivated hundreds of
detailed studies of the potential of linear colliders for measuring
supersymmetry parameters. Typically, however, there is little
opportunity to review in broad terms the underlying physics questions
and overall goals of this program.  In this overview, we therefore
begin with a brief introduction to supersymmetry in
\secref{Ssec:MSSM}, focusing on the most salient points for collider
physics.  We then describe several outstanding successes and puzzles
driving work in supersymmetry in \secref{Ssec:successes} and some of
the resulting models in \secref{Ssec:models}.  Against this backdrop,
we then review what linear colliders may tell us in the remaining
sections.  An index of key terms is provided at the end.  Throughout
this overview, we consider an $e^+e^-$ linear collider with
center-of-mass energy between $300-1000~\gev$, a luminosity of ${\cal
O}(100)~\ifb/\yr$, highly polarizable electron (and possibly also
positron) beams, and the capability of running in alternative modes,
such as $e^-e^-$, $e^- \gamma$.\index{collider parameters}

This overview is neither a thorough introduction to supersymmetry nor
a comprehensive review of the linear collider literature.
Supersymmetry and its phenomenological implications have been
systematically presented in a number of beautiful introductions.
(See, for example,
Refs.~\cite{Bagger:1990qh,Haber:1993wf,Bagger:1996ka,Drees:1996ca,%
Tata:1997uf,Martin:1997ns,Weinberg:cr,Polonsky:2001pn}.) Detailed
supersymmetry studies at linear colliders and exhaustive
bibliographies may be found in recent reports of the large linear
collider
collaborations~\cite{Abe:2001nn,Aguilar-Saavedra:2001rg,Abe:2001gc}.

\section{The Minimal Supersymmetric Standard Model}
\label{Ssec:MSSM}

The {\em minimal supersymmetric standard model} (MSSM)\index{MSSM
(minimal supersymmetric standard model)!definition} is the
supersymmetric extension of the standard model with minimal field
content, conserved $R$-parity, and no additional theoretical
assumptions.  It is the underlying framework for most
``model-independent'' collider studies, and we therefore begin by
describing its particle content and interactions.
  
\subsection{Particle Content}

Supersymmetry is, under general assumptions, the unique extension of
the Poincare algebra,\index{Poincare algebra} the algebra of spacetime
translations $P$, rotations $J$, and boosts $K$.  In its simplest
form, supersymmetry extends the Poincare algebra by introducing two
additional operators $Q$ and $\bar{Q}$,\index{supersymmetry!algebra}
which satisfy
\begin{equation}
\{Q_{\alpha}, \bar{Q}_{\dot{\beta}} \} = 
2 \sigma^{\mu}_{\alpha\dot{\beta}} P_{\mu}\  , \quad 
[Q_{\alpha}, P^{\mu}] = [\bar{Q}_{\dot{\beta}}, P^{\mu}] = 0 \ ,
\label{Salgebra}
\end{equation}
where the braces denote anti-commutation, $\sigma^0$ is the $2\times
2$ identity matrix, and the $\sigma^i$ are the Pauli sigma matrices.
$Q$ and $\bar{Q}$ are two-component spinors and fermionic, and so
supersymmetry transformations relate fermions to bosons.

If supersymmetry is a fundamental symmetry of nature, all particles
must lie in representations of the supersymmetry algebra.  In
conventional quantum field theory, particles are represented by
fields, functions of the four spacetime coordinates.  The Poincare
algebra is represented by differential operators acting on these
fields.  For example,
\begin{equation}
P_x = i \frac{\partial}{\partial x} \ , \
J_x = - i \left(y \frac{\partial}{\partial z} - 
z \frac{\partial}{\partial y} \right) \ , \
K_x = i \left(t \frac{\partial}{\partial x} + 
x \frac{\partial}{\partial t} \right) \ .
\label{Spoincareops}
\end{equation}
To formulate supersymmetric field theories, we must extend this
formalism to include supersymmetry.  Not surprisingly, given the
anti-commutator appearing in \eqref{Salgebra}, this requires an
extension of the usual spacetime coordinates to include new
anti-commuting, or Grassmanian, coordinates\index{Grassmanian
coordinates} $\theta_1$, $\theta_2$, $\bar{\theta}_1$ and
$\bar{\theta}_2$.  The supersymmetry
operators\index{supersymmetry!operators} may then be identified with
the differential operators
\begin{equation}
Q_{\alpha} = \frac{\partial}{\partial \theta^\alpha} - 
i \sigma_{\alpha \dot{\beta}}^{\mu} \bar{\theta}^{\dot{\beta}}
\partial_{\mu} \ , \
\bar{Q}_{\dot{\beta}} = - \frac{\partial}{\partial 
\bar{\theta}^{\dot{\beta}}} + 
i \theta^\alpha \sigma_{\alpha \dot{\beta}}^{\mu}
\partial_{\mu} \ . \
\label{Ssusyops}
\end{equation}
It is not hard to show that the operators of
\eqsref{Spoincareops}{Ssusyops} satisfy the relations of
\eqref{Salgebra}.  These operators then act on {\em
superfields},\index{superfields} functions of $x^{\mu}$, $\theta$, and
$\bar{\theta}$, which are the natural representations of the
supersymmetry algebra.

To determine the particle content of these theories, we expand
superfields in powers of $\theta$ and $\bar{\theta}$.  Conventional
four-dimensional fields then appear as the coefficients in this
expansion.\index{supersymmetry!as an extra dimension} This is
completely analogous to the case of a field in higher spacetime
dimensions, which may be decomposed into four-dimensional fields
through a Fourier expansion in the extra dimensional coordinates.
Four-dimensional fields emerge as coefficients in this expansion,
producing a zero-mode field and a tower of Kaluza-Klein states.  The
case of supersymmetry is similar, with one important difference: the
Grassmanian properties of the $\theta$ coordinates imply that the
expansion terminates, and so superfields contain only a finite set of
four-dimensional particles.

The simplest superfields are {\em chiral
superfields},\index{superfields!chiral} independent of $\bar{\theta}$
(and anti-chiral fields, independent of $\theta$).  Expanding in
$\theta$, ones finds
\begin{equation}
\Psi(x,\theta) = \phi(x)+\sqrt{2}\, \theta^\alpha \psi_{\alpha}(x)
+ \theta \theta F(x) \ ,
\end{equation}
where $\theta \theta \equiv \theta^\alpha \theta_{\alpha}$.  A chiral
superfield then contains a chiral fermion $\psi$ and a complex scalar
$\phi$.  The $\theta$ coordinates have mass dimension $-1/2$.  The
field $F$ therefore has mass dimension 2 and no renormalizable kinetic
term.  It contains no physical degrees of freedom and is known as an
{\em auxiliary field}.\index{auxiliary fields} It may be removed by
applying its equation of motion, but it is often convenient to retain
it for reasons to be discussed below.
 
The quarks and leptons of the standard model are contained in chiral
superfields, and so supersymmetry requires the existence of scalar
superpartners, the squarks\index{squarks} and
sleptons.\index{sleptons} Similarly, the Higgs boson of the standard
model is in a chiral superfield, and so supersymmetry requires a
fermionic superpartner, the Higgsino.  Anomaly cancellation\index{MSSM
(minimal supersymmetric standard model)!anomaly cancelation} requires
that this new fermion be accompanied by additional fermions.  In the
MSSM this is accomplished by introducing an additional Higgs doublet
and the accompanying superpartners.  The fields of the chiral
supermultiplets in the MSSM, along with their quantum numbers, are
listed in Table~\ref{Stable:particles}.  Note that the gauge quantum
numbers of particles in the same supermultiplet are identical, since
standard model gauge symmetries are `internal symmetries,' independent
of spacetime symmetries.

\begin{table}[tb]
\tbl{\footnotesize Particles of the Minimal Supersymmetric Standard
Model and their SU(3), SU(2), and U(1)$_Y$ quantum numbers.  All
chiral fermions are left-handed, and the superscript $\protect c$
denotes charge conjugation.  We will also use the common alternative
notations $\tilde{e}_L \equiv \tilde{e}^-_L \equiv \tilde{e}$,
$\tilde{e}^*_R \equiv \tilde{e}^+_R \equiv \tilde{e}^c$, and so
forth. }
%\begin{center}
{\begin{tabular}{|c|l|l|}
\hline
Chiral supermultiplets  & Quarks & Squarks\\
\hline
$Q$\ $(3,2, 1/6)$ &$q =(u, d)$ 
&$\tilde{q} = (\tilde{u}, \tilde{d} )$ \rule[0.0in]{0in}{.15in}\\
$U^c$\ $(3, 1, -2/3)$ & $u^c$ & $\tilde{u}^c$ \\
$D^c$\ $(3, 1, 1/3)$  & $d^c$ & $\tilde{d}^c$ \\
\hline
& Leptons    &   Sleptons \\
\hline
$L$\ $(1,2, -1/2)$ & $l = (\nu ,e)$ &
$\tilde{l} = (\tilde{\nu}, \tilde{e} )$\rule[0.0in]{0in}{.15in}\\
$E^c$\ $(1,1,1)$ & $e^c$ & $\tilde{e}^c$\\
\hline
 & Higgs bosons & Higgsinos \\
\hline
$H_d$\ $(1,2, -1/2)$ & $(H_d^0, H_d^-)$
&  $(\tilde{H}_d^0, \tilde{H}_d^-)$ \rule[0.0in]{0in}{.15in}\\ 
$H_u$\ $(1,2, 1/2)$ & $(H_u^+ , H_u^0)$
&  $(\tilde{H}_u^+, \tilde{H}_u^0)$ \\ 
\hline
\hline
Vector supermultiplets & Gauge bosons & Gauginos \\
\hline
$(8,1,1)$ & $g$ & $\tilde{g}$ (gluino) \rule[0.0in]{0in}{.15in}\\
$(1,3,1)$ & $W^{\pm}, Z$ & $\tilde{W}^{\pm}$, $\tilde{W}^0$ (Winos)  \\
$(1,1,1)$ & $\gamma$  & $\tilde{B}$ (Bino) \\
\hline
\hline
Gravity supermultiplet & Graviton    &   Gravitino \\
\hline
$(1,1, 1)$&  $g_{\mu\nu}$ & $\tilde{G}$ \rule[0.0in]{0in}{.15in}\\
\hline
\end{tabular}}
%\end{center}
\label{Stable:particles}
\index{MSSM (minimal supersymmetric standard model)!particle content}
\index{sleptons} \index{squarks} \index{gravitino} 
\index{Higgsinos} \index{gauginos}
\end{table}

{\em Vector supermultiplets}\index{superfields!vector} are superfields
that are functions of both $\theta$ and $\bar{\theta}$.  While there
are in general nine terms in the expansion, many of them are not
physical, as they may be removed by a supersymmetric generalization of
gauge transformations.  In the end, we are left with only
\begin{equation}
V(x,\theta, \bar{\theta}) = 
\theta \sigma^{\mu}\bar{\theta} v_\mu (x) 
- i \bar{\theta} \bar{\theta} 
\theta^\alpha \lambda_\alpha (x)
+ i \theta \theta
\bar{\theta}_{\dot{\alpha}} \bar{\lambda}^{\dot{\alpha}} (x)
+ \frac{1}{2} \theta \theta
\bar{\theta} \bar{\theta} D(x) \ .
\end{equation}
A vector superfield contains a vector boson $v$ and a Majorana fermion
$\lambda$.  The field $D$ has mass dimension 2 and so, as with $F$, is
an auxiliary field with no physical degrees of freedom.  The standard
model gauge bosons are contained in vector supermultiplets, and so the
MSSM also contains their fermionic partners, the gauginos. These are
included in Table~\ref{Stable:particles}.

Finally, although not strictly a particle of the standard model, the
graviton $g_{\mu\nu}$ is the gauge boson of gravity, coupling to the
energy-momentum tensor.  Its superpartner, the gravitino
$\tilde{G}$,\index{gravitino} is a spin 3/2 particle that couples to
the supercurrent $J^{\mu}_{Q}$,\index{supercurrent} which generates
supersymmetry.  The gravitino plays an important role in the
phenomenology of some supersymmetric models.

\subsection{Supersymmetric Matter Interactions}
\label{Ssec:matter}

Given the particles of Table~\ref{Stable:particles}, we may then
construct the most general supersymmetric Lagrangian. To construct the
matter interactions,\index{interactions!matter} it is convenient to
define the {\em superpotential}\index{superpotential}
\begin{equation}
W = \frac{1}{2} \mu_{ij} \Psi_i \Psi_j 
+ \frac{1}{3} y_{ijk} \Psi_i \Psi_j \Psi_k \ ,
\end{equation}
a gauge-invariant function of chiral superfields.  The most general
matter coupling is then compactly expressed in superfield notation as
\begin{equation}
{\cal L}_{\text{matter}} = 
\Psi^\dagger_i \Psi_i |_{\theta \theta \bar{\theta} \bar{\theta}} 
+ \left[ W|_{\theta \theta} + \text{H.c.} \right] ,
\label{Scomponents}
\end{equation}
where one takes the components in the $\theta$, $\bar{\theta}$
expansion as indicated.  The trilinear superpotential terms generate
the Yukawa couplings of the standard model with dimensionless
couplings $y_{ijk}$, along with other interactions related by
supersymmetry.  The parameters $\mu_{ij}$ have mass dimension
1.\index{$\mu$ parameter, definition}

In component fields, \eqref{Scomponents} becomes
\begin{eqnarray}
\lefteqn{{\cal L}_{\text{matter}} = 
\partial^\mu \phi_i^* \partial_\mu \phi_i
+ i \bar{\psi}_i \bar{\sigma}^{\mu} \partial_{\mu} \psi_i
+ F_i^* F_i} \nonumber \\
&& + \left[\mu_{ij} \left( \phi_i F_j - \frac{1}{2} \psi_i \psi_j \right)
+ y_{ijk} (\phi_i \phi_j F_k - \phi_i \psi_j \psi_k ) + \text{H.c.}
\right] .
\end{eqnarray}
The first line includes canonical kinetic terms for every $\psi_i$ and
$\phi_i$.  As discussed above, however, there are no corresponding
terms for the auxiliary fields $F_i$.  These may therefore be removed
by their equations of motion
\begin{equation}
F_i^* = -\mu_{ij} \phi_j - y_{ijk} \phi_j \phi_k \ .
\end{equation}
With this substitution, the complete matter Lagrangian becomes
\begin{eqnarray}
\lefteqn{{\cal L}_{\text{matter}} = 
\partial^\mu \phi_i^* \partial_\mu \phi_i
+ i \bar{\psi}_i \bar{\sigma}^{\mu} \partial_{\mu} \psi_i
- \frac{1}{2} \mu_{ij} \psi_i \psi_j 
- \frac{1}{2} \mu_{ij}^* \bar{\psi}_i \bar{\psi}_j} \nonumber \\
&& - y_{ijk} \phi_i \psi_j \psi_k  
   - y_{ijk}^* \phi_i^* \bar{\psi}_j \bar{\psi}_k 
     - \left| \mu_{ij} \phi_j + y_{ijk} \phi_j \phi_k \right| ^2 \ .
\label{SLmatter}
\end{eqnarray}
Terms such as the last one in \eqref{SLmatter} are called {\em
$F$-terms},\index{$F$-terms} as they arise from $|F_i|^2$.

{}From \eqref{SLmatter} we find that both scalar and fermion masses
are determined by $\mu$, and they are
degenerate.\index{superpartners!mass degeneracy of} This is a generic
prediction of exact supersymmetry.  The absence of a boson with the
couplings and mass of the electron implies that supersymmetry must be
broken.\index{supersymmetry breaking!necessity of} In addition, we see
that there are trilinear and quartic scalar couplings.  The quartic
couplings are completely determined by the Yukawa couplings $y$.  For
this reason, the quadratically-divergent contributions to the Higgs
boson mass shown in Fig.~\ref{Sfig:Higgs} exactly cancel, as required
for supersymmetry to stabilize the gauge
hierarchy.\index{problems!gauge hierarchy}

\begin{figure}[tbp]
\begin{minipage}[t]{0.49\textwidth}
\postscript{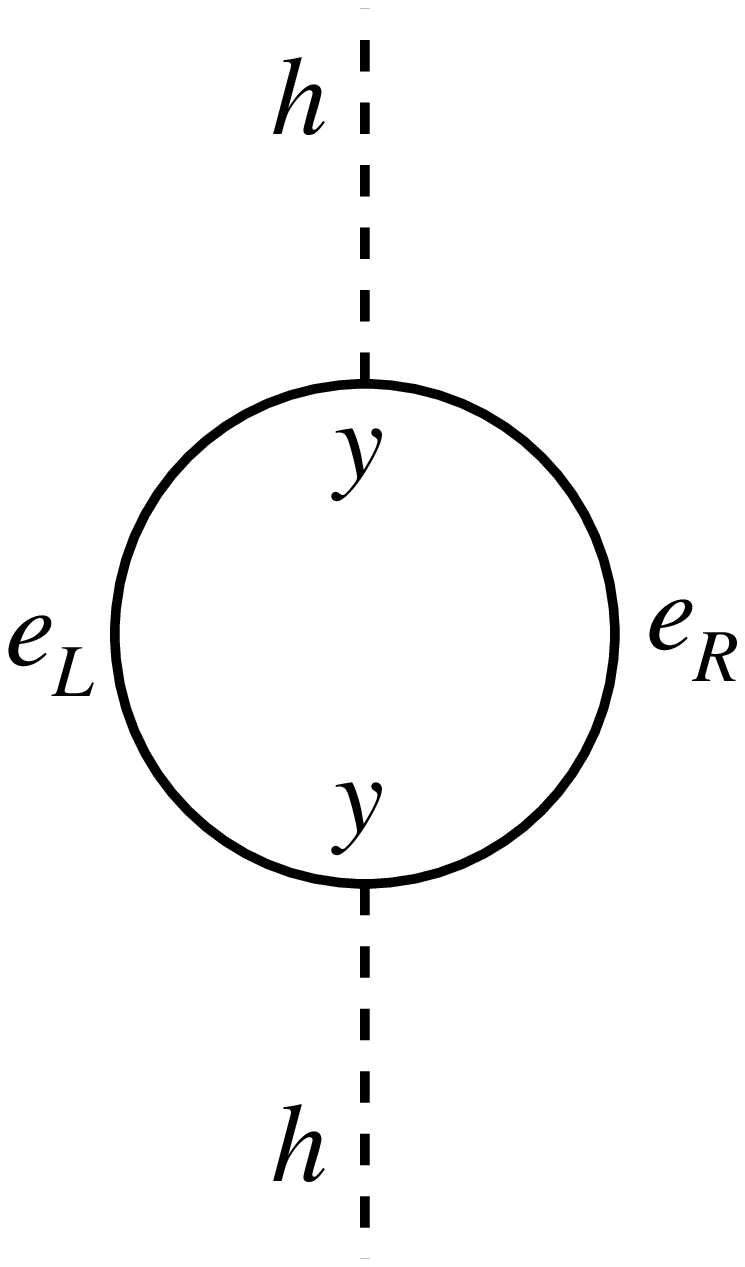}{0.50}
\end{minipage}
\hfill
\begin{minipage}[t]{0.49\textwidth}
\postscript{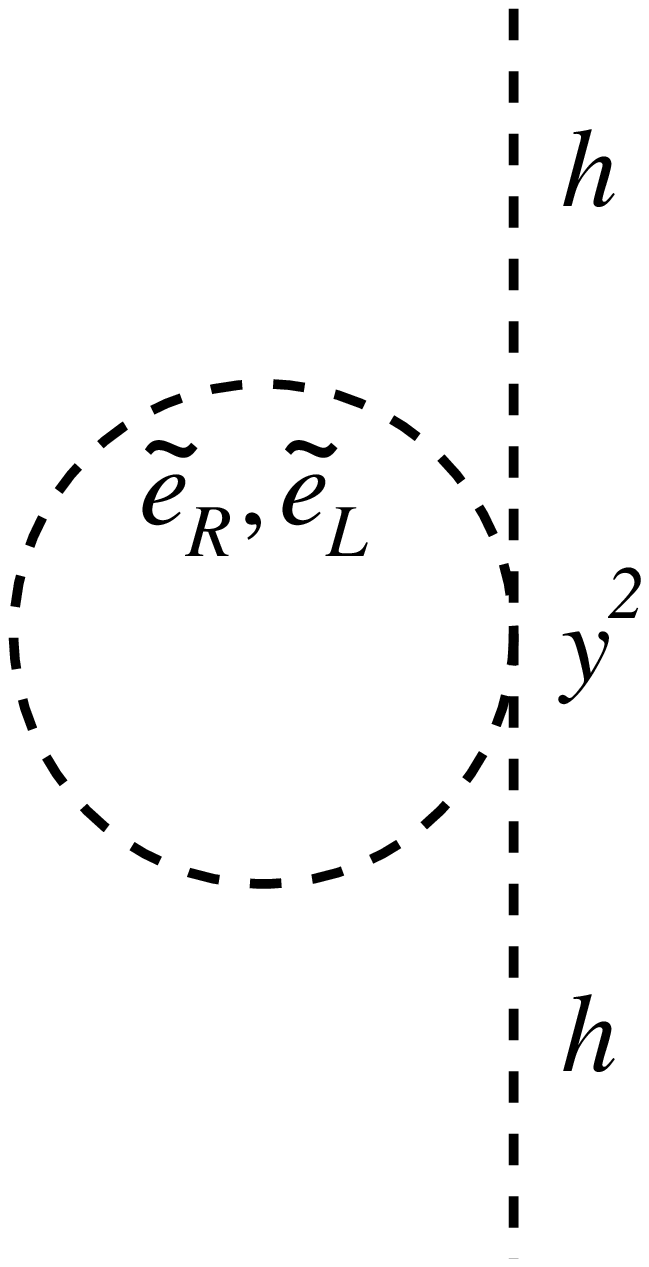}{0.42}
\end{minipage}
\caption{\footnotesize Quadratically-divergent contributions to the
Higgs boson mass in supersymmetry.  These cancel exactly in
the limit of exact supersymmetry.
\label{Sfig:Higgs}}
\end{figure}

The superpotential of the MSSM\index{MSSM (minimal supersymmetric
standard model)!superpotential} is
\begin{equation}
W= \epsilon_{ab} \left[ 
- y^u_{ij} H^a_u Q^b_i U^c_j + y^d_{ij} H^a_d Q^b_i D^c_j 
+ y^e_{ij} H^a_d L^b_i E^c_j - \mu H_u^a H_d^b \right] \ ,
\label{SW}
\end{equation}
where $a$, $b$ are SU(2) indices contracted by the anti-symmetric
$\epsilon$ tensor with $\epsilon_{12} = 1$, and $i$, $j$ are
generational indices.  As in Table~\ref{Stable:particles}, the
superscript $c$ denotes charge conjugation.  Gauge invariance forbids
bilinear terms in the standard model, but the additional Higgs doublet
of the MSSM, $H_u$, may have bilinear couplings.  Linear terms are
forbidden by gauge invariance in the MSSM, but may be present in
extensions of the MSSM that contain complete gauge singlets.
Additional non-renormalizable couplings may also be included in
extensions of the MSSM with quartic and higher order superpotential
terms suppressed by some heavy mass scale.

\eqref{SW} is {\em not} the most general gauge invariant
superpotential.  For example, as evident in
Table~\ref{Stable:particles}, the superfields $L_i$ and $H_d$ have
identical quantum numbers, leading to additional possible terms.  The
complete list of omitted terms is\index{$R$-parity!violating terms}
\begin{eqnarray}
W_L &=&  \lambda_{ijk} L_i L_j E^c_k + \mu_i L_i H_u \nonumber \\
W_{LB} &= &\lambda'_{ijk} L_i Q_j D^c_k \nonumber \\
W_B &=& \lambda''_{ijk} U^c_i D^c_j D^c_k \ .
\label{SRp}
\end{eqnarray}
$W_L$ violates lepton number $L$, $W_B$ violates baryon number $B$,
and $W_{LB}$ violates both $L$ and $B$.  If all terms are present,
protons decay through renormalizable operators, and present bounds
from proton decay are extremely stringent.\index{$R$-parity!and proton
decay} All of these terms may be eliminated by assigning to each
particle an {\em $R$-parity}\index{$R$-parity!definition} $R =
(-1)^{3(B-L)+2S}$, where $S$ is spin, and requiring that $R$-parity be
multiplicatively conserved.  All standard model particles have $R=1$
and all superpartners have $R=-1$, and so all terms of \eqref{SRp} are
eliminated.  $R$-parity conservation\index{$R$-parity!conservation}
implies that the lightest supersymmetric particle (LSP)\index{LSP
(lightest supersymmetric particle)} is stable, with strong
implications for collider studies and dark matter.  We will focus on
$R$-parity conserving supersymmetry throughout this review.

\subsection{Supersymmetric Gauge Interactions}
\label{Ssec:gauge}

The interactions of the MSSM that involve gauge fields fall into
several categories.  The terms involving only gauge
fields\index{interactions!pure gauge} are
\begin{equation}
- \frac{1}{4} F^a_{\mu\nu} F^{a\, \mu\nu} 
+ i \bar{\lambda}^a \bar{\sigma}^{\mu} \partial_{\mu}
 \lambda^a + \frac{1}{2} {D^a}^2
\end{equation}
for each gauge group, where $a$ labels gauge group generators.  The
first two terms are the canonical kinetic terms for gauge bosons and
gauginos.  As with the $F$ field, $D$ has no kinetic term and may be
eliminated by its equation of motion.  This leads to the quartic
scalar couplings
\begin{equation}
V_D=\frac{g_s^2}{2} \Big[ \sum_{a, ij} \phi^{\dagger}_i 
T^a_{3, ij} \phi_j \Big]^2
+ \frac{g^2}{2} \Big[ \sum_{a, ij} \phi^{\dagger}_i 
T^a_{2, ij} \phi_j \Big]^2
+ \frac{g'^2}{2} \Big[ \sum_i \phi^{\dagger}_i Y_i \phi_i \Big]^2 \ ,
\label{SDterms}
\end{equation}
where $T_3$ and $T_2$ are the generators of SU(3) and SU(2), and $Y$
is hypercharge.  These are known as {\em $D$-terms}.\index{$D$-terms}
As in \secref{Ssec:matter}, these quartic interactions are completely
determined by standard model couplings --- in this case, the gauge
couplings.  A host of quadratically-divergent contributions to the
Higgs mass that involve gauge interactions are thereby exactly
canceled by supersymmetric contributions, allowing supersymmetry to
completely stabilize the gauge hierarchy.\index{problems!gauge
hierarchy}

The gauge-matter couplings\index{interactions!gauge (gaugino)-matter}
include $v_\mu \phi \partial^{\mu}\phi^*$ couplings, as well as
four-point $v^{\mu} v_{\mu} \phi \phi^*$ interactions, where $v^{\mu}$
denotes gauge bosons.  Finally, and perhaps of greatest significance
for linear collider studies, there are sfermion-fermion-gaugino
couplings, the supersymmetric analogues to the fermion-fermion-gauge
boson couplings.  These are of the form
\begin{eqnarray}
\sqrt{2} g \phi_i T^a_{ij} \psi_j \lambda^a + \text{H.c.} \ ,
\label{Sgauginoterms}
\end{eqnarray}
where the coupling $g$ is a standard model gauge coupling.
Supersymmetry relates the couplings of gauginos to the couplings of
gauge bosons.\index{superpartners!coupling equivalence of} As we will
discuss in \secref{Ssec:testsusy}, this fact may be exploited to
definitively identify new particles as superpartners, as well as to
determine the mass scales of undiscovered superpartners.

\subsection{Supersymmetry-breaking Terms}
\label{Ssec:soft}

So far we have been discussing the purely supersymmetric part of the
MSSM Lagrangian. However, as noted above, supersymmetry predicts that
superpartners have equal mass.\index{superpartners!mass degeneracy of}
This is also evident from the supersymmetry algebra:~the supersymmetry
generator commutes with the momentum operator, and so
\begin{equation}
(H Q - QH) | \phi \rangle = (E_\psi-E_\phi) | \psi \rangle = 0 \ .
\end{equation}
Because we have not found superpartners degenerate with the known
particles, supersymmetry must be broken.\index{supersymmetry
breaking!necessity of}

Supersymmetry may be broken in many ways.  However, to preserve the
strong phenomenological motivation of solving the gauge hierarchy
problem,\index{problems!gauge hierarchy} it is imperative that the
quadratic divergences, so elegantly removed by exact supersymmetry,
are not re-introduced.  When we require this, the allowed terms, known
as {\em soft supersymmetry-breaking terms},\index{supersymmetry
breaking!soft terms} are limited to four classes:~gaugino masses $M
\lambda \lambda$,\index{gaugino masses} scalar masses $m^2 \phi^*
\phi$,\index{scalar masses} trilinear scalar couplings $A \phi \phi
\phi$ ({\em $A$-terms}),\index{interactions!trilinear
scalar}\index{$A$-terms} and bilinear scalar couplings $B \phi \phi$
({\em $B$-terms}).\index{interactions!bilinear scalar}\index{$B$-terms}
No new dimensionless couplings are introduced, preserving the
cancellations discussed in Secs.~\ref{Ssec:matter} and
\ref{Ssec:gauge}.  Also note, however, that not all dimensionful
contributions are allowed; for example, terms of the form $\phi_i^*
\phi_j \phi_k$ are not soft.\index{interactions!dimension 3 but not
soft}

In the MSSM, the full set of soft terms\index{MSSM (minimal
supersymmetric standard model)!soft terms} is
\begin{eqnarray}
\lefteqn{- {\cal L}_{\text{soft}} = \frac{1}{2} ( 
M_1 \tilde{B} \tilde{B} + M_2 \tilde{W}^a \tilde{W}^a +
M_3 \tilde{g}^a \tilde{g}^a ) } \nonumber \\
&& + m^2_{Q\, ij} ( \tilde{u}_{Li}^* \tilde{u}_{Lj}
+ \tilde{d}_{Li}^* \tilde{d}_{Lj} )
+ m^2_{U\, ij} \tilde{u}_{Ri}^* \tilde{u}_{Rj}
+ m^2_{D\, ij} \tilde{d}_{Ri}^* \tilde{d}_{Rj} \nonumber \\
&& + m^2_{L\, ij} \left(\tilde{\nu}_{Li}^* \tilde{\nu}_{Lj}
+ \tilde{e}_{Li}^* \tilde{e}_{Lj} \right)
+ m^2_{E\, ij} \tilde{e}_{Ri}^* \tilde{e}_{Rj} \nonumber \\
&& + \epsilon_{ab} [ 
- \tilde{A}^u_{ij} H^a_u \tilde{q}^b_i \tilde{u}_{Rj}^*
+ \tilde{A}^d_{ij} H^a_d \tilde{q}^b_i \tilde{d}_{Rj}^*
+ \tilde{A}^e_{ij} H^a_d \tilde{l}^b_i \tilde{e}_{Rj}^* \nonumber \\
&& \qquad \quad
- B H_u^a H_d^b + \text{H.c.} ] \ . 
\label{Ssoft}
\end{eqnarray}
Gauge invariance requires that scalar particles in the same
electroweak doublet receive the same soft masses.  The $A$- and
$B$-terms mimic the Yukawa and $\mu$ terms of the superpotential.  If
$R$-parity is broken, there are additional soft terms corresponding to
the superpotential terms of \eqref{SRp}.\index{$R$-parity!violation}

What is the origin of these soft terms?  At first sight, supersymmetry
has just introduced a bewildering array of new parameters.  If
supersymmetry is to take us closer to a fundamental theory, we would
hope that these parameters are not completely arbitrary.  In fact,
however, blind hope is not required! Experimental constraints already
guarantee that, if weak-scale supersymmetry exists, the soft terms
cannot be generic.  These constraints, some of which will be reviewed
in \secref{Ssec:successes}, imply that some pattern must emerge.
Further, we will see in \secref{Ssec:models} that the soft terms
contain an imprint of physics at even more microscopic scales, and
their determination may shed light on energy scales as high as the
Planck scale.  Just as discovering the mechanism of spontaneous gauge
symmetry breaking is one of the key issues in standard model physics,
uncovering the mechanism of supersymmetry breaking is the central
issue of supersymmetric physics.  The promise of model-independent
measurements of the soft parameters is the prime supersymmetric
motivation for the linear collider.

\subsection{Sleptons}
\label{Ssec:sleptons}

Given both the supersymmetric and the soft supersymmetry-breaking
terms, we are now ready to determine the mass eigenstates of the
MSSM.\index{sleptons} For lack of space, we focus on those with
electroweak quantum numbers only, although squarks\index{squarks} and
gluinos\index{gluinos} may also be produced at linear colliders with
many interesting results (see, for example,
Refs.~\cite{Feng:1994sd,Baer:1996vd,Bartl:1997yi,Drees:1999yz,%
Bartl:2000kw,Kitano:2002ss}).

We first discuss the scalar superpartners.  There are six charged
slepton states.  These receive masses from soft terms and also, after
electroweak symmetry breaking, from Yukawa couplings, $F$-terms, and
$D$-terms.  The full set of slepton mass terms\index{sleptons!mass
matrix!$6\times 6$ form} is $\tilde{l}^\dagger {\cal
M}_{\tilde{l}}^2\, \tilde{l}$, where $\tilde{l}^T = (\tilde{e}_L,
\tilde{\mu}_L, \tilde{\tau}_L, \tilde{e}_R, \tilde{\mu}_R,
\tilde{\tau}_R)$, and the $6\times 6$ mass matrix is
\begin{equation}
{\cal M}_{\tilde{l}}^2 = 
\left( \begin{array}{cc}
m^2_{LL}             &m^2_{LR} \\
m^{2\, \dagger}_{LR} & m^2_{RR} \end{array}
\right) .
\label{Sfullslepton}
\end{equation}
The diagonal $3\times3$ blocks\index{sleptons!mass matrix!$3\times 3$
form} are
\begin{eqnarray}
\! \! \! \! \! \! \! \! \! \! 
m^2_{LL} \! \! \! \! \! &=& \! \! \! \! \! 
\left( \begin{array}{ccc}
\! \! m_{L\, 11}^2+m_e^2+\Delta_L^2 \! \! \! 
& m_{L\, 12}^2 & m_{L\, 13}^2 \\
m_{L\, 21}^2 & \! \! \! m_{L\, 22}^2+m_{\mu}^2+\Delta_L^2 \! \! \!
& m_{L\, 23}^2 \\
m_{L\, 31}^2 & m_{L\, 32}^2 
& \! \! \! m_{L\, 33}^2+m_{\tau}^2+ \Delta_L^2 \! \!
\end{array}
\right) \nonumber \\
\! \! \! \! \! \! \! \! \! \! 
m^2_{RR} \! \! \! \! \! &=& \! \! \! \! \! 
\left( \begin{array}{ccc}
\! \! m_{E\, 11}^2+m_e^2+\Delta_R^2 \! \! \! 
& m_{E\, 12}^2 & m_{E\, 13}^2 \\
m_{E\, 21}^2 & \! \! \! m_{E\, 22}^2+m_{\mu}^2+\Delta_R^2 \! \! \!
& m_{E\, 23}^2 \\
m_{E\, 31}^2 & m_{E\, 32}^2 
& \! \! \! m_{E\, 33}^2+m_{\tau}^2+ \Delta_R^2 \! \!
\end{array}
\right) \! ,
\label{Smsleptondiagonal}
\end{eqnarray}
where $\Delta_L^2 = m_Z^2 (-\frac{1}{2} + \sin^2\theta_W) \cos 2\beta$
and $\Delta_R^2 = m_Z^2 (- \sin^2\theta_W) \cos 2\beta$.  The
$\Delta_{L,R}^2$ contributions originate from gauge interactions, the
$D$-terms of \eqref{SDterms}, and are therefore flavor-diagonal.
While they contain no supersymmetry-breaking parameters, they require
electroweak symmetry breaking, which in turn relies on
supersymmetry-breaking soft masses for the Higgs scalars.  The
off-diagonal blocks are given by
\begin{equation}
m^2_{LR} = - \left( \begin{array}{ccc}
\! \! \! \tilde{A}^e_{11} v_d + \mu y^e_{11} v_u \!
& \! \tilde{A}^e_{12} v_d + \mu y^e_{12} v_u \!
& \! \tilde{A}^e_{13} v_d + \mu y^e_{13} v_u \! \! \! \\
\! \! \! \tilde{A}^e_{21} v_d + \mu y^e_{21} v_u \! 
& \! \tilde{A}^e_{22} v_d + \mu y^e_{22} v_u \! 
& \! \tilde{A}^e_{23} v_d + \mu y^e_{23} v_u \!  \! \! \\
\! \! \! \tilde{A}^e_{31} v_d + \mu y^e_{31} v_u \!
& \! \tilde{A}^e_{32} v_d + \mu y^e_{32} v_u \! 
& \! \tilde{A}^e_{33} v_d + \mu y^e_{33} v_u \! \! \! 
\end{array}
\right) .
\end{equation}
Finally, the sneutrino masses are $\tilde{\nu}^\dagger {\cal
M}_{\tilde{\nu}}\, \tilde{\nu}$, where $\tilde{\nu}^T =
(\tilde{\nu}_e, \tilde{\nu}_\mu, \tilde{\nu}_\tau)$,
and\index{sneutrinos!mass matrix}
\begin{equation}
{\cal M}^2_{\tilde{\nu}} =
\left( \begin{array}{ccc}
m_{L\, 11}^2 + \Delta_{\nu}^2  & m_{L\, 12}^2 & m_{L\, 13}^2 \\
m_{L\, 21}^2 & m_{L\, 22}^2 + \Delta_{\nu}^2 & m_{L\, 23}^2 \\
m_{L\, 31}^2 & m_{L\, 32}^2 & m_{L\, 33}^2 + \Delta_{\nu}^2 
\end{array}
\right) ,
\label{Smsnu}
\end{equation}
where $\Delta_{\nu}^2 = m_Z^2 \frac{1}{2} \cos 2\beta$.

In general, we see that the sleptons masses are complicated.  Note
that, while it is always possible to rotate to a basis where the
Yukawa couplings $y^e$ are diagonal, there is no reason for these
rotations to simultaneously diagonalize the soft term matrices
$m^2_L$, $m^2_E$, and $\tilde{A}^e$.\index{flavor violation!is
generic} Studies of slepton flavor mixing may shed light on
fundamental issues, such as the masses of standard model quarks and
leptons, and will be discussed in
\secref{Ssec:lfv}.\index{sleptons!mixing!flavor}

Ultimately one hopes to study the sleptons in full generality at
colliders.  However, in studies so far, simplifying assumptions are
typically made to reduce the problem to a more manageable form.  A
common assumption is flavor conservation, so that $y^e$, $m^2_L$,
$m^2_E$, and $\tilde{A}^e$ are all simultaneously diagonalizable.  The
$6 \times 6$ matrix then reduces to three $2\times
2$\index{sleptons!mass matrix!$2\times 2$ form} blocks of the
form\index{staus!mass matrix}
\begin{equation}
m^2_{\tilde{\tau}} = 
\left(\begin{array}{cc}
m^2_{L\, 33} + m_\tau^2 + \Delta_L^2 
& -m_{\tau} (A_\tau + \mu \tan\beta) \\
-m_{\tau} (A_\tau + \mu \tan\beta) 
&m^2_{E\, 33} + m_\tau^2 + \Delta_R^2  
\end{array}\right) ,
\label{staumassmatrix}
\end{equation}
and similarly for the first and second generations.  The $\tilde{A}^e$
parameters have been replaced by $A$ parameters, following convention.
A common assumption is that $A_e$, $A_\mu$, and $A_\tau$ are of the
same order.  This is motivated by some forms of supersymmetry
breaking, but is by no means a universal requirement.  With this
assumption, bounds on $|A_\tau|$ from the requirement that there be no
tachyonic staus typically imply that left-right mixing in selectrons
and smuons is insignificant.  However, left-right
mixing\index{sleptons!mixing!left-right} may be important for the
staus.  In fact, although suppressed by $m_{\tau}$, the factor
$\mu\tan\beta$ leads to large mixing in many models.  In general, the
stau mass eigenstates and eigenvalues are\index{staus!mass eigenstates}
\begin{eqnarray}
\tilde{\tau}_1&=&\tilde{\tau}_L\cos\theta_{\tau}
+ \tilde{\tau}_R \sin\theta_{\tau} \nonumber \\
\tilde{\tau}_2&=&-\tilde{\tau}_L\sin\theta_{\tau}+
 \tilde{\tau}_R \cos\theta_{\tau}
\end{eqnarray}
and 
\begin{equation}
m^2_{\tilde{\tau}_1,\tilde{\tau}_2}=\frac{1}{2}
\left[
m^2_{\tilde{\tau}\, LL} + m^2_{\tilde{\tau}\, RR} \mp
\sqrt{ (m^2_{\tilde{\tau}\, LL} - m^2_{\tilde{\tau}\, RR})^2
+ 4 (m^2_{\tilde{\tau}\, LR})^2 } \right] .
\end{equation}
As we will see, when the slepton masses are assumed to be unified at a
high scale, renormalization group effects imply that staus are the
lightest sleptons and so are of great importance in collider studies.

Finally, in the flavor conserving case,
\eqsref{Smsleptondiagonal}{Smsnu} imply
\begin{equation}
m^2_{\tilde{e}_L}-m^2_{\tilde{\nu}_e} = - m_W^2 \cos 2\beta \ .
\label{sleptonsplitting}
\end{equation}
This mass relation\index{sleptons!$\tilde{l}_L$-$\tilde{\nu}_l$ mass
splitting} is an extremely robust prediction, resulting only from
gauge symmetry and supersymmetry in the MSSM.  It therefore provides a
non-trivial test that newly-discovered scalars must pass to be
identified as left-handed sleptons.\index{tests!of
$\tilde{l}_L$-$\tilde{\nu}_l$ mass relations}

\subsection{Charginos and Neutralinos}

We now turn to the fermionic superpartners.  The charged fermions are
Winos and Higgsinos.\index{charginos}\index{neutralinos} Without
electroweak breaking, the Winos have soft mass $M_2$ from
\eqref{Ssoft}, and the Higgsinos get mass through the superpotential
term $-\mu H_u H_d$ of \eqref{SW}.  Once electroweak symmetry is
broken, however, the Higgs-Wino-Higgsino interactions of
\eqref{Sgauginoterms} generate Wino-Higgsino mixing. In the standard
model, proper electroweak symmetry breaking\index{electroweak symmetry
breaking} requires a Higgs vacuum expectation value (vev) $\langle h
\rangle^2 = v^2/2$, where $(g^2+g'^2) v^2 / 2 = m_Z^2$.  As noted
above, however, the MSSM is a two Higgs doublet model, and so the vev
$v$ may be shared between both neutral Higgses: $\langle H_u^0
\rangle^2 + \langle H_d^0 \rangle^2 = v^2/2$.  This freedom is
typically parametrized by $\tan\beta$,\index{$\tan\beta$!definition}
where
\begin{equation}
\langle H^0_u \rangle = \frac{v}{\sqrt{2}} \sin\beta \ , \quad
\langle H^0_d \rangle = \frac{v}{\sqrt{2}} \cos\beta \ .
\end{equation}
With this parametrization, the charged mass terms are ${\psi^-}^T
{\cal M}_C\, \psi^+ + \text{H.c.}$, where ${\psi^{\pm}}^T =
(-i\tilde{W}^{\pm}, \tilde{H}^{\pm})$ and
\begin{equation}
\label{Schamass}
{\cal M}_C = \left( \begin{array}{cc}
 M_2                    &\sqrt{2} \, m_W\sin\beta \\
\sqrt{2} \, m_W\cos\beta   &\mu     \end{array} \right) .
\end{equation}
${\cal M}_C$\index{charginos!mass matrix} may be diagonalized by
unitary matrices $U$ and $V$ through $M_D = U^* {\cal M}_C\,
V^{-1}$. If ${\cal M}_C$ is real, $U$ and $V$\index{charginos!mixing
matrices} may also be chosen real and may be parametrized as
\begin{equation}
U=\left(\begin{array}{cc}
\cos\phi_- &\sin\phi_-  \\
-\sin\phi_- &\cos\phi_-
\end{array}\right),\ \ \
V=\left( \begin{array}{cc}
\cos\phi_+ &\sin\phi_+    \\
-\sin\phi_+ &\cos\phi_+
\end{array}
\right) .
\label{UVmixingangles}
\end{equation}
The resulting mass eigenstates are called {\em
charginos}\index{charginos} and are $\tilde{\chi}^-_i = U_{ij}
\psi^-_j$ and $\tilde{\chi}^+_i = V_{ij}\psi^+_j$.  In the limit where
$M_2, |\mu| \gg m_W$ the mass eigenstates become nearly pure Winos and
Higgsinos.\footnote{We assume that gaugino masses $M_i$ are positive
and real, unless otherwise stated.}  The mixing angles are sensitive
to $\tan\beta$ when $\tan\beta \sim 1$.

The neutral fermionic superpartners are the Bino, the neutral Wino,
and the two neutral Higgsinos.  As with charginos, electroweak
symmetry breaking generates gaugino-Higgsino mixing.  The resulting
mass eigenstates are called {\em neutralinos}.\index{neutralinos} The
neutralino mass matrix\index{neutralinos!mass matrix} is $\frac{1}{2}
{\psi^0}^T {\cal M}_N\, \psi^0 + \text{H.c.}$, where ${\psi^0}^T =
(-i\tilde{B},-i\tilde{W}^3, \tilde{H}_d^0, \tilde{H}_u^0)$ and
\begin{equation}
\label{Sneumass}
{\cal M}_N =
\left( \begin{array}{cccc}
M_1             &0            &\! \! -m_Z \cosb\, s_W & m_Z \sinb\, s_W \\
0               &M_2          & \! \! m_Z \cosb\, c_W &-m_Z \sinb\, c_W \\
-m_Z \cosb\, s_W  & \! \! m_Z \cosb\, c_W &0            &-\mu           \\
 m_Z \sinb\, s_W  &\! \! -m_Z \sinb\, c_W &-\mu         &0     \end{array}
\right) ,
\end{equation}
and we have introduced the shorthand notation $s_W \equiv
\sin\theta_W$, $c_W \equiv \cos\theta_W$, $s_\beta \equiv \sin\beta$,
and $c_\beta \equiv \cos\beta$.  The neutralino mass eigenstates are
$\chn_i = N_{ij} \psi^0_j$, where $N$ diagonalizes ${\cal
M}_N$.\index{neutralinos!mixing matrix}

In order of increasing mass, the two charginos are labeled $\chc_1$
and $\chc_2$, and the four neutralinos are $\chn_1$, $\chn_2$,
$\chn_3$, and $\chn_4$.  There are several phenomenologically
interesting limits for the chargino and neutralino mass spectrum:
\begin{itemize}
\item $M_1 < M_2 < |\mu|$:~``Bino LSP scenarios.'' As we will discuss
in \secref{Ssec:msugra}, this case is realized in much, but not all,
of minimal supergravity parameter space.  The lightest state is
$\chn_1 \approx \tilde{B}$, followed by a nearly degenerate triplet,
$\tilde{\chi}^0_2 \approx \tilde{W}^0$ and $\tilde{\chi}^{\pm}_1
\approx \tilde{W}^{\pm}$.  There is typically a large spacing between
the lightest state and all others.
\item $|\mu| < M_1, M_2$:~``Higgsino LSP scenarios.''
$\tilde{\chi}^0_1$, $\tilde{\chi}^0_2$, $\tilde{\chi}^+_1$ are all
fairly degenerate in mass and Higgsino-like.  Radiative corrections
are important in determining the mass differences.
\item $M_2 < M_1, |\mu|$:~``Wino LSP scenarios.''  These are realized
in a variety of models, including some anomaly-mediation models.  (See
\secref{Ssec:amsb}.)  $\tilde{\chi}^0_1$ and $\tilde{\chi}^{\pm}_1$
are Wino-like and extremely mass degenerate, and the charginos may
have an observable lifetime before decaying to the neutralino.  In
this case, radiative corrections are essential in determining the mass
difference.
\end{itemize}

\section{Successes and Puzzles}
\label{Ssec:successes}

The discovery of weak-scale supersymmetry will be accompanied by many
fundamental questions, such as those listed in
\secref{Ssec:intro}. Here we expand upon a number of these, both to
highlight some important goals of the linear collider program and also
to motivate the array of models to be described in
\secref{Ssec:models}.

\subsection{Unification}
\label{Ssec:unification}

The unification of forces requires that gauge couplings unify at some
scale.\index{unification!gauge coupling} The renormalization group
evolutions of gauge couplings in the standard model and the MSSM are
shown in Fig.~\ref{Sfig:running}.\index{renormalization
group!trajectories} For the MSSM, all sparticles are assumed to lie
between 250 GeV and 1 TeV.  In the standard model, the gauge couplings
obviously do not unify.  In the MSSM, however, they unify at the 1\%
level.  This is widely regarded as the strongest quantitative
motivation for any framework for physics beyond the standard model
proposed to date.

\begin{figure}[tb]
\postscript{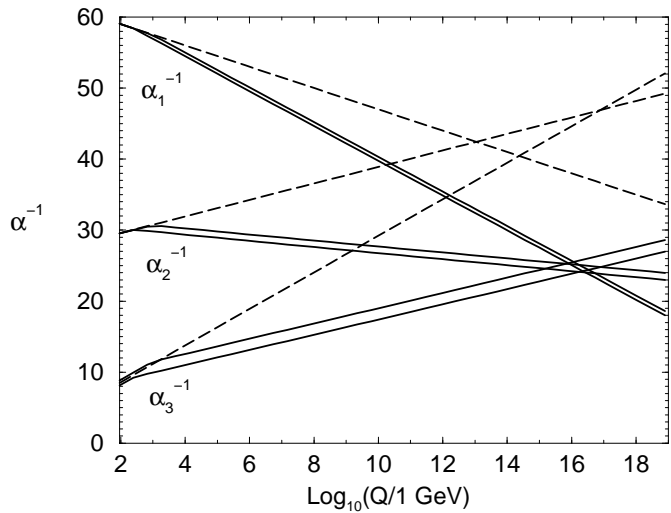}{0.80}
\caption{\footnotesize The renormalization group evolution of the
three gauge coupling constants in the standard model (dashed) and in
the MSSM (solid)~\protect\cite{Martin:1997ns}.
\label{Sfig:running}}
\end{figure}

The fact that the three couplings meet in supersymmetry is
significant, but so is where they meet.\index{unification!gauge
coupling!signficance of} The couplings unify with value
$\alpha_{\text{unif}} \approx 1/24$ at $\mgut \simeq 2\times
10^{16}~\gev$.  If the couplings had unified at $\alpha \agt 1$,
unification would occur beyond the perturbative regime.  At the same
time, unification at scales above $\mplanck \simeq 10^{19}~\gev$ would
be invalidated by quantum gravity effects, while unification at scales
$Q \alt 10^{16}~\gev$ would imply proton decay at levels in conflict
with experiment. (See \secref{Ssec:proton}.)  In fact, the proximity
of $\mgut$ to $\mplanck$ gives hope that the effective gravitational
coupling, $\alpha_{\text{grav}} \sim E^2/\mplanck^2$ also unifies with
the standard model gauge couplings near $Q \sim \mplanck$, as is
motivated in string theory.  The discrepancy between $\mgut$ and
$\mplanck$ may be resolved by a variety of
mechanisms~\cite{Dienes:1996du,Horava:1996ma}.\index{Planck
scale!discrepancy with GUT scale}

The unification of gauge couplings has immediate implications for
the linear collider.  Simple gauge group unification requires also
gaugino mass unification.  The quantities $M_i / g_i^2$ are invariants
of one-loop renormalization group evolution. (Two-loop effects
typically lead to small deviations~\cite{Martin:1997ns}.) If all gauge
couplings and gaugino masses are unified at any high scale, one
therefore expects\index{unification!gaugino mass}
\begin{equation}
M_1 : M_2 : M_3 = g_1^2 : g_2^2 : g_3^2 \approx 1 : 2 : 7
\label{Sgauginomasses}
\end{equation}
at lower scales, where $g_1$, $g_2$, and $g_3$ are the hypercharge,
weak, and strong couplings.  The model-independent measurement of
gaugino mass parameters is then of great importance, and the
verification of $2M_1 \approx M_2$ would provide strong supporting
evidence for unification.

Gauge group unification also implies that the particle content of the
MSSM lies in representations of a larger gauge
group.\index{unification!matter representations} There is already
tantalizing evidence for this --- although the quantum numbers of the
standard model particles shown in Table~\ref{Stable:particles} appear
rather arbitrary, they are consistent with their placement in
{\boldmath $\overline{5}$} + {\boldmath $10$} representations of
SU(5), or, with the addition of a right-handed neutrino, in the spinor
representation {\boldmath $16$} of SO(10). These unifications imply
relations between Yukawa couplings and soft supersymmetry-breaking
terms of particles in the same grand unified theory (GUT) multiplets,
providing additional motivations for model-independent measurements.

\subsection{Dark Matter}
\label{Ssec:dm}

As noted in \secref{Ssec:matter}, the most general gauge-invariant
supersymmetric standard model allows proton decay through
renormalizable operators, in gross violation of proton lifetime
limits.\index{dark matter} $R$-parity conservation is one elegant way
to prevent this.\index{$R$-parity!conservation}\index{$R$-parity!and
dark matter} This has the immediate consequence that the lightest
supersymmetric particle (LSP) is stable.\index{LSP (lightest
supersymmetric particle)}\index{dark matter!and LSP}\index{LSP
(lightest supersymmetric particle)!as dark matter} In many models,
this LSP is a neutralino, a weakly-interacting, weak-scale mass
particle that is a natural candidate for non-baryonic cold dark
matter~\cite{Goldberg:1983nd,Krauss:1983ik,Ellis:1983wd,Jungman:1995df}.

If supersymmetry is discovered at colliders, studies of the
superpartner spectrum will shed light on this hypothesis.  Of course,
collider experiments will never be able to prove that the LSP is dark
matter --- this requires knowing that the LSP is stable on
cosmological time scales.\index{dark matter!colliders and} However, if
the LSP is found to be charged, colored, or unstable, supersymmetric
dark matter will be largely disfavored.  On the other hand, if the LSP
is found to be a seemingly stable neutralino, there will be much work
to do.  One interesting test will be to determine the neutralino's
thermal relic density.\index{dark matter!thermal relic density} The
neutralino's relic abundance is determined by its annihilation cross
sections through processes such as those of Fig.~\ref{Sfig:annih}.
The processes $\chi \chi \to f \bar{f}$ are chirality suppressed, and
so the most relevant $t$-channel sfermions are typically stops,
sbottoms, and staus.  If the neutralino's mass and composition are
known, as well as the masses and compositions of these sfermions, its
thermal relic density may be accurately computed, assuming a standard
thermal history for the universe.  Values near the favored dark matter
density of $\Omega \approx 0.3$ will be strong evidence in favor of
neutralino dark matter.  Collider and conventional dark matter
experiments will also play complementary
roles.\index{complementarity!of particle and astroparticle physics}
The interactions of neutralinos with matter are given by interactions
such as those in Fig.~\ref{Sfig:scattering}.  By determining the
relevant supersymmetry parameters, collider experiments may guide dark
matter searches, or confirm their signals as being due to
supersymmetry.  We will discuss the relation of linear collider
experiments to cosmology in more detail in \secref{Ssec:cosmology}.

\begin{figure}[tbp]
\begin{minipage}[t]{0.49\textwidth}
\postscript{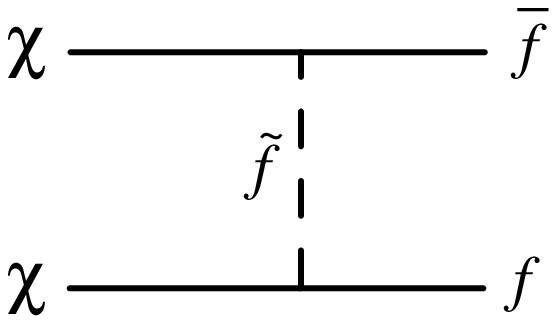}{0.85}
\end{minipage}
\hfill
\begin{minipage}[t]{0.49\textwidth}
\postscript{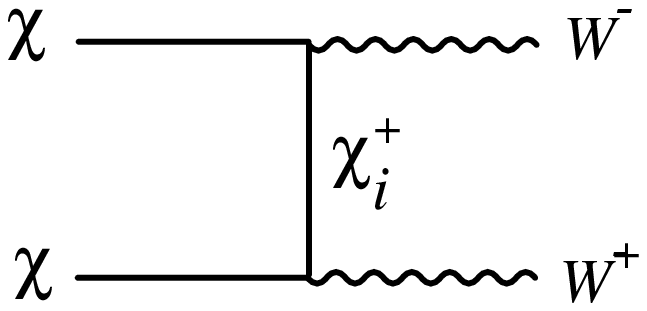}{0.99}
\end{minipage}
\caption{\footnotesize Neutralino dark matter annihilation processes.
\label{Sfig:annih}}
\index{dark matter!annihilation}
\end{figure}

\begin{figure}[tbp]
\begin{minipage}[t]{0.49\textwidth}
\postscript{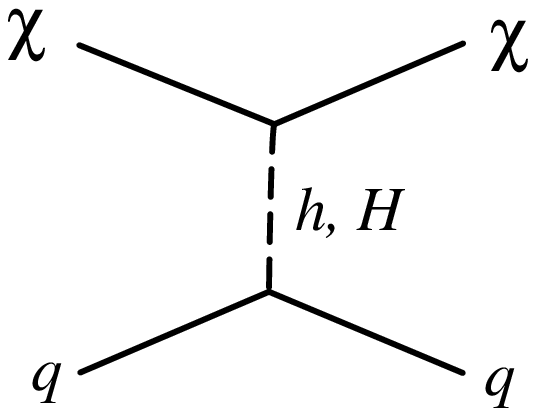}{0.80}
\end{minipage}
\hfill
\begin{minipage}[t]{0.49\textwidth}
\postscript{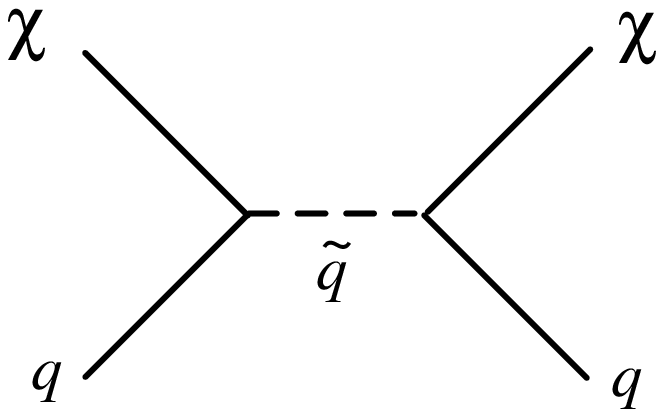}{0.95}
\end{minipage}
\caption{\footnotesize Neutralino dark matter scattering processes.
\label{Sfig:scattering}}
\index{dark matter!detection}
\end{figure}

\subsection{Flavor Violation}
\label{Ssec:flavor}

Weak-scale supersymmetry generically induces gross violations of
constraints on flavor-violating
observables~\cite{Gabbiani:1996hi}.\index{problems!supersymmetric
flavor} Among the best known is the supersymmetric contribution to the
$K_L$--$K_S$ mass splitting.\index{constraints!$\Delta m_K$}
Supersymmetry contributes to this through box diagrams, such as the
one involving squarks and gluinos shown in Fig.~\ref{Sfig:deltaK}.
The crosses on the squark propagators are flavor-violating mass
insertions $m_{\tilde{q}_{12}}^2$.  As noted in
\secref{Ssec:sleptons}, there is no reason {\em a priori} to assume
that squark and quark mass matrices are simultaneously diagonalizable,
and so such flavor-violating mass insertions generically exist.

\begin{figure}[tb]
\postscript{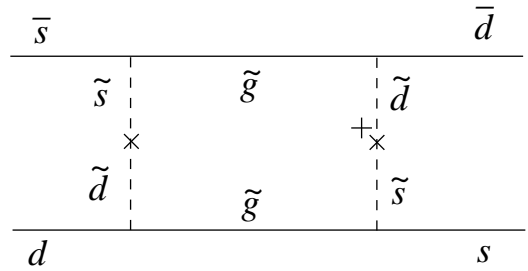}{0.60}
\caption{\footnotesize Supersymmetric contribution to the $K_L$--$K_S$
mass splitting.
\label{Sfig:deltaK}}
\end{figure}

The bound from the measured $\Delta m_K$ implies~\cite{Contino:1998nw}
\begin{equation}
\left[ \frac{10~\tev}{m_{\tilde{q},\tilde{g}}} \right]^2
\biggl[ \frac{\Delta m_{\tilde{q}_{12}}^2 / m_{\tilde{q}}^2}{0.1}
\biggr]^2 \alt 1 \ .
\end{equation}
For TeV-scale squarks and gluinos, the flavor-violating parameters
must be very small so that the supersymmetric generalization of the
GIM mechanism applies.  Alternatively, for ${\cal O}(1)$ flavor
mixing, squark and gluino masses must be of order 10 TeV, well above
the weak scale.

The supersymmetric flavor problem is among the most pressing
phenomenological problems of supersymmetry, and motivates many models
that naturally produce degenerate squarks, very heavy squarks and
gluinos, or quark-squark alignment.  Note that this is not simply a
problem in the hadronic sector --- bounds on lepton flavor
violation\index{constraints!lepton flavor violation} (LFV) from
$\mu$--$e$ conversion, $\mu \to e \gamma$, and other processes, cause
similar difficulties.\index{flavor violation!lepton!low energy} If
supersymmetry is discovered, one immediate question will be how the
supersymmetric flavor puzzle is resolved.

\subsection{CP Violation}
\label{Ssec:CP}

Weak-scale supersymmetry also generically violates bounds on CP
violation.\index{problems!supersymmetric CP} Bounds from
$\epsilon_K$\index{constraints!$\epsilon_K$} are numerically the most
stringent, but are typically satisfied in models that eliminate flavor
violation in some natural way.  However, CP-violating, but
flavor-conserving, constraints remain a problem.  A prime example of
these constraints is the electron's electric dipole moment
(EDM).\index{constraints!electron EDM} The supersymmetric
contribution to this observable is shown in Fig.~\ref{Sfig:EDMe}.
This contribution requires~\cite{Moroi:1995yh}
\begin{equation}
\left[ \frac{2~\tev}{m_{\tilde{e}}} \right]^2
\left[ \frac{\mu M_1}{m_{\tilde{e}}^2} \right]
\tb \sin \phi_{CP} \alt 1 \ ,
\end{equation}
that is, uncomfortably heavy selectrons or small CP-violating phases
$\phi_{CP} \ll 1$. As in the flavor-violating case, this problem is
not confined to one sector --- a similarly stringent constraint in the
hadronic sector follows from the neutron
EDM.\index{constraints!neutron EDM} Again, the discovery of
supersymmetry will raise the question of how these supersymmetric CP
problems are solved.

\begin{figure}[tb]
\postscript{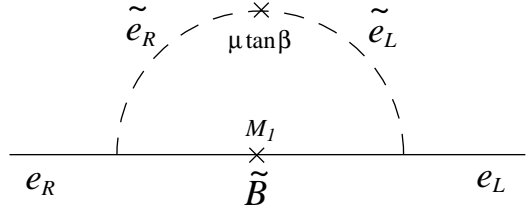}{0.60}
\caption{\footnotesize Supersymmetric contribution to the electron's
electric dipole moment.  A photon attached to the charged internal
line is implicit.
\label{Sfig:EDMe}}
\end{figure}

\subsection{Proton Decay}
\label{Ssec:proton}

In supersymmetric theories, proton decay through renormalizable
operators may be eliminated by $R$-parity
conservation.\index{constraints!proton decay} However, in grand
unified theories (GUTs), proton decay is again allowed through
non-renormalizable operators suppressed by the GUT
scale.\index{unification!and proton decay}

The leading contribution for large $\tan\beta$ in supersymmetric SU(5)
theories is given in
Fig.~\ref{Sfig:proton}~\cite{Goto:1998qg,Babu:1998ep,Goto:1999iz}.
For minimal SU(5), this contribution, normalized to the current bounds
from SuperKamiokande~\cite{Suzuki:2001rb}, implies
\begin{equation}
\biggl[ \frac{\tau(p\to K^+ \bar{\nu})}{1.6 \times 10^{33}\
{\rm yr}} \biggr]
\biggl[ \frac{(5~\tev)^2 \, \mu^2}{m_{\tilde{q}}^4} \biggr]
\biggl[ \frac{\tan\beta}{4} \biggr]^4
\biggl[ \frac{2 \times 10^{16}~\gev}{M_C}\biggr]^2 \alt 1 \ ,
\end{equation}
where $M_C$ is the colored Higgs mass, typically expected to be of the
order of the unification scale $\mgut$.  This operator relies only on
standard model flavor violation, and so is present even in
supersymmetric theories with degenerate scalars and CP conservation.

\begin{figure}[tb]
\postscript{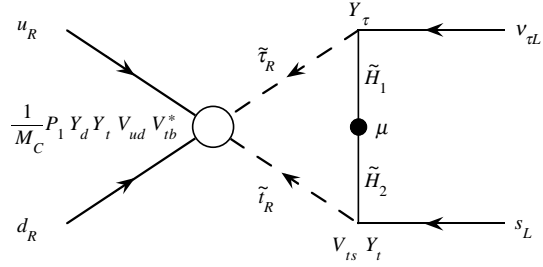}{0.70}
\caption{\footnotesize Contribution to proton decay in supersymmetric
grand unified theories~\protect\cite{Goto:1999iz}.
\label{Sfig:proton}}
\end{figure}

The absence of observed proton decay is becoming a severe constraint
for simple supersymmetric GUTs. It motivates models with naturally
heavy superpartners, extended GUT models, or even those with some
mechanism to unify gauge coupling without GUT gauge groups.  The
discovery of light squarks will immediately exclude the simplest GUT
models.

\section{Models}
\label{Ssec:models}

If weak-scale supersymmetry is found in nature, the promise of linear
colliders lies in the possibility of disentangling supersymmetry
without model-dependent assumptions.\index{models} However, without
actual data, studies of the potential of colliders require some model
framework to consider.  The models studied are typically motivated by
the desire to solve the problems without sacrificing the successes of
\secref{Ssec:successes}.  Here we describe several frameworks with
distinct collider phenomenology.

For collider phenomenology, models are largely specified by their soft
supersymmetry-breaking terms.  What is the origin of the soft
terms?\index{supersymmetry breaking!soft terms!origin of} It is
typically expected that they are generated by some dynamical mechanism
closely analogous to the mechanism of spontaneous gauge symmetry
breaking.  In electroweak symmetry breaking, for example, gauge
symmetry is broken when the Higgs field gets a non-vanishing vev.  For
supersymmetry breaking, the $F$ fields play the role of the Higgs
field.  To see this, note that supersymmetry transformations on chiral
superfields may be defined as $\delta_{\xi} \Psi = [\xi Q + \bar{\xi}
\bar{Q}, \Psi]$.  Expanding both sides in powers of $\theta$, we find
\begin{eqnarray}
\delta_\xi \phi &=& \sqrt{2} \xi \psi \\
\delta_\xi \psi &=& -i \sqrt{2} \sigma^{\mu} \bar{\xi}
\partial_{\mu} \phi + \sqrt{2} \xi F \\
\delta_\xi F &=& -i \sqrt{2} \bar{\xi} \bar{\sigma}^{\mu}
\partial_{\mu} \psi \ . 
\end{eqnarray}
If $F$ obtains a vev, 
\begin{equation}
\langle 0 | F | 0 \rangle = 
\langle 0 | [ \xi Q + \bar{\xi} \bar{Q}, \psi ] | 0 \rangle \ne 0 \ ,
\end{equation}
implying that $Q | 0 \rangle \ne 0$, and supersymmetry is broken in
the vacuum.

In most supersymmetric models, including those to be discussed below,
the terms of the Lagrangian may be divided into three sectors:~the
supersymmetry~breaking sector, containing only non-MSSM fields $Z$,
the mediation sector coupling the $Z$ and MSSM fields, and the MSSM
sector, containing only MSSM fields. Supersymmetry is broken when
dynamical effects\index{supersymmetry breaking!dynamical} generate a
vev for the $F$ component of a $Z$ field.  This vev is called the {\em
scale of supersymmetry breaking}.\index{supersymmetry breaking!scale
of} Terms in the mediation sector then generate soft terms for MSSM
fields.  For example, when $F_Z$ gets a vev, the soft scalar masses
and trilinear $A$-terms of \eqref{Ssoft} will be generated by
higher-dimensional operators suppressed by the mass scale $M$ of the
mediation sector:
\begin{eqnarray}
\lambda_{ij} \frac{Z^\dagger Z}{M^2} 
\Psi^{\dagger}_i \Psi_j |_{\theta \theta}
&\to& \lambda_{ij} \frac{|\langle F_Z \rangle|^2}{M^2}
\phi^*_i \phi_j \nonumber \\
\lambda_{ijk} \frac{Z}{M}\Psi_i \Psi_j \Psi_k |_{\theta \theta} 
&\to& \lambda_{ijk} \frac{\langle F_Z \rangle}{M}\phi_i\phi_j\phi_k 
\ .
\label{Ssoftgen}
\end{eqnarray}
Gaugino masses and $B$-terms may also be generated in this way.  The
sparticle masses and other weak-scale parameters therefore contains an
imprint of terms in the mediation sector which are determined by
physics at energies much higher than the weak scale.  Field
condensation does not alter the high energy behavior of the theory,
and terms generated from $F$ condensation, such as those above, do not
disturb the cancellation of quadratic divergences.

For most phenomenological applications, {\em how} supersymmetry is
broken is largely irrelevant.  Far more important is the
supersymmetry-mediation mechanism,\index{supersymmetry
breaking!mediation mechanism} that is, the terms of the mediation
sector, and the supersymmetry-breaking scale.  Both are crucial in
determining the signatures of supersymmetry.  The mediation terms
determine the pattern of soft terms and, consequently, the masses,
mixings, and interactions of the standard model superpartners.  The
supersymmetry-breaking scale\index{supersymmetry breaking!scale of}
determines the properties of the gravitino.  The gravitino's
mass\index{gravitino!mass} is
\begin{equation}
m_{\tilde{G}} = \frac{\fdsb}{\sqrt{3} \mplanck} \ , 
\label{Smgravitino}
\end{equation}
and its interactions\index{gravitino!interactions} are given by
\begin{equation}
{\cal L} \supset 
\frac{1}{\fdsb} \left[ m_{\tilde{f}}^2 \bar{f} \tilde{f} 
+ \frac{m_{\lambda}}{4\sqrt{2}}\bar{\lambda} \sigma^{\mu \nu}
F_{\mu\nu} \right] \tilde{G} \ ,
\label{Sgravitinointeractions}
\end{equation}
where $\fdsb\equiv [ \sum_i \langle F_i \rangle^2 ]^{1/2}$ is the
total $F$ vev generated by dynamical supersymmetry breaking.  For
simple cases where there is only one non-zero $\langle F \rangle$,
$\fdsb$ is equivalent to the $\langle F_Z \rangle$ appearing in
\eqref{Ssoftgen}, but in general, they need not be identical.

We will now describe several models, beginning with three supergravity
theories.  In supergravity,\index{models!supergravity} supersymmetry
breaking is mediated through non-renormalizable gravitational
interactions, and the large mass scale of \eqref{Ssoftgen} is the
Planck mass $\mplanck$.  These interactions are generically present in
all theories with supersymmetry and gravity, and supergravity theories
are therefore natural models to consider.  However, by their very
nature, the coefficients of these non-renormalizable terms cannot be
precisely determined without knowing the microscopic quantum theory of
gravity from which they presumably derive.  To proceed, various simple
and phenomenologically attractive assumptions are made.  These
assumptions then fix the soft terms, allowing one to conduct concrete
collider studies.

Another approach is to suppress the non-renormalizable supergravity
terms and find new sources for soft terms.  This requires new
structure involving additional fields and interactions, but has the
advantage that the soft terms are calculable in field theory.  We
describe models with gauge-\index{models!gauge-mediation!motivation}
and anomaly-mediated\index{models!anomaly-mediation!motivation}
supersymmetry breaking, two prominent examples of this approach.

Finally, we conclude with comments on the implications of GUT and
string models for soft supersymmetry-breaking parameters.

\subsection{Minimal Supergravity}
\label{Ssec:msugra}

In supergravity, soft terms are generated by terms such as those in
\eqref{Ssoftgen}, where the large mass scale is $M=\mplanck$.  Soft
scalar masses have the form $m_{ij}^2 = \lambda_{ij} m_0^2$, where
$m_0 \equiv \langle F \rangle / \mplanck$, and $i,j$ are generational
indices.  These terms generically mediate flavor-changing transitions
that violate the constraints discussed in \secref{Ssec:flavor}.

In {\em minimal supergravity}~\cite{Chamseddine:jx,Barbieri:1982eh,%
Hall:iz},\index{models!minimal supergravity} one assumes that all of
the soft scalar masses are given by a single parameter $m_0$ at some
high scale, typically taken to be $\mgut \simeq 2 \times
10^{16}~\gev$.  Similarly, one assumes a unified gaugino mass
$\mgaugino$, and a universal $A$-term parameter $A_0$.  At $\mgut$,
the model is completely specified by standard model parameters and
$m_0$, $\mgaugino$, $A_0$, $B$ and $\mu$.

The weak-scale MSSM Lagrangian is then determined by evolving all
parameters to the weak-scale through renormalization group evolution.
Renormalization group evolution of supersymmetry parameters is
described by a complicated system of differential equations.
Schematically, however, the 1-loop renormalization group
equations\index{renormalization group!equations} are
\begin{eqnarray}
\frac{dg}{dt} &\sim& \frac{1}{16\pi^2} g^3 \\
\frac{dy}{dt} &\sim& \frac{1}{16\pi^2} \left[ g^2 y - y^3 \right]
\\
\frac{dM}{dt} &\sim& \frac{1}{16\pi^2} g^2 M \\
\frac{dA}{dt} &\sim& \frac{1}{16\pi^2} \left[ - g^2 M - y^2 A
\right] \\
\frac{dm^2}{dt} &\sim& \frac{1}{16\pi^2} \left[ g^2 M^2 - y^2 A^2
- y^2 m^2 \right] \ , \label{SscalarRGE}
\label{SRGEs}
\end{eqnarray}
where $t\equiv \ln (Q_0/Q)$ with $Q_0$ some fixed renormalization
scale, and {\em positive} numerical coefficients and gauge and flavor
indices have been neglected.  In evolving down from a high scale,
gauge interactions raise $m^2$, while Yukawa interactions lower $m^2$.
Unique among scalar mass parameters, $m_{H_u}^2$ is driven down by the
large top quark Yukawa without a compensating positive contribution
from the strong gauge interactions.  It is typically driven negative,
breaking electroweak symmetry.  This feature of {\em radiative
electroweak symmetry breaking}\index{electroweak symmetry
breaking!radiative} is often regarded as another success of
supersymmetry.

At the weak scale, the supersymmetry parameters are constrained by the
following tree-level relations:\index{electroweak symmetry
breaking!tree relations}
\begin{eqnarray}
\frac{1}{2} m_Z^2 &=& \frac{m_{H_d}^2 - m_{H_u}^2 \tan^2\beta }
{\tan^2\beta -1} - |\mu|^2  \label{SEWSB1} \\
2 B &=& (m_{H_u}^2 + m_{H_d}^2 + 2 |\mu|^2)\sin2\beta \ .
\label{SEWSB2}
\end{eqnarray}
Since $m_Z$ is measured, these relations are used to exchange $B$ and
$\mu$ at the high scale for $\tan\beta$ and $\Arg(\mu)$.  In minimal
supergravity, $\mu$ is assumed real, and so the fundamental parameters
of minimal supergravity\index{models!minimal supergravity!fundamental
parameters} are
\begin{equation}
m_0,\, \mgaugino,\, A_0,\, \tan\beta,\, \sign(\mu) \ .
\end{equation}

The scale of supersymmetry breaking is set by the requirement that the
soft terms are of the order of the weak scale, so $\langle F \rangle
\sim m_W \mplanck \sim 10^{20}~\gev^2$.  By \eqref{Smgravitino}, the
gravitino mass is also of the order of the weak
scale.\index{gravitino!mass!in supergravity} It is typically assumed
to be heavier than the standard model superpartners.  In this case,
the gravitino is irrelevant for colliders, as its interactions are
gravitational and so extremely weak.

The masses of the rest of the superpartners are determined by
renormalization group evolution.  At the weak scale, the gaugino mass
parameters are given by
\begin{equation}
\frac{M_i}{\alpha_i} = \frac{M_{1/2}}{\alpha_{\text{unif}}} \ ,
\end{equation}
in accord with \eqref{Sgauginomasses}.
Numerically, this implies
\begin{equation}
M_1 \approx 0.4 M_{1/2} \ .
\label{Smsugra1}
\end{equation}
For the scalar masses, neglecting the effects of Yukawa couplings, we
find the weak-scale values
\begin{eqnarray}
m_Q^2 &=& m_0^2+ 6.3 M^2_{1/2} \nonumber \\
m_U^2 &=& m_0^2+ 5.8 M^2_{1/2} \nonumber \\
m_D^2 &=& m_0^2+ 5.8 M^2_{1/2} \nonumber \\
m_L^2 &=& m_0^2+ 0.5 M^2_{1/2} \nonumber \\
m_E^2 &=& m_0^2 + 0.15 M^2_{1/2} \ .
\label{Smsugra2}
\end{eqnarray}
If $m_0$ and $M_{1/2}$ are of the same order, the squark masses are
dominated by $M_{1/2}$, while the slepton masses are dominated by
$m_0$.  Generically the renormalization group equations predict heavy
squarks and light sleptons.  The lightest scalars are
$\tilde{\tau}_R$, followed by $\tilde{\mu}_R$, $\tilde{e}_R$, and the
left-handed sleptons.

Minimal supergravity has a number of virtues.  Grand unified boundary
conditions are automatically incorporated.  In addition, as is evident
from \eqsref{Smsugra1}{Smsugra2}, in much of parameter space, the LSP
is the lightest neutralino, and so is a possible dark matter
candidate.  The nature of the LSP in various regions of parameter
space is presented in Fig.~\ref{Sfig:gfraction}.  For most of the
region with $m_0 \alt 1~\tev$, the LSP is a Bino-like
neutralino.\index{dark matter!Bino-like} Minimal supergravity also
solves the supersymmetric flavor and CP problems, but only through
{\em ad hoc} assumptions.  {}From a universal scalar mass at $\mgut$,
renormalization group evolution generates mass splittings and flavor
mixing, but these are generally small.  The assumption of a real $\mu$
parameter also guarantees the absence of supersymmetric CP violation,
suppressing dangerous contributions to electric dipole moments.

\begin{figure}[tb]
\postscript{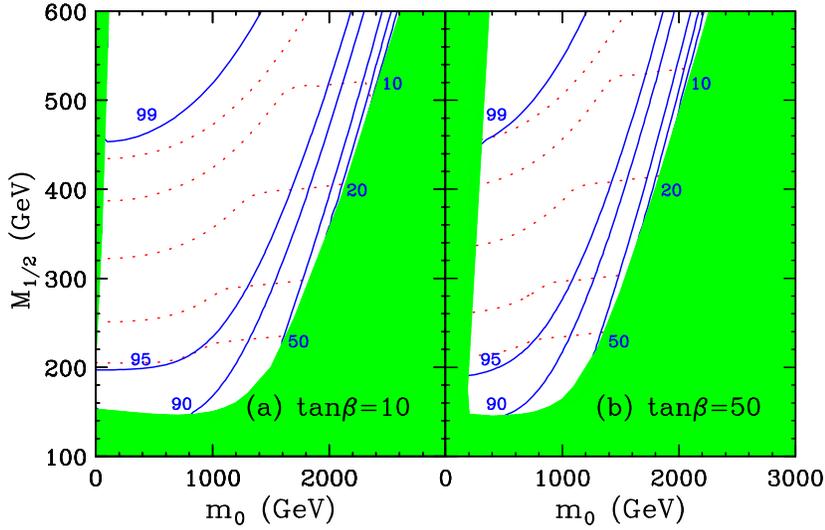}{0.95}
\caption{\footnotesize The character of the LSP in minimal
supergravity with $A_0 = 0$ and $\mu < 0$~\protect\cite{Feng:2000gh}.
In the unshaded regions, the LSP is the lightest neutralino, and
contours of the gaugino-ness of the LSP $R_{\chi} \equiv |N_{11}|^2 +
|N_{12}|^2$ are given (in percent). In the narrow upper left shaded
regions, the LSP is the lighter stau.  The lower right shaded regions
are excluded by bounds on the chargino mass.
\label{Sfig:gfraction}}
\index{LSP (lightest supersymmetric particle)!in minimal supergravity}
\end{figure}

To summarize, a typical minimal supergravity scenario
predicts\index{models!minimal supergravity!typical spectrum}
\begin{eqnarray}
\text{NLSP\ candidates:} && \tilde{l}_R \ \text{or} \  
\tilde{\chi}^0_2 , \tilde{\chi}^{\pm}_1 \approx \tilde{W}^0,
\tilde{W}^{\pm} \nonumber \\
\text{LSP:} && \tilde{\chi}_1^0 \approx \tilde{B} \ ,
\end{eqnarray} 
where NLSP denotes the next-to-lightest supersymmetric particle, which
is the lightest detectable in colliders.  One therefore generally
expects neutralinos, charginos, and sleptons to be produced in the
greatest numbers at linear colliders.

\subsection{Focus Point Supersymmetry}
\label{Ssec:focuspoint}

Perhaps the most straightforward solution to the flavor and CP
constraints of \secsref{Ssec:flavor}{Ssec:CP} is to assume heavy
superpartners.\index{models!focus point supersymmetry} However, large
supersymmetry parameters typically re-introduce the gauge hierarchy
and so are considered unnatural.\index{problems!gauge hierarchy} As an
example, consider \eqref{SEWSB1}: if $|m_{H_u}^2| \gg m_Z^2$, the
parameter $\mu$ must be highly fine-tuned to cancel most of
$m_{H_u}^2$ to give the correct $Z$ mass.

Naturalness does not require all supersymmetry-breaking parameters to
be small, however. In focus point
models~\cite{Feng:1999mn,Feng:1999zg}, large fundamental supersymmetry
parameters do not imply large cancellations in \eqref{SEWSB1}.  A
simple example is provided by the minimal supergravity model just
discussed.  Rewriting \eqref{SEWSB1} in terms of the fundamental
parameters, we find through numerical analysis
\begin{equation}
\frac{1}{2} m_Z^2 = -0.04\, m_0^2 + 8.8 M_1^2 - |\mu|^2 \ .
\label{SfpEWSB}
\end{equation}
The small $m_0^2$ coefficient implies that even large values of $m_0$
do not result in large fine-tuning in the electroweak potential.  The
small coefficient may be understood in terms of a focus point in the
$m_{H_u}^2$ renormalization group trajectories, shown in
Fig.~\ref{Sfig:focuspoint}.  For a large range of initial values
$m_0$, these trajectories focus to a point with $m_{H_u}^2 \approx 0$
at the weak scale. Naturalness bounds $m_{H_u}^2$ at the weak scale,
but constrains the GUT-scale parameter $m_0$ only loosely.  While
$m_{H_u}^2$ is insensitive to $m_0$, the masses of all sleptons and
squarks are of order $m_0$, and so very heavy scalars are consistent
with naturalness in minimal supergravity, while at the same time
alleviating flavor and CP constraints~\cite{Feng:2000bp}.

\begin{figure}[tb]
\postscript{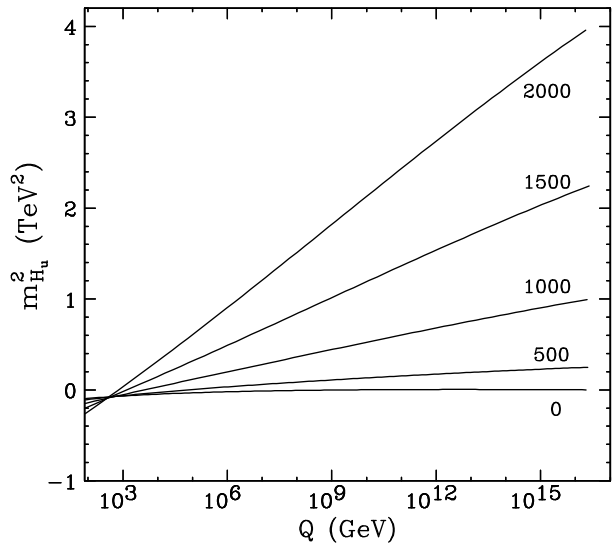}{0.70}
\caption{\footnotesize Renormalization group trajectories of
$m_{H_u}^2$ in minimal supergravity for various $m_0$ as labeled and
$\mgaugino = 300~\gev$, $A_0= 0$, $\tan\beta =
10$~\protect\cite{Feng:1999zg}. These focus to a point at $Q \sim
m_W$.
\label{Sfig:focuspoint}}
\index{renormalization group!trajectories!in focus point
supersymmetry}
\end{figure}

The focus point region in minimal supergravity is the region of $m_0
\agt 1~\tev$ in Fig.~\ref{Sfig:gfraction}.  {}From \eqref{SfpEWSB}, we
see that, for fixed $\mgaugino$, large $m_0$ requires small $|\mu|$.
This is also evident in Fig.~\ref{Sfig:gfraction}, where, for large
$m_0$ there is a region with $\mgaugino$ and $|\mu|$ comparable, and
therefore significant Higgsino content in the lightest neutralino.
The prediction of focus point supersymmetry, then, is that, while
sleptons and squarks are typically heavy, possibly even with multi-TeV
masses beyond the LHC, the electroweak gauginos and Higgsinos are all
light.

In fact, constraints on the dark matter relic density typically
require significant gaugino-Higgsino mixing in focus point
supersymmetry.  The requirement that dark matter not overclose the
universe implies that the processes of Fig.~\ref{Sfig:annih} be
sufficiently efficient.  In focus point supersymmetry, the processes
mediated by sleptons and squarks are highly suppressed, and so dark
matter annihilation relies on the processes with gauge boson final
states.  These processes are, however, absent for Bino-like LSPs, as
they rely on gaugino-Higgsino mixing.  As a result, dark matter
constraints require gaugino-Higgsino mixing,\index{dark
matter!gaugino-Higgsino}\index{dark matter!in focus point
supersymmetry} and $|\mu|$ not too far above the gaugino mass
parameters.

In focus point supersymmetry, then, all 6 charginos and neutralinos
may be within reach of the linear collider, allowing detailed studies of
these systems.  A typical focus point spectrum is\index{models!focus
point supersymmetry!typical spectrum}
\begin{eqnarray}
\text{Heavy:} && \tilde{l}, \tilde{q} \nonumber \\
\text{NLSP\ candidates:} && \tilde{\chi}^0_2, \tilde{\chi}^{\pm}_1 
 \approx \text{\ gaugino-Higgsino\ mixtures} \nonumber \\
\text{LSP:} && \tilde{\chi}^0_1 \approx \tilde{B}, \tilde{H}^0\ 
\text{mixture}\ .
\end{eqnarray} 

\subsection{Superheavy Supersymmetry}
\label{Ssec:superheavy}

Superheavy supersymmetry scenarios also alleviate the flavor and CP
problems by decoupling.\index{models!superheavy supersymmetry} The
soft Higgs masses $m_{H_u}^2$ and $m_{H_d}^2$ are constrained to the
weak scale by naturalness.  They are determined by the scalar mass
renormalization group equation of \eqref{SscalarRGE}.  Note, however,
that the other scalar mass parameters enter this renormalization group
equation proportional to their Yukawa couplings.

This then naturally suggests a split spectrum with light third
generation sfermions, and superheavy sfermions of the first two
generations~\cite{Drees:1986jx}.  The light third generation sfermions
preserve naturalness and do not induce violations of existing bounds,
since low energy bounds on third generation observables are relatively
weak.  At the same time, superheavy sleptons and squarks of the first
two generations, with masses of the order of 10 TeV, alleviate the
most stringent flavor- and CP-violating constraints, such as those in
from the kaon system, and $\mu$-$e$ transitions, but do not sacrifice
naturalness, since their effect on the $m_{H_u}^2$ and $m_{H_d}^2$
renormalization group equations are suppressed by tiny Yukawa
couplings.  The split spectrum of superheavy supersymmetry may be
generated immediately upon supersymmetry-breaking~\cite{Dvali:1996rj}
or by renormalization group evolution effects~\cite{Feng:1999iq}.

A typical superheavy supersymmetry spectrum is\index{models!superheavy
supersymmetry!typical spectrum}
\begin{eqnarray}
\text{Superheavy:} && \tilde{e}, \tilde{\mu}, \tilde{u}, \tilde{d}, 
\tilde{c}, \tilde{s} \nonumber \\
\text{NLSP\ candidates:} && \tilde{\tau}, \tilde{\nu}_{\tau},
 \tilde{b}, \tilde{t}, \tilde{\chi}^0_2, \tilde{\chi}^{\pm}_1
 \nonumber \\
\text{LSP:} && \tilde{\chi}^0_1 \ .
\end{eqnarray} 

\subsection{Gauge Mediation}
\label{Ssec:gmsb}

In supergravity scenarios, the soft terms are assumed to arise from
non-renormalizable gravitational
interactions.\index{models!gauge-mediation} As noted above, however,
these are not calculable and {\em ad hoc} assumptions must be made to
avoid constraints from flavor and CP bounds.  In gauge-mediated
supersymmetry breaking, one assumes that $\langle F \rangle \ll
10^{20}~\gev^2$, so that these troublesome gravitational contributions
to soft masses are highly suppressed.

Of course, weak-scale soft terms must still be generated.  This is
achieved by introducing $N$ ``messenger
fields''\index{messenger!fields} with mass
$\mmess$~\cite{Dine:1993yw,Giudice:1998bp}.\index{messenger!scale}
These have standard model gauge quantum numbers and generate MSSM soft
masses at the scale $M_{mess}$ of the form
\begin{eqnarray}
M_i &=& N \frac{g_i^2|_{Q=\mmess}}{16\pi^2} 
\frac{\langle F \rangle}{\mmess} \\
m^2_{\tilde{f}} &=& 2 N \sum_i C_{if} 
\left[ \frac{g_i^2|_{Q=\mmess}}{16 \pi^2}\right]^2 \left[
\frac{\langle F \rangle}{\mmess} \right]^2 
\label{Sgmsbmasses}
\end{eqnarray}
through diagrams such as those of Fig.~\ref{Sfig:ssi_gmsbfeyn}.  Here
$C_{if}$ are coefficients determined by the sfermions' gauge
representations.  Yukawa effects are typically small and may be
neglected.  These soft masses are therefore dependent only on standard
model gauge quantum numbers to an excellent approximation, and gauge
mediation therefore elegantly predicts flavor-blind soft terms, which
suppress contributions to flavor violation.  Generically, CP-violating
EDMs are still problematic, however.

\begin{figure}[tb]
\postscript{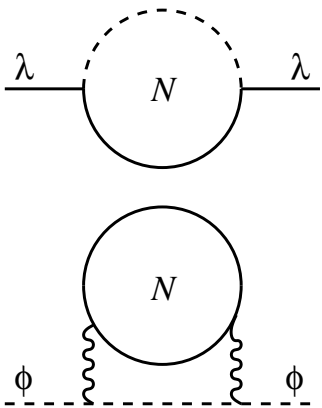}{0.40}
\caption{\footnotesize Contributions to scalar and gaugino masses in
gauge-mediated supersymmetry breaking.
\label{Sfig:ssi_gmsbfeyn}}
\end{figure}

The gaugino and sfermion masses of \eqref{Sgmsbmasses} are then
evolved from the messenger scale to the weak scale.  The gaugino
masses are in the same proportion as in minimal supergravity, even
though they are never unified, but the scalar masses are very
different.  The scalar spectrum is determined by only three parameters
in the simplest models:\index{models!gauge-mediation!fundamental
parameters} $\langle F \rangle $, $\mmess$, and $N$.  The possibility
of distinguishing such spectra from supergravity predictions will be
discussed in \secref{Ssec:extraplanck}.

For colliders, a key feature of gauge-mediated supersymmetry breaking
is the prediction of $m_{\tilde{G}} \sim \langle \fdsb
\rangle/\mplanck \ll m_{\text{Weak}}$.\index{gravitino!mass!in
gauge-mediation} The gravitino is therefore always the LSP.  The NLSP
is determined by a number of factors.  For example, although the
gaugino masses grow linearly with $N$, the scalar masses grow only as
$\sqrt{N}$.  For low $N$, the NLSP is typically a neutralino, but for
larger $N$, the NLSP is typically a
slepton~\cite{Dimopoulos:1997ww}. This may be seen in
Fig.~\ref{Sfig:n5}, where the character of the NLSP in various regions
of parameter space is shown.  In general, then, we have the
possibilities\index{models!gauge-mediation!typical spectrum}
\begin{eqnarray}
\text{NLSP\ candidates:} && \tilde{\chi}^0_1 \ , \ \tilde{l} 
\nonumber \\
\text{LSP:} && \tilde{G}  \ .
\end{eqnarray} 

\begin{figure}[tb]
\postscript{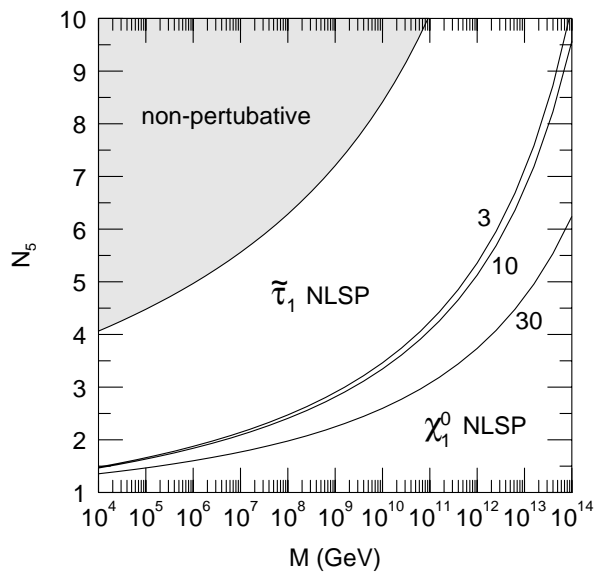}{0.70}
\caption{\footnotesize The NLSP in various regions of minimal
gauge-mediation model parameter space~\protect\cite{Feng:1997zr}.  $M$
is the messenger mass or scale, and $N_5$ is the number of messenger
multiplets.\index{messenger!scale} In the region above the labeled
contours, the NLSP is a stau; in the region below, the NLSP is a
neutralino.  The boundary between the two regions depends slightly on
$\tan\beta$; the boundaries for $\tan\beta = 3$, 10, and 30 are shown.
In the shaded region, gauge coupling constants become non-perturbative
below the GUT scale under two-loop renormalization group
evolution. \label{Sfig:n5}} \index{models!gauge-mediation!NLSP in}
\end{figure}

The NLSP decays to the gravitino with decay length
\begin{equation}
L\sim 0.1\ \text{mm} \times 
\biggl[\frac{E_{\text{NLSP}}^2 - m_{\text{NLSP}}^2}
{m_{\text{NLSP}}^2} \biggr]^{\frac{1}{2}}
\biggl[\frac{\sqrt{\fdsb}}{10^5~\gev}\biggr]^{4} \left[\frac{100
~\gev}{m_{\text{NLSP}}}\right]^5 \ .
\end{equation}
$\sqrt{\fdsb}$ may range from $10^5~\gev$ to $10^9~\gev$, where the
lower limit is set by lower bounds on superpartner and messenger
masses and the upper limit follows from requiring the supergravity
contributions to be sub-dominant.  For low values of $\sqrt{\fdsb}$ in
this range, the NLSP decays within collider detectors, producing
signals with energetic photons and leptons, depending on what
sparticle is the
LSP~\cite{Stump:1996wd}.\index{models!gauge-mediation!signals}
However, for large $\sqrt{\fdsb}$, the NLSP may travel macroscopic
distances before decaying.  In the case of neutralino NLSPs, this
produces the conventional missing energy signals of supersymmetry.
For slepton NLSPs, however, non-relativistic sleptons may produce
highly-ionizing charged tracks and tracks with unusual time-of-flight
signatures.\index{highly-ionizing charged tracks}\index{time-of-flight
signatures}

\subsection{Anomaly Mediation}
\label{Ssec:amsb}

In anomaly-mediated supersymmetry
breaking~\cite{Randall:1998uk,Giudice:1998xp}, the dangerous
supergravity contributions are suppressed geometrically ---
supersymmetry-breaking fields are placed on one 3-brane, with MSSM
fields on another, as shown in
Fig.~\ref{Sfig:sequesteredbranes}.\index{models!anomaly-mediation} The
usual supergravity contributions are then suppressed because the
supersymmetry-breaking and MSSM wave functions are separated in space.

\begin{figure}[tb]
\postscript{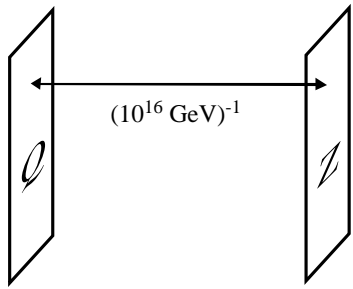}{0.40}
\caption{\footnotesize Fields in anomaly-mediation: MSSM fields $Q$
are separated from supersymmetry-breaking fields $Z$ in extra
dimensions.
\label{Sfig:sequesteredbranes}}
\end{figure}

In contrast to gauge mediation, however, in anomaly mediation, the
soft terms are generated by loop-suppressed supergravity contributions
of order $\sim (1/16\pi^2) \langle F\rangle/\mplanck$.  These arise
from the Weyl rescaling anomaly.  They are always present, but are
typically dominated by tree-level contributions in conventional
four-dimensional supergravity.  The required supersymmetry-breaking
scale is therefore larger than in conventional supergravity, and the
gravitino mass is of order 100 TeV.\index{gravitino!mass!in
anomaly-mediation} As with gauge mediation, the soft terms generated
are essentially determined by gauge couplings, and so solve the
supersymmetry flavor problem through degeneracy.  The minimal models
predict tachyonic sleptons --- this catastrophe may be avoided either
by assuming an additional contribution to scalar
masses~\cite{Gherghetta:1999sw} or by postulating other additional
structure~\cite{Pomarol:1999ie}.\index{models!anomaly-mediation!and
tachyonic sleptons} Also, as in gauge-mediated models, additional
structure is typically required to naturally satisfy EDM constraints.
 
The most striking phenomenological feature for colliders is the
gaugino mass spectrum.  The gaugino masses generated by the Weyl
anomaly are $M_i \propto |b_i g_i^2|$, where $b_i$ is the one-loop
$\beta$-function coefficient for the corresponding gauge group in the
MSSM.  In the MSSM, $b_i = (-33/5, -1, 3)$, and the Wino mass is
therefore predicted to be the smallest.  The lightest states are then
a triplet of Winos, and the typical spectrum
is\index{models!anomaly-mediation!typical spectrum}

\begin{eqnarray}
\text{NLSP:} && \tilde{\chi}^{\pm}_1 \approx \tilde{W}^{\pm} 
\nonumber \\
\text{LSP:} && \tilde{\chi}^0_1 \approx \tilde{W}^0 \ .  
\end{eqnarray} 
These Wino states are extraordinarily degenerate --- at tree-level,
their splitting is typically of order 1 MeV.  One-loop contributions
raise this to $150~\mev$ or more, allowing $\tilde{\chi}^{\pm}_1 \to
\tilde{\chi}^0_1 \pi^\pm$~\cite{Chen:1995yu}.  However, the decay length
may be of order 1 cm, and the resulting pion is very soft, leading to
a peculiar and qualitatively new supersymmetry signature at colliders.

\subsection{GUT and Planck Scale Frameworks}
\label{Ssec:GUTmodels}

In the discussion above, we noted that the soft terms are in principle
not predictable in
supergravity.\index{models!GUT}\index{models!string} Properties such
as scalar universality appeared as phenomenological assumptions,
motivated by simplicity and experimental constraints.  This is not the
complete story, however.  As also noted above, one strong motivation
for supersymmetry is the unification of forces, including the three
standard model forces and also gravity.  The circle of ideas
surrounding GUTs and string theory therefore naturally have
implications for the GUT scale boundary conditions.

As an example, in string theory, the low energy effective theory is
expected to be supersymmetric.  In addition to the MSSM fields, this
theory also contains many light scalar fields, including the dilaton
and moduli, with gravitationally-suppressed couplings.  The dilaton
superfield's scalar component is the source of gauge and gravitational
couplings, and its $F$ term may break supersymmetry if it condenses.
If this is the main source of supersymmetry breaking, the resulting
models, called {\em dilaton-dominated supersymmetry breaking
models}~\cite{Kaplunovsky:1993rd},\index{models!dilaton-dominated}
predict mass spectra similar to minimal supergravity, but with
specific relations among the average scalar mass, $\mgaugino$, and
$A_0$ at the Planck scale.  Small splittings of scalar masses are
introduced by moduli fields, and there are sum rules for scalar
masses.

At slightly lower scales, consider GUTs motivated by the unification
of couplings described in \secref{Ssec:unification}. In these
theories, the MSSM gauge multiplets are merged into GUT multiplets,
and have GUT interactions. For example, in SO(10) GUTs, the $\tau$ and
$t$ superfields are unified in one GUT
representation.\index{unification!matter multiplet} As a result, above
the GUT scale, renormalization group evolution of the $\stau$ mass
includes radiative corrections arising from the triplet Higgs boson
with strength proportional to the top Yukawa
coupling~\cite{Barbieri:1994pv}. This effect is not present for
selectrons and smuons, leading to a substantial splitting of slepton
masses, even if they are initially degenerate.  GUT symmetries may
also lead to alternative parameterizations of the GUT scale boundary
conditions.  For example, in SO(10) theories, the Higgs and matter
multiplets are members of separate multiplets, the {\boldmath $10$}
and {\boldmath $16$} representations, respectively.  Gauge symmetry
does not require their scalar masses to be unified.  Furthermore, even
if unified, SO(10) breaking to SU(5) as an intermediate step will
split these scalar masses by
$D$-terms~\cite{Drees:1986vd,Kawamura:1994ys}.  As a result, the Higgs
and matter scalar masses are more generally parameterized by two
independent parameters at the GUT scale, opening up many new
possibilities in the low energy MSSM spectrum.

While we cannot describe all models investigated to date here, we can
extract important lesson from the above discussion.  The soft terms of
the MSSM may carry the imprint of interactions that are not amenable
to direct experimental investigation.  Measurements of soft mass
relations may therefore shed light on physics at the highest energy
scales.

\section{Slepton Studies}
\label{Ssec:slstudy}

We now turn to specific studies of supersymmetry at the linear
collider.  We begin with sleptons, which are the lightest observable
superpartners in many models. As we will see, slepton studies may shed
light on a large number of important questions.  They also provide an
arena for exploiting many of the most powerful techniques available to
study new physics the linear collider.

\subsection{Signal and Background}
\label{Ssec:slsgbg}

Slepton pair production proceeds through $s$-channel gauge bosons,
$\epem \to \gamma, Z \to \tilde{l}^+ \tilde{l}^-$ and $\epem \to Z \to
\tilde{\nu} \tilde{\nu}^\ast$.\index{sleptons!production} In the case
of 1st generation sleptons, there are also contributions from
$t$-channel neutralino and chargino exchange.  In the presence of
supersymmetric LFV, $t$-channel contributions are also present for the
2nd and 3rd generations.  We will assume lepton flavor is conserved
here, deferring discussion of LFV to \secref{Ssec:lfv}.

Let us focus on supergravity scenarios with $R$-parity
conservation.\footnote{In general, $R$-parity violating theories are
not difficult to study at linear colliders, as the relevant
backgrounds are not overwhelming and all superpartner decay products
are visible, allowing full momentum and energy
reconstruction~\cite{Huitu:1997qu}.\index{$R$-parity!violation}} When
sleptons are heavier than neutralinos or charginos, they decay
through\index{sleptons!decay}
\begin{eqnarray}
\tilde{l}&\to& \tilde{\chi}^0_i l, \ \tilde{\chi}^-_i \nu_l  \\
\tilde{\nu}&\to& \tilde{\chi}^+_i l,\ \tilde{\chi}^0_i\nu_l \ . 
\end{eqnarray}
Sleptons have spin 0 and so decay isotropically in their rest frames.
Right-handed sleptons decay dominantly to the lighter neutralino with
the largest $\tilde{B}$ component.  Left-handed sleptons decay
primarily to the lighter neutralinos and charginos with the largest
$\tilde{W}$ components.  When decay into heavy charginos or
neutralinos is possible, sleptons may initiate cascade decays ending
in the LSP with many visible particles produced along the way.

We now focus on the simplest case in which a charged slepton is the
NLSP, and the LSP is the lightest neutralino $\tilde{\chi}^0_1$.  The
signal is then\index{sleptons!signal}
\begin{equation}
\epem \to \tilde{l}^+ \tilde{l}^- \to l^+ l^-
\tilde{\chi}^0_1 \tilde{\chi}^0_1 \ .
\end{equation}
The momenta of the two leptons may be measured precisely, but the two
LSPs escape detection. The signal is same flavor, opposite sign
leptons with missing transverse momentum $\sla{p_T}$.

The major standard model background to this process is $W$ pair
production, leading to $\epem \to W^+ W^- \to l^+ \nu_l l'^-
\bar{\nu}_{l'}$.\index{sleptons!background} The total $W^+W^-$ cross
section is enormous compared to typical supersymmetry cross sections,
but it proceeds largely through $t$-channel exchange of the light
electron, and so is very forward-peaked.  The total $W^+W^-$ cross
section may therefore be reduced from about 10 pb at $\sqrt{s}=
500~\gev$ to less than 2 pb with the requirement $\vert\cos\theta\vert
< 0.7$, where $\theta$ is the polar angle of the $W$ relative to the
beam axis. Of this, only 1\% produces a given same flavor final
state. For comparison, the typical $\tilde{\mu}^+_R\tilde{\mu}^-_R$
production cross section is $50~\ifb$ not far above threshold, and
such processes are well-represented in the central region.

Beam polarization may also greatly reduce the $W^+W^-$
background~\cite{unknown:1992mg,Tsukamoto:1993gt}.\index{beam
polarization!use for sleptons} An electron's spin along its direction
of motion is essentially determined by its chirality.  We define
polarizations for the electron and positron beams through
\begin{eqnarray}
P_{e^-} &=& \frac{N\left(h=\frac{1}{2}\right) -
N\left(h=-\frac{1}{2}\right)}
{N\left(h=\frac{1}{2}\right) + 
N\left(h=-\frac{1}{2}\right)}   
\nonumber \\
P_{e^+}& = &\frac{N\left(\bar{h}=
-\frac{1}{2}\right)-N\left(\bar{h}=\frac{1}{2}\right)}
{N\left(\bar{h}=\frac{1}{2}\right)+N\left(\bar{h}=
-\frac{1}{2}\right)} \ ,
\end{eqnarray}
where $h$ and $\bar{h}$ are the $e^-$ and $e^+$ helicities,
respectively.  The cross section for pair production of any particles
may then be expressed as
\begin{eqnarray}
\lefteqn{\frac{d}{d\cos\theta}\sigma(h, \bar{h})
= \frac{1}{32\pi s} \beta_f
\sum_{h,\bar{h}} \vert {\cal M} (h, \bar{h})
\vert^2} \nonumber \\
&& \times \frac{1}{4} \left[1+(-)^{-1/2+h}P_{e^-}\right]
\left[1+(-)^{1/2+\bar{h}} P_{e^+}\right] \ . 
\label{Spolarizedsigma}
\end{eqnarray}
Here $\beta_f=(1-m_f^2/\ebeam^2)^{1/2}$ is the velocity of the final
state particles, where $m_f$ is their mass, and $\ebeam = \sqrt{s}/2$
is the beam energy.  ${\cal M} (h, \bar{h})$ is the amplitude for
incoming electrons and positrons with helicities $h$ and $\bar{h}$,
respectively, and depends on $\cos\theta$.

Much of $W$ pair production proceeds through $\bar{\nu} \gamma_\mu
P_Le W^{\mu}$ couplings.  For right-polarized beams, then, the
production cross section is reduced by the factor $(1-P_{e^-})
(1-P_{e^+})$ relative to the unpolarized case. For $P_{e^-}=0.8$ and
unpolarized positrons, the cross section is reduced by a factor of 5.
In the limit where beam polarization is perfect ($P_{e^-} = P_{e^+} =
1$) and $\sqrt{s} \gg m_W$, the production cross section of transverse
$W$ bosons goes to zero and the cross section is dominated by the
process with longitudinal $W$ bosons in the $s$-channel.\footnote{In
this limit, the cross section reduces to that of Goldstone boson
production, as required by the equivalence theorem.}

Another important background is $e^+e^-\to e^+e^- l^+l^-$.  This is a
background to slepton pair production when the two photons radiated
off from the initial-state $e^+$ and $e^-$ collide to produce a
$l^+l^-$ pair, while the initial-state $e^+$ and $e^-$ escape into the
beam pipe.  Note that beam polarization cannot reduce this background.
In current linear collider detector designs, electrons and positrons
traveling within $\theta_{\rm min}\sim 5$ mrad are undetected, as this
region is not covered by luminosity monitors. The maximum $\sla{p_T}$
carried by these electrons and positrons is then
\begin{equation} 
\sla{p_T}^{\text{max}} = 2.5~\gev \frac{\theta_{min}}{5\ {\rm
mrad}} \frac{\ebeam}{500~\gev} \ .
\end{equation}
Supersymmetric events with $\sla{p_T} \alt 2.5~\gev$ are therefore
obscured by this background.  Of course, the two-photon background may
be greatly reduced with $\sla{p_T}$ and acoplanarity cuts, with little
impact on most supersymmetry signals.\index{two-photon background}

\subsection{Slepton Masses}
\label{Ssec:slmasses}

\vspace*{.10in} \noindent {\em Kinematic Endpoints} \vspace*{.08in} 

At linear colliders, the parton energy is fixed, neglecting the
effects of initial state radiation, beam energy spread, and
beamstrahlung.\index{sleptons!mass measurement!by kinematic endpoints} To
a reasonable approximation, then, sleptons are produced with energy
equal to the beam energy $\ebeam=\sqrt{s}/2$, and velocity
\begin{equation}
\beta_{\tilde{l}} = \frac{p}{\ebeam} = \sqrt{1-\frac{ m^2_{\tilde{l}} }
{\ebeam^2}} \ .
\end{equation}
For sleptons produced not too far above threshold, as will typically
be the case at least initially, $\beta_{\tilde{l}}$ is substantially
smaller than 1.

The slepton then decays to $l+\chn_1$.  In the slepton's rest frame,
the lepton's energy is
\begin{equation}
E^*_l =\frac{m_{\tilde{l}}^2 -m_{\chi}^2+ m_l^2}{2 m_{\tilde{l}}} \ ,
\end{equation}
where the lepton's mass may be neglected.  Boosting to the lab frame,
the lepton's energy is
\begin{equation}
E_l (\cos\theta) = E^*_l \frac{1 + \beta_{\tilde{l}} \cos\theta }
{\sqrt{1-\beta^2_{\tilde{l}}}} \ ,
\label{Sdecaykin}
\end{equation}
where $\theta$ is the polar angle of $l$ in the slepton rest frame
measured relative to the slepton's boost direction.  In the lab frame,
the lepton's energy distribution is therefore flat.

{}From \eqref{Sdecaykin}, one can see that, provided
$\beta_{\tilde{l}}$ is significantly less than 1, the $\tilde{l}$ and
$\chn_1$ masses may be well-determined by measuring the two endpoints
of the lepton energy
distribution~\cite{unknown:1992mg,Tsukamoto:1993gt}. The relations
between the masses and endpoints are
\begin{eqnarray}
m_{\tilde{l}}^2 &=& \frac{s E_{\text{max}} E_{\text{min}}}
{(E_{\text{max}}+E_{\text{min}})^2} \\
1 - \frac{m^2_{\tilde{\chi}^0_1}}{m^2_{\tilde{l}}} &=& 
2 \frac{E_{\text{max}}+E_{\text{min}}}{\sqrt{s}}\ .
\end{eqnarray}

In Fig.~\ref{Sfig:M1} we show the muon energy distribution from
$\tilde{\mu}_R$ pair production, along with the dominant
backgrounds~\cite{Martyn:1999tc}.  The distribution is not perfectly
flat:~this distortion is caused by initial state radiation, beam
energy spread, and beamstrahlung, along with acoplanarity cuts.  Such
effects smear the energy distribution near the edges and must be
corrected through measurements of beam properties and Monte-Carlo
simulations.

\begin{figure}[tb]
\begin{center}
\includegraphics[angle=90,width=9cm]{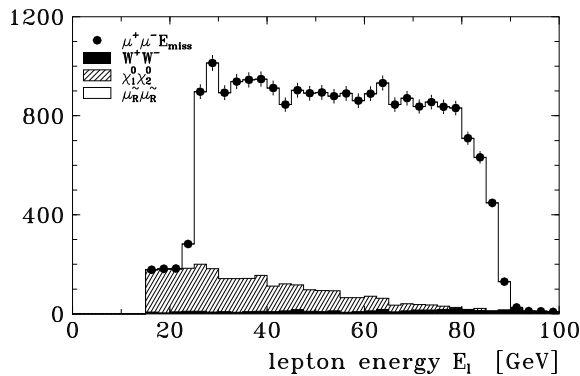}
\caption{\footnotesize Muon energy distribution from the signal $e^-_R
e^+_L \to \tilde{\mu}^+_R\tilde{\mu}^-_R \to \mu^+ \tilde{\chi}^0_1
\mu^- \tilde{\chi}^0_1$ (open), and the dominant backgrounds (shaded),
for $\sqrt{s} = 320~\gev$ and an integrated luminosity of
$160~\ifb$~\protect\cite{Martyn:1999tc}. The underlying supersymmetric
masses are $m_{\tilde{\mu}_R}=132~\gev$ and $m_{\chn_1}=71.9~\gev$.
\label{Sfig:M1} }
\end{center}
\end{figure}

Doing so, one finds that $m_{\tilde{\mu}_R}$ can be measured to within
0.4 GeV from kinematic endpoints, assuming an integrated luminosity of
$160~\ifb$ for this case.  Of course, the same measurement can be done
for selectrons, and the precision of the mass measurement is of the
same order~\cite{Dima:2001jr}.  This provides an immediate and
stringent test of scalar mass relations, such as the universality of
scalar masses at the GUT scale discussed in
\secref{Ssec:msugra}.\index{tests!of scalar mass unification}

\vspace*{.10in} \noindent {\em Minimum Mass Method} \vspace*{.08in} 

In the kinematic endpoint method, the fact that leptons come in pairs
in slepton events is never used.\index{sleptons!mass measurement!by
minimum mass method} An analysis that incorporates this pairing
information is the minimum mass method~\cite{Feng:1994sd}, where for
each event, one determines the minimum slepton mass consistent with
the measured lepton momenta and a postulated $\chn_1$ mass. The
minimum mass distribution peaks sharply at the actual
mass~\cite{Feng:1994sd,Wagner_1,Lykken:1997np,Aguilar-Saavedra:2001rg},
as may be seen in Fig.~\ref{Sfig:minmass}, making possible mass
measurements that are more precise than those from the kinematic
endpoint method discussed above. For scenarios in which the $\chn_1$
mass is well-known from some other process, such as chargino
production, the minimum mass method provides another avenue for
precise slepton mass determination.

\begin{figure}[tb]
\begin{center}
\includegraphics[angle=0,width=8cm]{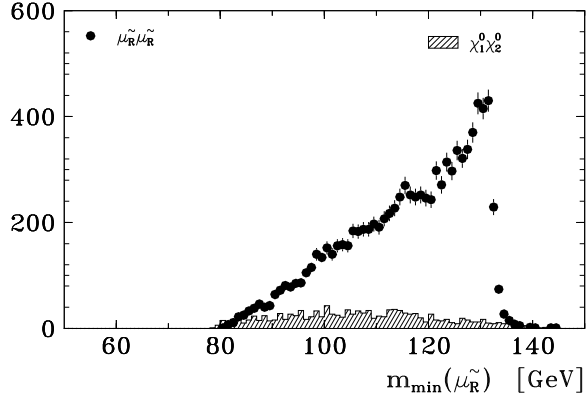}
\caption{\footnotesize Minimum mass distribution for smuons from the
process $e^-_R e^+_L \to \tilde{\mu}^+_R\tilde{\mu}^-_R \to \mu^+
\tilde{\chi}^0_1 \mu^- \tilde{\chi}^0_1$, for $\sqrt{s} = 320~\gev$
and an integrated luminosity of
$160~\ifb$~\protect\cite{Aguilar-Saavedra:2001rg}. The underlying
supersymmetry parameters are $m_{\tilde{\mu}_R}=132~\gev$ and
$m_{\chn_1}=71.9~\gev$.
\label{Sfig:minmass} }
\end{center}
\end{figure}

\vspace*{.10in} \noindent {\em Threshold Scans} \vspace*{.08in} 

Slepton masses may also be determined through threshold
scans~\cite{Feng:1998ud,Martyn:1999tc,Martyn:1999xc,Feng:2001ce,%
Freitas:2001zh,Mizukoshi:2001nc,Blochinger:2002zw}.%
\index{sleptons!mass measurement!by $\epem$ threshold scan} For
pair-production of right-handed sleptons, the required initial state
has angular momentum $J_z=1$. Since the final state particles have
spin 0, they must be produced in a $P$ wave, and the production cross
section grows as $\beta^3$ near threshold, in contrast to fermion pair
production, where the cross section grows as $\beta$.  The measurement
of this behavior is a telling check that sleptons are, in fact,
scalars.\index{sleptons!spin measurement}\index{tests!of slepton spin}
An example of this threshold behavior is given in Fig.~\ref{Sfig:M2}.
In view of the $\beta^3$ suppression, such measurements require high
luminosity, and finite width effects, threshold corrections, and
sub-dominant diagrams must all be carefully controlled.  However,
$\epem$ threshold scans may allow precise mass measurements at the
$100~\mev$ level, assuming luminosities of $\sim
100~\ifb$~\cite{Martyn:1999tc,Feng:2001ce}.

In $e^-e^-$ collisions, the relevant initial state for $\tilde{l}_R^+
\tilde{l}_R^-$ production has angular momentum $J_z = 0$, and so
slepton pair production has $\beta$ threshold
behavior~\cite{Feng:1998ud,Feng:2001ce}.\index{sleptons!mass
measurement!by $\emem$ threshold scan}\index{$\emem$ option, uses
for!mass measurements} Cross sections for $\tilde{e}_R$ pair
production in $e^-e^-$ and $e^+e^-$ modes are compared in
Fig.~\ref{Sfig:M2}.  Mass measurements with $100~\mev$ precision can
be achieved with a total luminosity of ${\cal O}(1)~\ifb$, two orders
of magnitude less than required in $e^+e^-$ collisions for similar
precision~\cite{Feng:2001ce}. The full arsenal of linear collider
modes may even allow one to extend this mass measurement to the rest
of the first-generation sleptons through a series of $\beta$ threshold
scans: $e^-e^- \to \tilde{e}^-_R \tilde{e}^-_R$ yields
$m_{\tilde{e}_R}$; $e^+e^- \to \tilde{e}^{\pm}_R \tilde{e}^{\mp}_L$
yields $m_{\tilde{e}_L}$; $e^+e^- \to \tilde{\chi}^+_1
\tilde{\chi}^-_1$ yields $m_{\tilde{\chi}^{\pm}_1}$; and $e^- \gamma
\to \tilde{\nu}_e \tilde{\chi}^-_1$ yields
$m_{\tilde{\nu}_e}$~\cite{Barger:1998qu}.  If completed, this would
yield a high precision test of
\eqref{sleptonsplitting},\index{tests!of $\tilde{l}_L$-$\tilde{\nu}_l$
mass relations} a robust prediction of supersymmetry and gauge
invariance, and possibly also a high precision measurement of
$\tan\beta$.\index{$\tan\beta$!measurement!from
$\tilde{l}_L$-$\tilde{\nu}_l$ splitting}

\begin{figure}[tb]
\begin{center}
\includegraphics[width=6cm]{scan.eps} \hspace*{-2cm}
\includegraphics[width=5.5cm]{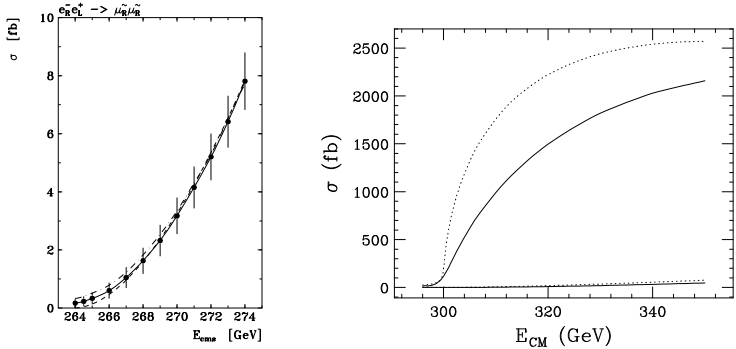}
\end{center}
\caption{\footnotesize Left: Threshold cross sections for $\epem \to
\tilde{\mu}_R^+ \tilde{\mu}_R^-$, with $m_{\tilde{\mu}}$ = 132 GeV and
smuon width $\Gamma_{\tilde{\mu}_R}=0.3$ (solid) and 0 and $0.6~\gev$
(dashed)~\protect\cite{Martyn:1999xc}. The error bars are for a 10
point scan with total integrated luminosity ${\cal L}
=100~\ifb$. Right: Threshold behavior for $\sigma(e^-e^- \to
\tilde{e}^-_R\tilde{e}^-_R)$ (upper two contours) and $\sigma(e^+e^-
\to \tilde{e}^+_R \tilde{e}^-_R)$ (lower two contours) for
$m_{\tilde{e}_R}=150$ GeV and $M_1=100$ GeV and beam polarizations
$P_{e^-} = 0.8$ and $P_{e^+} = 0$~\protect\cite{Feng:2001ce}. In each
pair, the dotted curve neglects all beam effects, and the solid curve
includes the initial state radiation, beamstrahlung, and beam energy
spread of flat beams. }
\label{Sfig:M2}
\end{figure}

\vspace*{.05in} \noindent {\em Mass Measurements at the LHC} 
\vspace*{.05in} 

The LHC may also measure superpartner masses with some precision.%
\index{sleptons!mass measurement!at LHC} At hadron
colliders, the parton-parton center-of-mass energy is unknown.  In
supersymmetric events, typically two LSPs are missing, but only the
total missing momentum can be reconstructed, and there is a large
uncertainty in the longitudinal momentum of the system. Nevertheless,
it is still possible to obtain reliable mass measurements from cascade
decays if one can isolate an invariant mass distribution from a
particular step in a cascade
decay~\cite{Hinchliffe:1996iu,Bachacou:1999zb}. For example, in the
process $\tilde{\chi}^0_2\to\tilde{l}_R l \to \tilde{\chi}^0_1
l^+l^-$, the $l^+l^-$ invariant mass distribution is predicted to be
proportional to $m_{ll}$, with maximum
\begin{equation}
m_{ll}^{\text{max}} = \sqrt{
\frac{ (m^2_{\tilde{\chi}^0_2}-m^2_{\tilde{l}_R})
(m^2_{\tilde{l}_R}-m^2_{\tilde{\chi}^0_1})}
{m^2_{\tilde{l}_R}}} \ .
\end{equation}
The endpoint of the $m_{ll}$ distribution is therefore determined by
the three masses involved in the cascade decay. An example of this
mass distribution is shown in Fig.~\ref{Sfig:M3}.  When the jets from
the parent decay $\tilde{q} \to \tilde{\chi}^0_2 q$ are also
identified, the number of identified end points increases to four: the
upper endpoints of the $m_{jl}$ and $m_{jll}$ distributions, and the
lower endpoint of the $m_{jll}$ distribution for $m_{ll} >
m_{ll}^{\text{max}}/\sqrt{2}$.  In principle, then, all 4 masses
entering the cascade may be reconstructed. The resulting mass
determinations for two sample points are also shown in
Fig.~\ref{Sfig:M3}.  The mass difference $m_{\tilde{l}_R} -
m_{\chn_1}$ is determined rather precisely. Note, however, that the
absolute sparticle masses are poorly constrained, with uncertainties
of the order of several tens of GeV.

\begin{figure}[tb]
\begin{center}
\includegraphics[width=5cm]{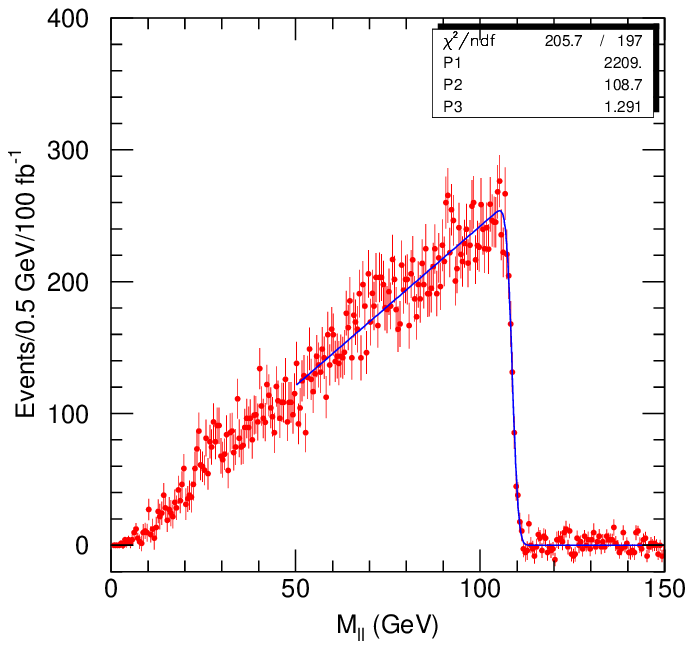}
\includegraphics[width=5cm]{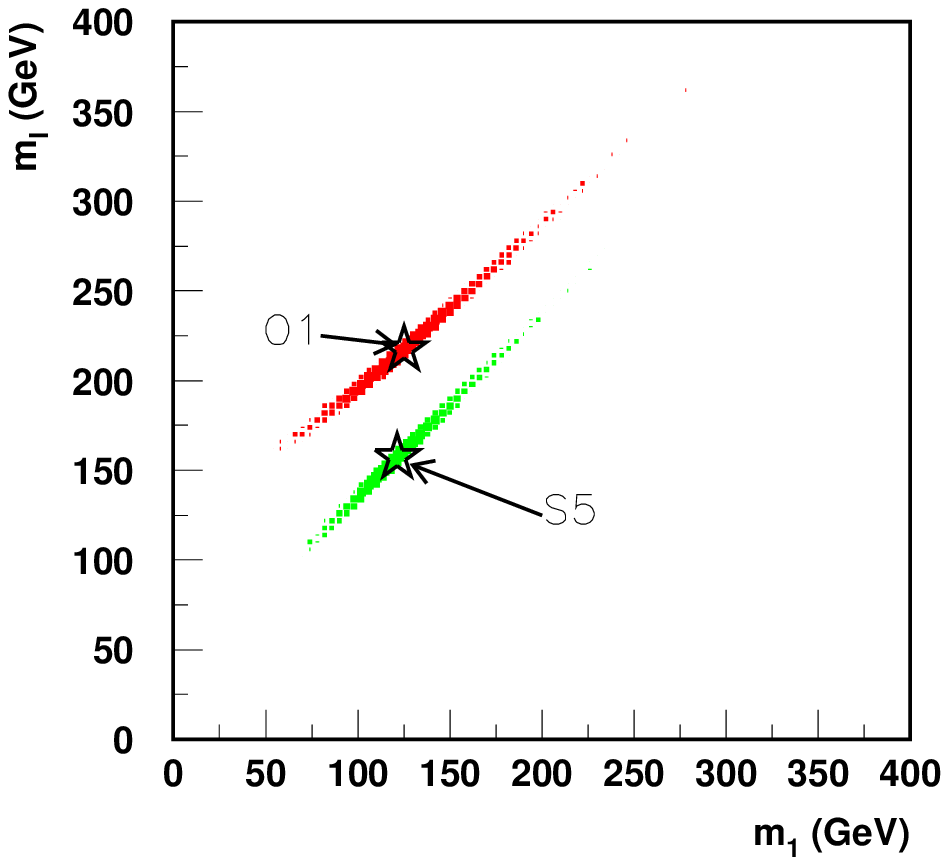}
\end{center}
\caption{\footnotesize Left: The $e^+e^- + \mu^+\mu^- -e^{\pm}
e^{\mp}$ mass distribution for LHC minimal supergravity Point 5 with
$\tilde{\chi}^0_2\to\tilde{l}l \to \tilde{\chi}^0_1
l^+l^-$~\protect\cite{AtlasTDR:1999fq}.  Right: Scatter plot of
reconstructed values of the $\tilde{l}_R$ and $\tilde{\chi}^0_1$
masses for LHC point 5 (S5) and for a different model (O1) using the
decay chain $\tilde{q}_L \to \tilde{\chi}^0_2 q \to \tilde{l}_R l q
\to \tilde{\chi}^0_1 llq$~\protect\cite{Allanach:2000kt}. }
\label{Sfig:M3}
\end{figure}

The example above illustrates the complementarity of the LHC and the
linear collider.\index{complementarity!of LC and LHC} At the LHC,
squarks and gluinos are produced with large cross sections, and their
cascade decays typical include many charginos and neutralinos, as well
as charged sleptons and sneutrinos.  Distributions within these
cascade decays may provide important kinematic information.  The LHC
therefore provides a ``top view'' of the superparticle spectrum.  On
the other hand, the linear collider produces the lighter sparticles
first and determines their properties systematically from the bottom
up.  Remarkably, the uncertainties of masses and other parameters may
be determined at each step with percent level uncertainty.  Combining
the data from the linear collider with that available at the LHC, all
sparticle masses may be determined with a precision of the order of a
few percent.

It is also noteworthy that, while the LHC determines decay patterns,
detailed determinations of the spins and couplings of sleptons and
other particles is difficult.  Using polarized beams, the linear
collider may determine these slepton properties, as we now describe.

\subsection{Polarized Cross Sections}
\label{Ssec:slcrosssection}

We now consider selectron pair
production.\index{sleptons!production!polarized} As noted in
\eqref{Spolarizedsigma}, cross sections for various beam polarizations
are the sum of four independent polarized cross sections determined by
the amplitudes ${\cal M}(h, \bar{h})$.  Each ${\cal M}(h, \bar{h})$ is
a function of gauge-sfermion-sfermion and gaugino-sfermion-fermion
couplings. For $\tilde{e}_R^+ \tilde{e}_R^-$ production,
\begin{eqnarray}\label{b1}
i{\cal M}(h, \bar{h})&=&
-i\lambda e^{i\lambda\phi}\sin\theta 
s\beta_f \left[\frac{g^2}{\cos^2 \theta_W} \frac{A_{h}A_{\frac{1}{2}}}
{s-m_Z^2+i\Gamma_Z}\right.\nonumber\\
&&\left.+\frac{e^2}{s}+\frac{(1\pm (-)^{\bar{h}+\frac{1}{2}})}{2}
\sum_{j}\frac{1}{2}
\frac{\vert A_{j R}^2\vert}{t-m_{\tilde{\chi}_j}^2}\right],
\end{eqnarray}
where $h = \pm 1/2$ is the helicity of the initial state electron,
$\bar{h} = \pm 1/2$ is the helicity of the initial state positron, and
$\lambda \equiv h - \bar{h}$.  The angles $\theta$ and $\phi$
specify the $\se^-_R$ production direction in polar coordinates,
$\beta_f$ is the selectron velocity, and $t = - \frac{s}{4} (1 -
2\cos\theta\beta_f + \beta_f^2)$.  The couplings $A_{h}$ and $A_{jR}$
are given by
\begin{eqnarray}
A_{\frac{1}{2}} &=& \sin^2\theta_W \ , \
A_{-\frac{1}{2}}=-\frac{1}{2}+\sin^2 \theta_W \\ 
A_{jR} &=& -\sqrt{2} g N_{j1} \tan\theta_W \ . 
\end{eqnarray}
The amplitude for $\tilde{e}_L^+ \tilde{e}_L^-$ pair production 
is obtained from \eqref{b1} by the replacements
\begin{eqnarray}
A_{h} &\to& A_{-h} \\
1\pm (-)^{\bar{h} + 1/2} &\to& 1\mp (-)^{\bar{h}+1/2} \\
A_{jR} &\to& A_{jL}= \frac{g}{\sqrt2} (N_{j2}+
N_{j1} \tan\theta_W ) \ .
\end{eqnarray}

The contribution of $t$-channel neutralino exchange to the amplitudes
above is easy to understand.  For $\tilde{e}_R$ pair production, the
$t$-channel amplitude contributes only to ${\cal M}(1/2,-1/2)$; for
$\tilde{e}_L$ pairs, it contributes only to ${\cal M}(-1/2,1/2)$.
This follows from the fact that gaugino interactions preserve
chirality, and so only the $\tilde{e}_{R} e_{R} \tilde{B}$ and
$\tilde{e}_{L} e_{L} \tilde{B}$ couplings are non-vanishing.
 
The $s$-channel amplitude contributes to both amplitudes ${\cal
M}(1/2,-1/2)$ and ${\cal M}(-1/2, 1/2)$.  In the limit $s \gg m_Z^2$,
and for right-polarized $e^-$ beams with $h = 1/2$, one finds
\begin{eqnarray}
e^+e^-_R\to \se^+_R\se^-_R && i {\cal M}(1/2, -1/2) = -i C g'^2
\label{SselectronR} \\
e^+e^-_R\to \se^+_L\se^-_L && i {\cal M}(1/2,-1/2) = -i C \frac{1}{2}
g'^2 \ , \label{SselectronL}
\end{eqnarray}
where $C= \lambda e^{i\lambda\phi} \sin\theta \beta_f$.  In this
high-energy limit, the $Z$ mass may be neglected, and the $\gamma$ and
$Z$ mass eigenstates may be replaced by the gauge eigenstates $B$ and
$W^0$.  Since $e_R$ does not couple to $W^0$, only the hypercharge
gauge boson contributes, and the amplitudes are proportional to the
hypercharge coupling $g'$, with hypercharges $1$ and $1/2$ for
$\tilde{e}_R$ and $\tilde{e}_L$.
 
This example illustrates the power of beam polarization.\index{beam
polarization!power of} By polarizing the beam, we can switch off some
Feynman diagrams `by hand.' One may then isolate particular amplitudes
and determine fundamental parameters
precisely~\cite{unknown:1992mg,Tsukamoto:1993gt}.  For example, if we
use right-polarized $e^-$ beams and observe $\tilde{e}_R$ and
$\tilde{e}_L$ pair production with the rates predicted by
\eqsref{SselectronR}{SselectronL}, we can establish the
chirality-preserving nature of the supersymmetric gaugino
interactions.\index{tests!of chirality preservation in gaugino
interactions}

For $e^+e^-\to \tilde{e}_L^+\tilde{e}^-_R$ production, the $s$-channel
amplitude is absent because gauge interactions preserve chirality.
The same is true for $e^-e^- \to \tilde{e}^-_{R(L)}
\tilde{e}^-_{R(L)}$, where the $s$-channel amplitude is forbidden by
total lepton number conservation.  In this case, the $t$-channel
amplitude has a new feature:~it is proportional to the mass of the
neutralino exchanged in the $t$-channel, which is required to flip
chirality.  For example,
\begin{eqnarray}
e^+e^-\to \se^+_L\se^-_R &&
i{\cal M}(h, \bar{h})
= i \sqrt{s} \delta_{\lambda,0 }
\left[1+(-)^{\bar{h} - \frac{1}{2}}\right] \nonumber \\
&& \qquad \qquad \times \sum_j \frac{1}{2} m_{\tilde{\chi}^0_j} 
\frac{A_{jL} A_{jR}}{t-m_{\tilde{\chi}^0_j}^2} \ .
\end{eqnarray}
The existence of the $t$-channel amplitude for $e^-e^- \to \tilde{e}^-
\tilde{e}^-$ proves that the exchanged $t$-channel particle is a
Majorana fermion.  Note also that the amplitude is now not
proportional to the velocity $\beta$, and so the cross section now
grows like $\beta$ near threshold.  This behavior was discussed in
\secref{Ssec:slmasses} in the context of threshold mass measurements.
Finally, the amplitude is proportional to $M_1$ in both of the limits
$M_1\ll M_2, |\mu|$ and $M_1\gg M_2, |\mu|$.  The total cross section
is therefore directly related to fundamental mass parameters.

At high enough beam energies, selectrons will be produced through all
mechanisms ($\se^+_L\se^-_L$, $\se^+_R\se^-_R$, and
$\se^{\pm}_L\se^{\mp}_R$). All of these processes may result in the
signal $e^+e^- + \sla{p_T}$, as discussed above in
\secref{Ssec:slsgbg}.  However, the various signals may often be
disentangled, both kinematically, as the energy ranges of the
electrons and positrons differ for the different processes, and by
comparing the cross sections with various beam polarizations.

Slepton production cross sections have interesting dependences on the
fundamental supersymmetry parameters.  Fig.~\ref{Sfig:M4} shows cross
sections for $\tilde{e}_R$ pair production in the $(M_1, \tan\beta)$
plane. Here we fix the LSP $\tilde{\chi}^0_1$ mass. The GUT relation
for $M_1$ and $M_2$ is also assumed. Not surprisingly, there is almost
no dependence on $\tan\beta$.  This dependence enters only through the
neutralino mass matrix in the form of $\cos\beta$ and $\sin\beta$,
which are both fairly constant for moderate and large $\tan\beta$.
For a completely right-polarized beam, the only state exchanged in the
$t$-channel is the Bino.  This $t$-channel contribution dominates at
the left-hand side of the plot, and then decouples for large $M_1$.
At the left-hand side of the plot, the LSP is Bino-like, but becomes
Higgsino-like on the right.  However, the production proceeds through
the neutralino with largest $\tilde{B}$ component. The $\tan\beta$
independence of the cross section is even more prominent for the
process $e^- e^-\to \tilde{e}^-\tilde{e}^-$, where chirality requires
a gaugino mass insertion, which is independent of neutralino mixings.
This process also provides sensitivity to even very large $M_1$, and
may allow for a measurement of the Bino mass even when the Bino is far
from kinematically
accessible~\cite{Feng:2001ce,Blochinger:2002zw}.\index{tests!of
gaugino Majorana-ness}

\begin{figure}[tb]
\begin{center}
\includegraphics[width=8cm]{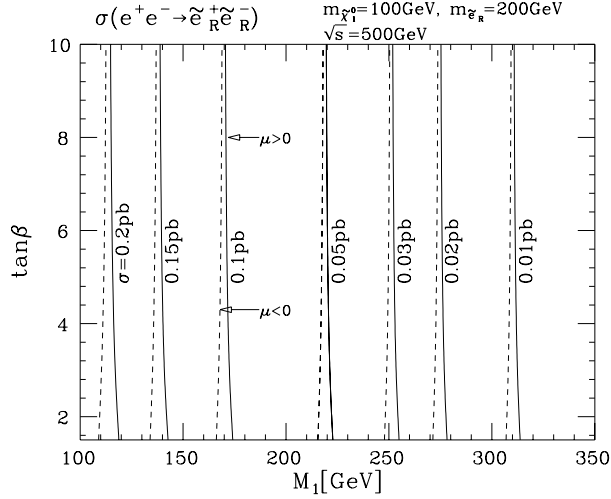}
\end{center}
\caption{\footnotesize Contours of $\sigma(e^+ e^-_R \to \tilde{e}_R^+
\tilde{e}_R^-)$ for $m_{\tilde{e}_R}=200~\gev$ and
$\protect\sqrt{s}=500~\gev$ in the $(M_1,\tan\beta)$
plane~\protect\cite{Nojiri:1996fp}.  At each point in the plane, $\mu$
is chosen so that $\mchi=100~\gev$. Solid lines correspond to a
$\mu>0$ solution and dashed lines to $\mu<0$.
\label{Sfig:M4}}
\end{figure}

We have concentrated on selectrons here, as they have many interesting
properties related to the existence of $t$-channel amplitudes.  It
should be stressed that smuon and stau studies are also interesting.
In particular, stau studies may be very important at the linear
collider. The lighter stau is often the lightest scalar superpartner
as a result of GUT-scale interactions (see \secref{Ssec:GUTmodels}),
renormalization group evolution between the GUT and weak scales (see
\secref{Ssec:msugra}), and left-right mixing (see
\secref{Ssec:sleptons}).\index{staus!lightness of} Note that staus may
also be studied at the LHC, but such studies are significantly
complicated by $\tau$ decay to low energy leptons or mesons.

The signal of $\tilde{\tau}$ production is very complicated and will
be considered in more detail in \secref{Ssec:stau}. Here we merely
note that the production amplitude for $\epem \to \tilde{\tau}^+_1
\tilde{\tau}^-_1$ in the limit $\sqrt{s}\gg m_Z$ for right-polarized
electrons is~\cite{Nojiri:1994it}\index{staus!production}
\begin{equation}
i {\cal M}(1/2, -1/2) = -i C g'^2 
\left(\frac{1}{2}\cos^2 \theta_{\tilde{\tau}}
+\sin^2 \theta_{\tilde{\tau}} \right) \ ,
\end{equation}
where $\theta_{\tilde{\tau}}$ is the stau left-right mixing
angle.\index{staus!left-right mixing} The measurement of the
$\tilde{\tau}$ production cross section for right-polarized electron
beams will determine $\theta_{\tilde{\tau}}$.  If one finds
$\tilde{\tau}_2$ in addition to $\tilde{\tau}_1$, the full
$\tilde{\tau}$ mass matrix may be reconstructed.  The reconstruction
of the $\tilde{\tau}$ mass matrix will provide yet another
opportunities to test the universality of scalar masses at the high
scale.\index{tests!of scalar mass unification} In addition, the
off-diagonal stau mass $m^2_{LR}$ is proportional to $\tan\beta$ and
$\mu$ as noted in \eqref{staumassmatrix}, and the decay distribution
also depends on $\tan\beta$, as we will see.  Stau studies may
therefore provide a rare opportunity to measure $\tan\beta$ for large
$\tan\beta$.\index{$\tan\beta$!measurement!from staus}

\subsection{Lepton Flavor Violation}
\label{Ssec:lfv}

As noted in \secref{Ssec:sleptons}, there is no reason for the fermion
and sfermion mass matrices to be simultaneously
diagonalizable.\index{flavor
violation!lepton}\index{sleptons!mixing!flavor} This implies that the
discovery of superpartners will lead to a whole new sector of flavor
physics to explore.  In the standard model, after the initial
discoveries and mass measurements, the focus has naturally turned to
measurements of flavor violation, such as in the quark and neutrino
sectors.  A similar progression may be expected if superpartners are
discovered.

At the moment, there is no standard explanation of fermion
masses.\index{problems!standard model flavor} Recently, our knowledge
of neutrino mixing has expanded tremendously, but no compelling theory
of flavor has emerged.  One might wonder if supersymmetric flavor
studies will also lead to a bewildering wealth of data without
furthering our fundamental understanding of flavor.  While possible,
there are important differences in these two cases.  In the case of
neutrinos, the newly discovered mixings are in principle unrelated to
those of the quarks.  For example, in theories that attempt to explain
flavor through horizontal flavor symmetries, neutrino mixing is
dependent on a whole new set of flavor representation assignments.
For superpartners, however, the new mixings are tied to the flavor
properties of standard model fermions by supersymmetry.  For example,
standard model particles and their superpartners must be governed by
the same flavor symmetries, since they are in the same supermultiplet.
In these frameworks, then, superpartners do not introduce additional
degrees of freedom, but rather provide new constraints on the same
flavor physics governing the standard model particles.  For this
reason, the careful investigation of superpartner flavor symmetries
provides a promising avenue for understanding not only superpartner
properties, but also the masses and mixings observed in the standard
model.

The full $6\times 6$ slepton mass matrix of \eqref{Sfullslepton} is
very complicated.  As a first step, one may begin by neglecting
left-right mixing, a reasonable approximation for selectrons and
smuons.  One can further specialize to two generation mixing and
consider, for example, $\tilde{e}_R$-$\tilde{\mu}_R$ mixing.  A
convenient basis to consider is the mass eigenstate basis for both
charged leptons and sleptons, in which all mixing is confined to
gaugino vertices.  Lepton flavor violation (LFV) then occurs in
slepton pair production through $t$-channel neutralino exchange and in
decay vertices~\cite{Arkani-Hamed:1996au,Arkani-Hamed:1997km}.  In
simple scenarios where the sleptons decay directly to a neutralino
LSP, the resulting signals of flavor-violating slepton pair production
at the linear collider are $e^+ e^- \to \tilde{l}^+ \tilde{l}^- \to
e^{\pm} \mu^{\mp} \tilde{\chi}^0_1 \tilde{\chi}^0_1$ and $e^- e^- \to
\tilde{l}^- \tilde{l}^- \to e^- \mu^- \tilde{\chi}^0_1
\tilde{\chi}^0_1$.\index{flavor violation!lepton!collider signals}

The probability for flavor-violating decay is
\begin{equation}
P(\tilde{e}_R \to \mu) = \frac{1}{2} \sin^2 2 \theta_R \,
\frac{(\Delta m^2_R)^2}{4 m_R^2 \Gamma^2 + 
(\Delta m_R^2)^2} \ ,
\end{equation}
where $\Delta m^2_R = m_{1}^{2} - m_{2}^{2}$ and $m_R = (m_{1} +
m_{2})/2$, with $m_1$ and $m_2$ the physical slepton masses, and
$\theta_R$ is the flavor mixing angle.  $\Gamma$ is the slepton decay
width.  If $m_R \Gamma \ll \Delta m_R^{2}$, the mixing
probability has its maximal value of $(\sin^2 2 \theta_R)/2$; however,
for $m_R \Gamma \gg \Delta m_R^{2}$, $P$ vanishes, since
sleptons decay before they have time to mix.

In the $\epem$ case, flavor violation requires a careful treatment of
$t$-channel and $s$-channel interference, as flavor violation is
present in $t$-channel diagrams, but absent in $s$-channel
processes.\index{sleptons!flavor mixing!at $\epem$ collider} However,
as with neutrino studies, results are conveniently presented in the
$(\sin 2\theta_R, \Delta m_R^2)$ plane.  In Fig.~\ref{Sfig:MLFV},
cross sections for the flavor-violating signal are given, along with
the discovery reach of the linear collider.  Beam polarization again
provides a useful tool to reduce background from processes such as
$W^+ W^-$ and $e^+ \nu W^-$.\index{beam polarization!use in flavor
measurements} Given an integrated luminosity of $50~\ifb$, $e$-$\mu$
flavor violation can be discovered for mixing angles of order
$\theta_R \sim 0.05$ and mass degeneracies at the 1\% level.  For the
same luminosity, the $e^-e^-$ mode provides an even better discovery
potential, as backgrounds such as $W^+ W^-$ are completely prohibited
by total lepton number conservation.\index{sleptons!flavor mixing!at
$\emem$ collider}\index{$\emem$ option, uses for!flavor measurements}

\begin{figure}[tb]
\begin{center}
\includegraphics[width=8cm]{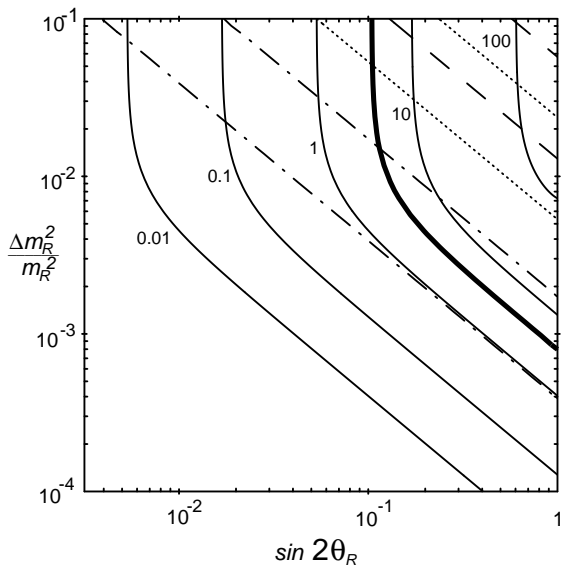}
\end{center}
\caption{\footnotesize The solid contours are for $\sigma (e^+e^-_R\to
e^{\pm} \mu^{\mp} \tilde{\chi}^0_1 \tilde{\chi}^0_1 )$ in fb for
$\protect\sqrt{s} = 500~\gev$, $m_{\tilde{e}_R}, m_{\tilde{\mu}_R}
\approx 200~\gev$, and $M_1 =
100~\gev$~\protect\cite{Arkani-Hamed:1996au}.  The thick contour
represents the LFV discovery reach given $50~\ifb$. Constant contours
of $B(\mu \to e\gamma)=4.9\times 10^{-11}$ and $2.5\times 10^{-12}$
are also plotted for degenerate left-handed sleptons with mass 350 GeV
and $-(A + \mu \tan\beta)/m_R = 0$ (dotted), 2 (dashed), and 50
(dot-dashed).
\label{Sfig:MLFV}}
\end{figure}

Supersymmetric flavor violation is already constrained by low energy
data, such as $\mu$-$e$ conversion and $\mu \to e
\gamma$.\index{flavor violation!lepton!low energy} In
Fig.~\ref{Sfig:MLFV}, these constraints are given by straight lines.
These are highly dependent on supersymmetry parameters, notably
$\tan\beta$, but are suppressed by $\Delta m_R^2 / m_R^2$ through the
supersymmetric analogue of the GIM mechanism. Note that the mass
splitting below which LFV is suppressed is set by the slepton width
$\Gamma$ for the collider signal, but by the slepton mass $m_R$ for
low energy constraints.  Since $\Gamma \ll m_R$, there is a large
range of mass splittings in which LFV is suppressed in low energy
experiments, but observable at colliders.

At the same time, because the high-energy and low-energy LFV rates
have different functional dependences on the mass splitting and mixing
angle, simultaneous measurements of LFV in both high- and low-energy
experiments will provide complementary
information.\index{complementarity!of low and high energy experiments}
The combined results may even allow the extraction of mass splittings
and mixing angles separately, providing valuable information for
attempts to identify the fundamental origin of quark and lepton masses
and mixings.

\subsection{Tau Polarization from Stau Decay}
\label{Ssec:stau}

The polarizations of all leptons resulting from slepton decay carry
information about the underlying supersymmetry parameters.  Typically,
of course, these polarizations are unobservable.  However, in the case
of taus, polarization can be measured at colliders.  Here we describe
the parameter dependence of the polarization of taus from stau decay
and show how this polarization can be reconstructed experimentally.

Tau leptons produced in stau decay have the simple energy distribution
of \eqref{Sdecaykin}. The $\tau$ decays into $A\nu_{\tau}$, where $A=
e\nu_e, \mu\nu_{\mu}, \pi, \rho, a_1$. The heavy mesons further decay
into pions through $\rho^{\pm}\to \pi^{\pm}\pi^0$ and $ a_1^{\pm}\to
\pi^{\pm}\pi^{\mp}\pi^{\pm}, \pi^{\pm}\pi^0\pi^0$. The signature of
$\tilde{\tau}$ pair production is therefore two acoplanar jets with
low multiplicity.\index{staus!signals}

Because $E_{\tau}\gg m_{\tau}$, the decay products maintain the
original direction of the parent $\tau$. However, the total energy of
the decay products is substantially reduced since the tau neutrinos
escape detection.  The $E_{\text{jet}(\tau)}$ distribution is
therefore not flat, but decreases to zero at the endpoint, degrading
the sharp edge expected in $\tilde{\mu}$ or $\tilde{e}$ pair
production.  Fits to the total $E_{\text{jet}}$ distribution yield
both $m_{\tilde{\tau}}$ and $m_{\tilde{\chi}^0_1}$, but the
sensitivity to $m_{\tilde{\tau}}$ is reduced.

On the positive side, however, this jet energy distribution contains
information about the $\tau$ polarization, which in turn depends on
the stau and neutralino mass eigenstates.  The chirality of a sfermion
is preserved in the resulting fermion in gauge, and therefore, gaugino
interactions.  On the contrary, Higgs, and therefore Higgsino,
interactions flip the chirality.  This is depicted in
Fig.~\ref{Sfig:M6}, where the arrows indicate the flow of
chirality. The two types of interaction also have different couplings,
with the $\tau \tilde{\tau} \tilde{H}$ coupling proportional to tau
Yukawa coupling $y_{\tau} \propto m_{\tau}/\cos\beta$ and the $\tau
\tilde{\tau} \tilde{B}$ and $\tau \tilde{\tau} \tilde{W}^0$ couplings
proportional to gauge couplings $g_1$ and $g_2$, respectively.

\begin{figure}[tb]
\begin{center}
\includegraphics[width=8cm]{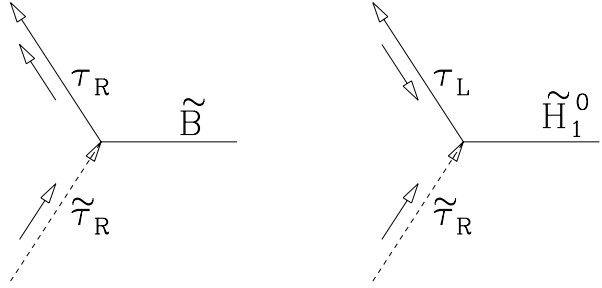}
\end{center}
\caption{\footnotesize The flow of chirality in interactions of
$\stau_R$ with gauginos and Higgsinos.
\label{Sfig:M6}}
\end{figure}

For a general stau mass eigenstate, in the limit where the lightest
neutralino is a pure Bino state, the $\tau$ polarization is given by
\begin{equation}
P_{\tau}(\tilde{\tau}_1\to \tilde{B}\tau)
= \frac{4\sin^2 \theta_{\tau} - \cos^2\theta_{\tau}}
{4\sin^2 \theta_{\tau} + \cos^2\theta_{\tau}} \ ,
\end{equation}
depending only on $\theta_{\tau}$, as expected.  On the other hand, if
the neutralino is a mixed object, there is a non-trivial dependence on
both gauge and Yukawa couplings.  For a general neutralino eigenstate,
in the limit of a pure $\stau_R$ state, the resulting tau polarization
is
\begin{equation}
P_{\tau}(\tilde{\tau}_R\to \tilde{\chi}^0_1\tau) 
= \frac{ (g\sqrt{2}N_{11}\tan\theta_W)^2-(y_{\tau}N_{13})^2 }
{(g\sqrt{2}N_{11}\tan\theta_W)^2+(y_{\tau}N_{13})^2} \ .
\end{equation}
Because $\theta_{\tilde{\tau}}$ can be measured through the
polarization dependence of the stau production cross section, as noted
in \secref{Ssec:slcrosssection}, a measurement of the $\tau$
polarization in this case will provide information on the tau Yukawa
coupling and $\tan\beta$.\index{$\tan\beta$!measurement!from staus} As
$y_{\tau}= g m_{\tau} / (\sqrt{2} m_W \cos\beta)$ is comparable to
gauge couplings only for large $\tan\beta$, the sensitivity to
$\tan\beta$ is high only for large $\tan\beta \agt 10$ when the
charginos and neutralinos involved have significant Higgsino
components.

The branching fraction of taus to $\rho$ mesons is large, with $B(\tau
\to \rho) \approx 23\%$, and the $\rho$ polarization is mostly
longitudinal for $\tau_R$ and transverse for $\tau_L$.  One can thus
determine $P_{\tau}$ by measuring $P_\rho$.  The pion energy
distributions in $\rho_{L(T)} \to \pi^{\pm}\pi^0$ decay are simple
functions of $z_c= E_{\pi^{\pm}}/E_{\text{jet}}$, where
$E_{\text{jet}}$ is the total energy of the jet to which the
$\pi^{\pm}$ belongs.  These distributions are
\begin{eqnarray}
\frac{d\Gamma(\rho_T\to 2\pi)}{dz_c} &\sim& 2 z_c(1-z_c) -
2 m^2_{\pi}/m^2_{\rho} \nonumber \\
\frac{d\Gamma(\rho_L\to 2\pi)}{dz_c} &\sim& (2z_c-1)^2 \ ,
\end{eqnarray}
where $(1-\beta_{\pi})/2 \le z_c\le (1+ \beta_{\pi})/2$ and $\beta_\pi
= \sqrt{1-4m^2_{\pi}/m^2_{\rho} }\,$.

A detailed Monte-Carlo study of tau polarization is done in
Ref.~\cite{Nojiri:1996fp}.  In this work, detector granularity is
incorporated in the detector simulation, and tracking information is
used to subtract the charged track fraction of the energy. The study
finds reasonable $\pi$, $\rho$, and $a_1$ separation through the
measurement of jet invariant mass, and $\rho$ candidates are used to
determine the $\tau$ polarization.  Determination of $P_{\tau}$ by
measuring $E_{\pi^+}/E_{jet}$ with the error of $\delta P_{\tau}=
0.08$ appears possible given $10^4$ stau pairs. (See
Fig.~\ref{Sfig:M8}.) However, conclusive results may require a full
detector simulation.

\begin{figure}[tb]
\begin{center}
\includegraphics[width=5cm]{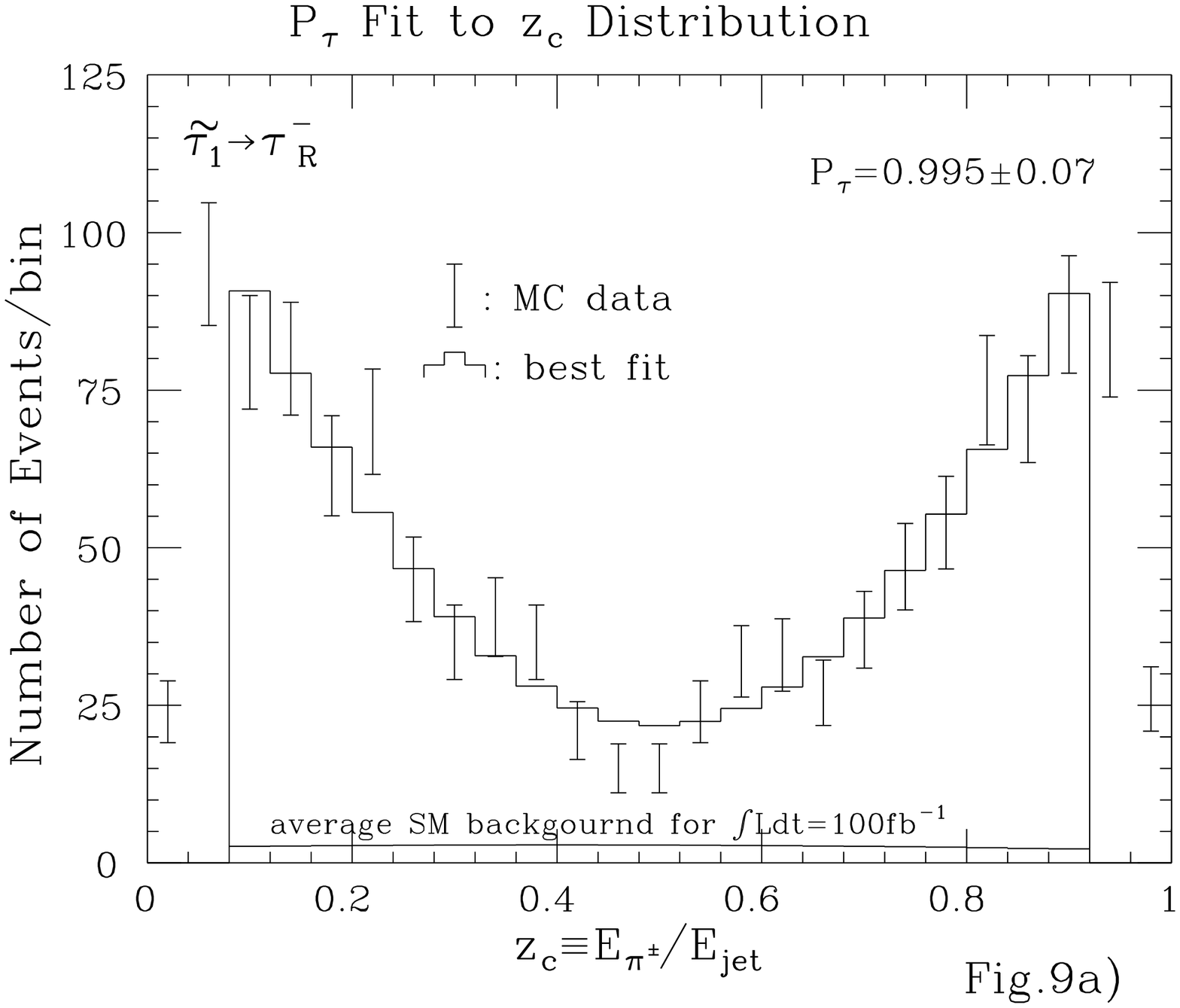}
\includegraphics[width=5cm]{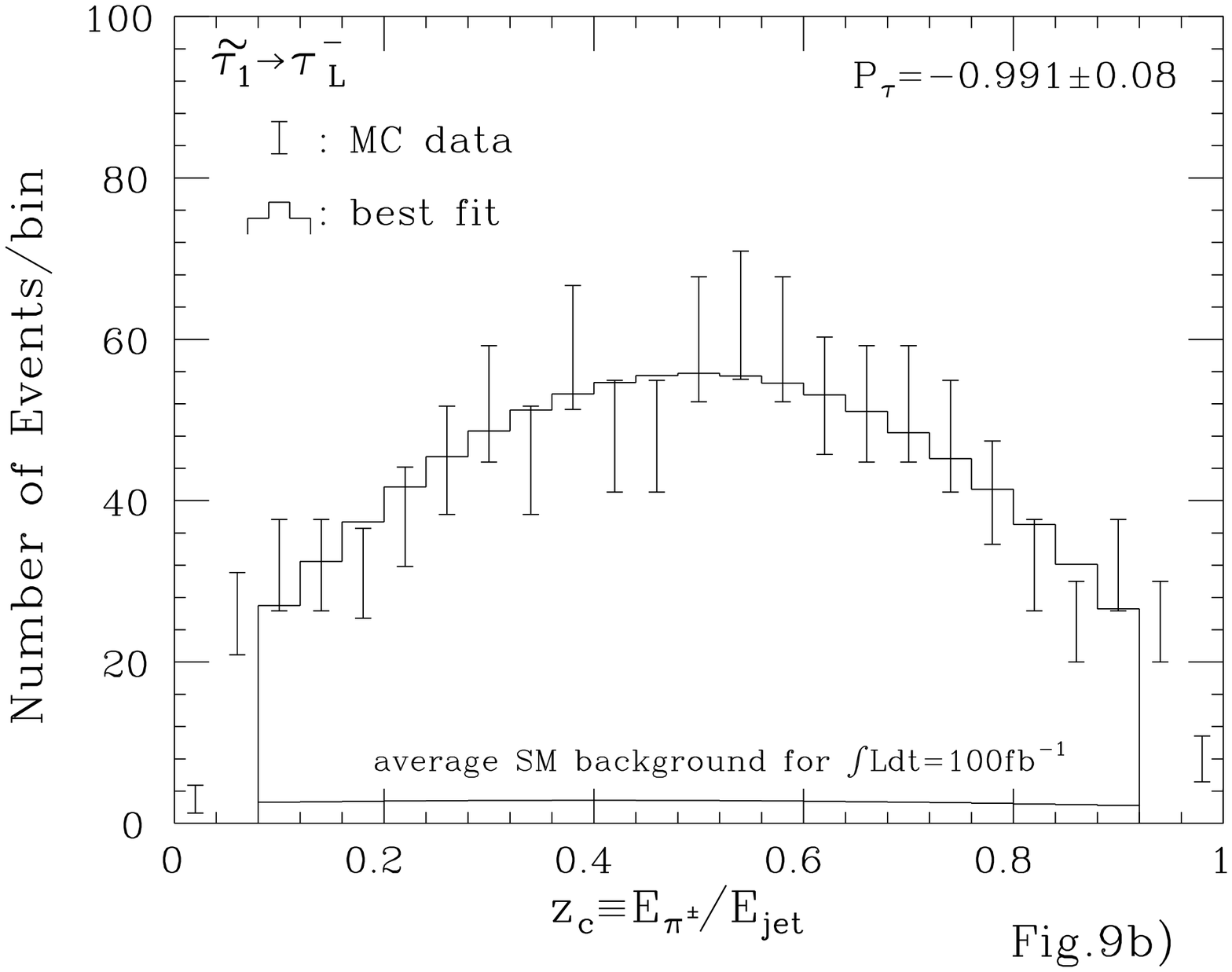}
\end{center}
\caption{\footnotesize Distributions in $z_c$ for $\rho$ candidates
selected from $10^4$ $\sti$ pairs decaying exclusively into $\tau_R$
(left) and $\tau_L$ (right), together with the best fit
histograms~\protect\cite{Nojiri:1996fp}.
\label{Sfig:M8} }
\end{figure}

\section{Chargino and Neutralino Studies}
\label{Ssec:chaneustudy}

\subsection{Signal and Background}
\label{Ssec:sigbg}

The gaugino mass relations $M_1 : M_2 : M_3 \simeq 1:2:7$ are
predicted in scenarios with unified gaugino masses at the GUT scale,
and also in gauge-mediated supersymmetry breaking scenarios.  This
hierarchy implies that Binos and Winos, or more generally, charginos
and neutralinos, are among the lightest superpartners, and so are
among the most likely to be produced at the linear collider.
 
The relevant production processes include
\begin{equation}
e^+e^- \to \tilde{\chi}^0_1\tilde{\chi}^0_1, \ 
\tilde{\chi}^0_1\tilde{\chi}^0_2, \
\tilde{\chi}^{\pm}_1\tilde{\chi}^{\mp}_1, \ 
\tilde{\chi}^0_2 \tilde{\chi}^0_2 \ .
\label{chaneuprocesses}
\end{equation}
In favorable scenarios where $|\mu| \sim M_1, M_2$, such as the focus
point models discussed in \secref{Ssec:focuspoint}, the
$\tilde{\chi}^{\pm}_2$, $\tilde{\chi}^0_3$, and $\tilde{\chi}^0_4$
states may also be produced, allowing access to the entire neutralino
and chargino systems.  Neutralino production proceeds through
$s$-channel $Z$ boson exchange and $t$-channel $\tilde{e}$
exchange.\index{neutralinos!production} Charginos are produced through
$s$-channel exchange of $\gamma$ and $Z$ and $t$-channel
$\tilde{\nu}_e$ exchange.\index{charginos!production}

In many models, the lightest neutralino $\tilde{\chi}^0_1$ is stable
and therefore invisible. In this case, the first process of
\eqref{chaneuprocesses} is unobservable. (It is a frustrating fact
that thousands of neutralinos may already have been produced in
colliders without our knowledge.)\index{frustration!of invisible
neutralinos} The heavier states typically decay visibly, however, with
decay modes such as\index{charginos!decay}\index{neutralinos!decay}
\begin{eqnarray}
\tilde{\chi}^+_1 &\to& W^+ \tilde{\chi}^0_1\cr
&\to& \tilde{\tau}^+ \nu,\  \tilde{l}^+ \nu,\  \tilde{\nu} l^+\cr 
&\to& \bar{q}'q \tilde{\chi}^0_1,\  l^+\nu\tilde{\chi}^0_1 \cr
\tilde{\chi}^0_2 &\to& h\tilde{\chi}^0_1,\   Z\tilde{\chi}^0_1\cr
&\to& \tau\tilde{\tau},\  l\tilde{l}\cr
&\to& ll\tilde{\chi}^0_1,\  \nu\bar{\nu}\tilde{\chi}^0_1,\  
 \bar{q}q\tilde{\chi}^0_1 \ .
\end{eqnarray}
The signal of chargino pair production is therefore missing momentum
accompanied by $4j$, $2j + l$, or $l^+l'^-$.\index{charginos!signals}
Neutralino pair signals are similar, but with each charged lepton
replaced by a same flavor lepton pair.\index{neutralinos!signals} Note
that if the lightest neutralino is stable, the second lightest
neutralino is also invisible if it decays to
$\nu\bar\nu\tilde{\chi}^0_1$.

As in the case of sleptons, the dominant backgrounds for chargino and
neutralino events are again various standard model gauge boson
processes, such as $WW$ and
$ZZ$.\index{charginos!backgrounds}\index{neutralinos!backgrounds} These
may be removed with cuts similar to those used for slepton production.
Note, however, that Wino-like states couple only to left-handed
(s)electrons. In the case of pair production of Wino-like states,
then, a right-polarized electron beam reduces both the signal and
backgrounds.\index{beam polarization!difficulty for Winos}

\subsection{Masses and Polarized Cross Sections}
\label{Ssec:chmass}

Chargino production is among the most well-studied supersymmetric
processes, and we will concentrate on charginos here, although
neutralinos also provide many insights.  The fundamental supersymmetry
parameters entering the chargino mass matrix of \eqref{Schamass} are
$\mu$, $M_2$, and $\tan\beta$.  These determine not only the chargino
masses, but also the chargino mixing matrices $U_{ij}$ and $V_{ij}$ of
\eqref{UVmixingangles}.  The production cross section also depends on
the sneutrino mass that enters through the $t$-channel process.  The
chargino pair production cross section is therefore determined by four
supersymmetry parameters:
\begin{equation}
\mu\, , \  M_2\, , \  \tan\beta\, , \  m_{\tilde{\nu}_e} \ .
\label{Schaparameters}
\end{equation}
In addition to these, the observables of chargino events depend on the
parameters entering chargino decay.  These include $M_1$ and other
sparticle masses.  Note that chargino production and decay are
intertwined through spin correlations, and so cannot be treated
separately in principle.

The flexibility of the linear collider provides the possibility of
disentangling all of these parameters even if only one chargino state
is kinematically accessible~\cite{Tsukamoto:1993gt,Feng:1995zd,%
Choi:1998ut,Moortgat-Pick:1999yj}.  First, for center-of-mass energies
reasonably far above threshold, the impact of decay parameters on many
observables may be reduced to low levels.  Second, beam polarization
allows a variety of cross section measurements with different
dependences on the fundamental parameters.\index{beam polarization!use
for charginos} Among the most important quantities are
\begin{equation}
m_{\tilde{\chi}^+_1}\, , \  \sigma_R\, , \  A^{FB}_R
\, , \ \sigma_L\, , \   A^{FB}_L \ ,
\label{Schaobservables}
\end{equation}
where $\sigma$ and $A^{FB}$ are the total cross section and
forward-backward asymmetry for chargino production, and the subscripts
$L$ and $R$ denote left- and right-polarized electron beams.  Each of
these quantities has an interesting and unique dependence on the
fundamental parameters.

The chargino mass $m_{\tilde{\chi}^+_1}$ depends in principle on
$M_2$, $|\mu|$, and $\tan\beta$.  However, when the off-diagonal
entries are small compared to the diagonal entries, as is often the
case, $m_{\tilde{\chi}^+_1}$ provides essentially a direct measurement
of $M_2$ or $|\mu|$, whichever is smaller.  The chargino mass can be
measured by kinematic endpoints through analyses similar to those
described in \secref{Ssec:slmasses}.  It may also be measured very
precisely through threshold scans, since the cross section is
proportional to $\beta_{\tilde{\chi}^+_1}$, as discussed in
\secref{Ssec:slmasses}.\index{charginos!mass measurements}

Although the chargino cross section has both $s$-channel and
$t$-channel contributions in general, the $t$-channel sneutrino
contribution is eliminated in the limit of purely right-polarized
electron beams.\index{beam polarization!power of} The right-polarized
cross section therefore contains only $s$-channel contributions
mediated by $\gamma$ and $Z$ gauge bosons.  As discussed in
\secref{Ssec:slcrosssection}, in the high energy limit $\sqrt{s} \gg
m_Z$ where the $Z$ mass is negligible, the exchanged gauge bosons may
be replaced by the gauge eigenstates $B$ and $W^0$ to an excellent
approximation.  The $W^0$ contribution is absent for right-polarized
electrons, while $B$ exchange is absent for Wino-like charginos.  The
right-polarized cross section $\sigma_R$ is therefore highly
suppressed for Winos and is a sensitive measure of the Higgsino
content of the chargino.\index{charginos!mixing measurements} This can
be seen in Fig.~\ref{Sfig:fengch}.  For $M_2 \ll |\mu|$ where the
lighter chargino is Wino-like, $\sigma_R$ is all but absent, while for
$M_2 \gg |\mu|$, where the lighter chargino is Higgsino-like,
$\sigma_R$ is large.  Note that $\sigma_R$ becomes near maximal even
very close to threshold (the hatched region) as a result of the
$\beta$ threshold behavior.  $A^{FB}_R$ is also sensitive to chargino
mixing, with a dependence different from $\sigma_R$. The
right-polarized quantities $\sigma_R$ and $A^{FB}_R$ therefore provide
information on chargino mixing that is highly complementary to the
information provided by
$m_{\tilde{\chi}^+_1}$.\index{complementarity!of polarized and
unpolarized measurements}

\begin{figure}[tb]
\begin{center}
\includegraphics[width=8.0cm]{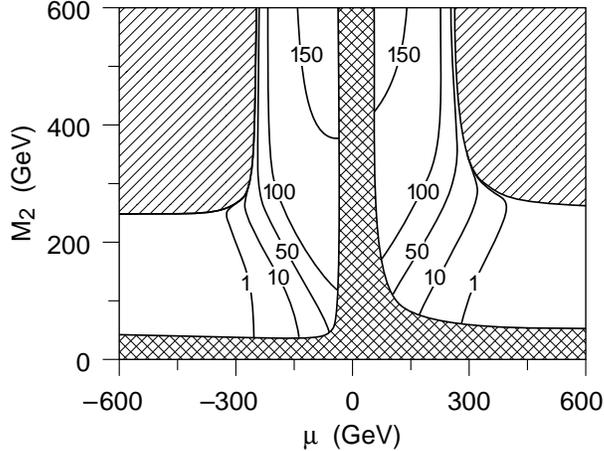}
\end{center}
\caption{\footnotesize Contours of $\sigma(e^+e^-_R \to
\tilde{\chi}^+_1 \tilde{\chi}^-_1)$ (in fb) in the $(\mu, M_2)$ plane
for fixed $\tan\beta=4$ and $\sqrt{s}=500~\gev$.  The cross-hatched
region is excluded by current bounds, and charginos are kinematically
inaccessible in the hatched region~\protect\cite{Feng:1995zd}.
\label{Sfig:fengch}}
\end{figure}

Finally, the left-polarized quantities $\sigma_L$ and $A^{FB}_L$ are
sensitive to the mass of the exchanged sneutrino.  The $t$-channel
amplitude is proportional to $1/(t-m^2_{\tilde{\nu}_e})$ where $t =
m_{\tilde{\chi}^+_1}-(1-\beta_{\tilde{\chi}^+_1}\cos\theta) s/2$ and
$\theta$ is the angle between the $\tilde{\chi}^-_1$ momentum and the
electron beam. The $s$- and $t$-channel contributions interfere
destructively, and $\sigma_L$ provides a measurement of
$m_{\tilde{\nu}_e}$ even when the sneutrino mass is several hundreds
of GeV and sneutrinos are too heavy to be produced
directly.\index{sneutrinos!mass measurement} In addition, the
forward-backward asymmetry is sensitive to the sneutrino mass.  This
asymmetry may be transferred to the decay products.  For example, in
Fig.~\ref{Sfig:chaafb}, the forward-backward asymmetry of the final
state $e$ in chargino events is shown as a function of
$m_{\tilde{\nu}_e}$ for various $m_{\tilde{e}_L}$, assuming a
left-polarized electron beam.  Assuming the MSSM
$m_{\tilde{\nu}_e}$-$m_{\tilde{e}_L}$ splitting relation of
\eqref{sleptonsplitting}, $A^{FB}_L$ provides another measurement of
$m_{\tilde{\nu}_e}$. Alternatively, one can check the validity of the
slepton splitting relation using the combined
measurements.\index{tests!of $\tilde{l}_L$-$\tilde{\nu}_l$ mass
relations}\index{sleptons!$\tilde{l}_L$-$\tilde{\nu}_l$ mass
splitting!measurement}

\begin{figure}[tb]
\begin{center}
\includegraphics[width=8cm]{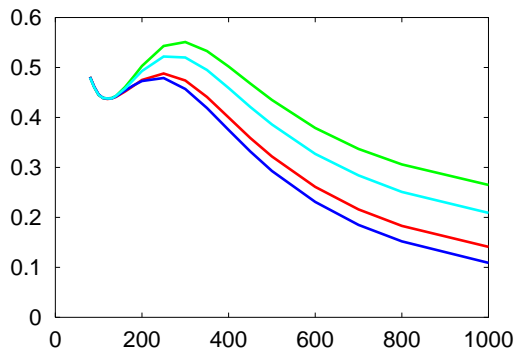}
\end{center}
\caption{\footnotesize The forward-backward asymmetry of the final
state electron from chargino decay in chargino pair events as a
function of the sneutrino mass (in
GeV)~\protect\cite{Moortgat-Pick:1999yj}.  The curves are for
$m_{\tilde{e}_L} = 130~\gev$, $150~\gev$, $200~\gev$, and
$m_{\tilde{e}_L}^2 = m_{\tilde{\nu}_e}^2 - m_W^2 \cos 2 \beta$, from
above.  The other underlying parameters are fixed to $M_2 = 152~\gev$,
$\mu = 316~\gev$, $\tan\beta = 3$, and $M_1 = 79~\gev$, and the
center-of-mass energy is $\sqrt{s} = 500~\gev$, with beam
polarizations $P_{e^-}= -0.85$ and $P_{e^+}= -0.6$.
\label{Sfig:chaafb}}
\end{figure}

Chargino studies therefore provide a rich arena for supersymmetry
studies.  Many important implications follow from such studies.  For
example, if the charginos are determined to be Wino-like, the chargino
and neutralino masses become highly correlated with $M_2$ and $M_1$,
respectively, allowing a model-independent measurement of these
parameters and providing a test of GUT or gauge-mediated
predictions.\index{tests!of gaugino mass unification} On the other
hand, if they are determined to have significant Wino-Higgsino mixing,
upper bounds on the heavier chargino and neutralinos may be
obtained.\index{upper bounds!from Wino-Higgsino mixing} Chargino
studies may also determine, or set upper bounds on, sneutrino masses
even when sneutrinos are far beyond the kinematic reach of the
collider, providing another target energy for future supersymmetry
searches.\index{upper bounds!on sneutrino mass}

\subsection{CP Violation}
\label{Ssec:CP1}

Because the linear collider allows one to overconstrain the chargino
system, as discussed above, it also allows one to explore additional
degrees of freedom assumed absent in the simplest models.  For
example, one may explore supersymmetric sources of CP violation, or
even the basic supersymmetry relations between particle and sparticle
couplings.\index{CP violation} We discuss the former here and the
latter in \secref{Ssec:testsusy}.
 
Many supersymmetric parameters are in general complex.  For example,
the gaugino and Higgsino mass parameters are, in general,
\begin{equation}
\mu=|\mu| e^{i\phi_{\mu}}\, , \ M_i=|M_i| e^{i\phi_{i}} \ .
\end{equation}
As we already discussed in \secref{Ssec:CP}, CP-violating phases of
supersymmetry parameters are strongly constrained by neutron and
electron EDMs.\index{problems!supersymmetric
CP}\index{constraints!electron EDM}\index{constraints!neutron EDM}
Large ${\cal O}(1)$ phases therefore require fine-tuning among
parameters~\cite{Ibrahim:1998je}.  However, baryogenesis requires some
source of CP violation beyond the standard model,\index{CP
violation!and baryogenesis} and it is natural to ask how CP violation
in supersymmetry may be discovered.\footnote{Large CP-violating phases
are also motivated by string theory, where all couplings, including
the standard model Yukawa couplings, are expected to result from the
dynamical condensation of moduli fields.  If soft
supersymmetry-breaking terms arise from the same source, large phases
in soft supersymmetry parameters are also expected.}

The CP phases of supersymmetry parameters may be measured through
their impact on CP-{\em conserving} quantities such as masses and
decay distributions.\index{CP violation!measurement!through
CP-conserving quantities} For example, allowing the $\mu$ parameter to
be complex, the two chargino masses are given by~\cite{Choi:1998ut}
\begin{equation}
m^2_{\tilde{\chi}^{\pm}_{1,2}}
=\frac{1}{2}\left[M_2^2 +\vert \mu\vert^2 2 m_W^2 \mp \Delta_C \right]
\ ,
\end{equation}
where 
\begin{eqnarray}
\Delta_C^2 &=& (M_2^2 - |\mu|^2)^2+ 4 m_W^4 \cos^2 2 \beta 
+ 4 m_W^2(M_2^2 + |\mu|^2) \nonumber \\
&&+8m_W^2 M_2 |\mu| \sin 2\beta \cos\phi_{\mu} \ .
\end{eqnarray}
If the full chargino system is available and the two chargino masses
and two mixing angles $\phi_+$ and $\phi_-$, suitably modified to
diagonalize the complex chargino mass matrix~\cite{Choi:1998ut}, are
constrained, deviations of $\phi_{\mu}$ from zero may be observed.
CP-violating phases also distort the neutralino mass
spectrum~\cite{Choi:2001ww}.

The dependence of various CP-conserving observables on the phases of
$\mu$ and $M_1$ are shown in Fig.~\ref{Sfig:M9}.  Significant
variations are clearly possible.  In these figures, however, the
magnitudes of all parameters are held fixed.  It is therefore not
clear whether the effects of CP-violating phases may be mimicked
simply by suitable adjustments of real parameters. To answer this
question, a CP-violating scenario was studied in
Ref.~\cite{Barger:2001nu}.  The predicted values for three cross
sections, $\sigma( \tilde{\chi}^0_1 \tilde{\chi}^0_2)$,
$\sigma(\tilde{\chi}^0_2 \tilde{\chi}^0_2)$, and
$\sigma(\tilde{\chi}^\pm_1 \tilde{\chi}^\mp_1)$, and three masses,
$m_{\tilde{\chi}^0_1}, m_{\tilde{\chi}^0_2}$, and
$m_{\tilde{\chi}^\pm_1}$ were then determined.  The measured values of
these observables were then assumed to lie in Gaussian distributions
around these central values, where the Gaussian widths were chosen to
simulate realistic experimental resolutions and statistical errors.
10,000 `pseudo-data sets' were formed by choosing values of these
observables within these distributions, and for each pseudo-data set,
the best fit underlying supersymmetry parameters were determined.

\begin{figure}[tb]
\includegraphics[width=11cm]{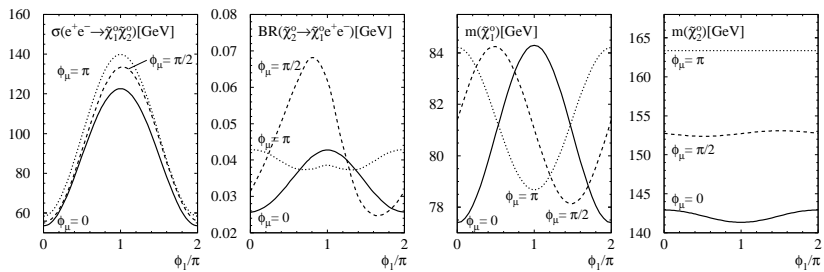}
\caption{\footnotesize The dependence of various observables on
CP-violating phases for the $\mu$ parameter and Bino mass, assuming
magnitudes of the supersymmetry parameters as given in minimal
supergravity with $m_0=100~\gev$, $\mgaugino=200~\gev$, $A_0=0$,
$\tan\beta=4$ and $\mu>0$~\protect\cite{Barger:2001nu}.
\label{Sfig:M9} }
\end{figure}

Fig.~\ref{Sfig:M10} shows the results of this study.  The input model
parameters are shown as arrows in the figures.  The left panel shows
best fit values of $M_1$ and $\phi_1$ for the 10,000 data sets,
assuming $\tan\beta = 4$ fixed to its underlying value (dark points)
and including $\tan\beta$ among the fitted parameters (light green
points).  The $\phi_1$ distribution is summarized in the right-hand
panel.  As can be seen, the hypothesis of a vanishing phase is
strongly disfavored.  In these examples, the underlying value of
$|\mu|= 310~\gev$ is assumed known from chargino studies.  However,
uncertainties of 5 GeV in the $\mu$ parameter do not alter these
conclusions.

\begin{figure}[tb]
\begin{center}
\includegraphics[width=11cm]{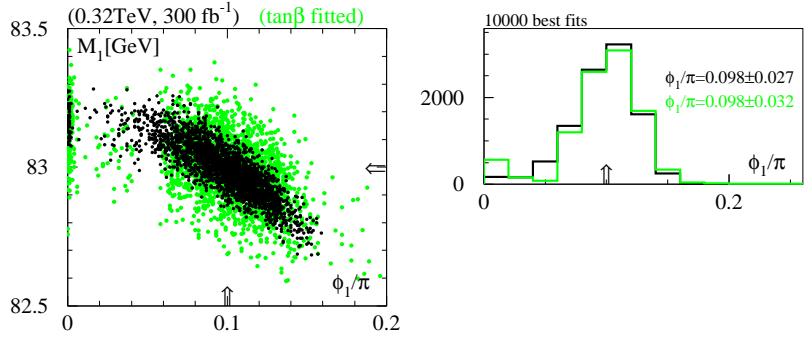}
\end{center}
\caption{\footnotesize The best fit values of $M_1$ and $\phi_1$ given
measurements of three cross sections, $\sigma( \tilde{\chi}^0_1
\tilde{\chi}^0_2)$, $\sigma(\tilde{\chi}^0_2 \tilde{\chi}^0_2)$, and
$\sigma(\tilde{\chi}^\pm_1 \tilde{\chi}^\mp_1)$, and three masses,
$m_{\tilde{\chi}^0_1}, m_{\tilde{\chi}^0_2}$, and
$m_{\tilde{\chi}^\pm_1}$~\protect\cite{Barger:2001nu}.  The effects of
experimental resolution and finite statistics have been included (see
text).  The parameter $\tan\beta$ is either fixed to its underlying
value (dark) or included among the fitted parameters (light green).
The right-hand panel summarizes the distribution of best-fit $\phi_1$
phases.
\label{Sfig:M10} }
\end{figure}

CP violation may also be measured directly through manifestly CP-{\em
violating} observables~\cite{Choi:1998ut}.\index{CP
violation!measurement!through CP-violating quantities} Such
observables are required to be sensitive to the polarization of
charginos and neutralinos relative to the plane of production.  These
observables typically require production of heavy charginos and
neutralinos, as in $\tilde{\chi}^{\pm}_1 \tilde{\chi}^{\mp}_2$ and
$\tilde{\chi}^0_i \tilde{\chi}^0_j$ ($i\ne j$) events, and are
challenging, as the dominant CP-even contribution becomes a background
for the extraction of a small CP-odd component.  However, measurements
of CP violation in this way are possible, and such observables could
provide direct and unambiguous evidence for supersymmetric CP
violation.

\section{Testing Supersymmetry}
\label{Ssec:testsusy}

Newly discovered particles need not conform to our preconceived
expectations --- the $\mu$-$\pi$ and $c$-$\tau$ puzzles are well-known
cautionary tales.\index{tests!of supersymmetry itself} A doubling of
the standard model spectrum is often thought to be the smoking gun
signal of supersymmetry.  However, there is no guarantee that all
superpartners will be discovered together.  In addition, even if all
of the appropriate degrees of freedom are discovered, the new physics
need not be supersymmetric.\footnote{For example, see
Ref.~\cite{Cheng:2002ab} for an extra-dimensional model where KK modes
provide a new particle spectrum identical to that predicted by
supersymmetry.  In addition, the colored KK particles are heavier than
the electroweak ones, and even the missing energy signals predicted in
many supersymmetry frameworks are mimicked by the existence of a
stable weakly-interacting lightest KK particle.}  Can new particles be
identified as supersymmetric through incisive tests?  Also, if some
superpartners are missing, can we constrain their properties, much as
precision electroweak data constrained the mass of the top quark
previously and constrains the Higgs boson mass now?

It turns out that these two questions may both be answered by testing
purely supersymmetric relations. We have already noted in
\secsref{Ssec:matter}{Ssec:gauge} that the gaugino-sfermion-fermion
and Higgsino-sfermion-fermion couplings are determined by the gauge
and Yukawa couplings of the standard model, respectively.  Tests of
the equivalence of these couplings are therefore model-independent
tests of supersymmetry~\cite{Feng:1995zd}.  At the same time,
measurements of small deviations from these identities are
measurements of supersymmetry breaking, and so constrain soft
parameters and the superparticle spectrum~\cite{Nojiri:1996fp,%
Cheng:1997sq,Cheng:1997vy,Nojiri:1997ma,Katz:1998br,Kiyoura:1998yt,%
Mahanta:1999hx}.

\subsection{Verifying Supersymmetry}
\label{Ssec:verifying}

Exact supersymmetry relates both dimensionful and dimensionless
couplings.  The dimensionful relations, such as the equivalence of the
electron and selectron masses, must be broken, and the new
contributions are the model-dependent soft supersymmetry-breaking
terms of \secref{Ssec:soft}.  In contrast, the relations between
dimensionless couplings are preserved in all attractive models, as
these relations are required if the gauge hierarchy is to be
preserved.  They therefore provide model-independent predictions that
may be exploited to confirm that newly-discovered particles are indeed
superpartners.

How well may the dimensionless supersymmetric identities be tested at
the linear collider?  To quantify this, we may treat the standard
model and supersymmetric couplings as independent parameters, and then
determine how well the supersymmetric couplings may be determined.
These new couplings must be determined along with all of the usual
unknown supersymmetry parameters, and the introduction of a yet
another degree of freedom implies that many independent measurements
of a given reaction are necessary.  Testing supersymmetry in this
model-independent way therefore makes full use of the flexibility and
potential of linear colliders.

Tests of supersymmetry are possible with many different superpartners
and many different coupling relations.  As an example, consider
$\tilde{e}_R$ pair production.  The $t$-channel neutralino exchange
process depends on the Bino coupling $g_{\tilde{B}\tilde{e}_R e_R}$.
To investigate the level of sensitivity to this coupling, we define
the parameter
\begin{equation}
Y_{\tilde{B}} \equiv \frac{g_{\tilde{B}\tilde{e}_Re_R}} 
{\sqrt{2} g'} \ .
\label{Snewpara}
\end{equation}
Supersymmetry predicts $Y_{\tilde{B}} = 1$.  In the limit $m_Z \ll
M_1, |\mu|$ and assuming 100\% right-handed electron beams, the
amplitude for $\tilde{e}_R$ pair production is approximately
\begin{equation}
{\cal M}\propto \beta_{\tilde e}
\left[1- \frac{4Y^2_{\tilde{B}}}{1-2\cos\theta\beta_{\tilde{e}_R} +
\beta^2_{\tilde{e}_R} + 4M_1^2/s}\right] \ ,
\end{equation}
where the first and second terms come from $s$- and $t$-channel
processes, respectively.  We may therefore constrain both
$Y_{\tilde{B}}$ and $M_1$ by measuring the differential cross section
$d\sigma/d\cos\theta$.

To determine the differential cross section, we must know the slepton
production angle.  This is, of course, not directly observable.
However, in any given event, we may determine it up to a two-fold
ambiguity~\cite{Tsukamoto:1993gt}.  The angle between an observed
lepton and its parent slepton is determined by the lepton energy
through \eqref{Sdecaykin}.  The slepton direction then lies on a cone
with known opening angle centered on the lepton direction.  The
direction of each slepton may be constrained in this way; because the
slepton pair must be produced back-to-back, we find two possible
solutions for the slepton production angle in each event, as shown in
Fig.~\ref{Sfig:M11}.  The distribution of the wrong solution turns out
to be more or less flat, and so the combined distribution plotted in
Fig.~\ref{Sfig:M11} allows one to determine the sparticle angular
distribution in a statistical manner.

\begin{figure}[tb]
\begin{center}
\includegraphics[width=4.5cm]{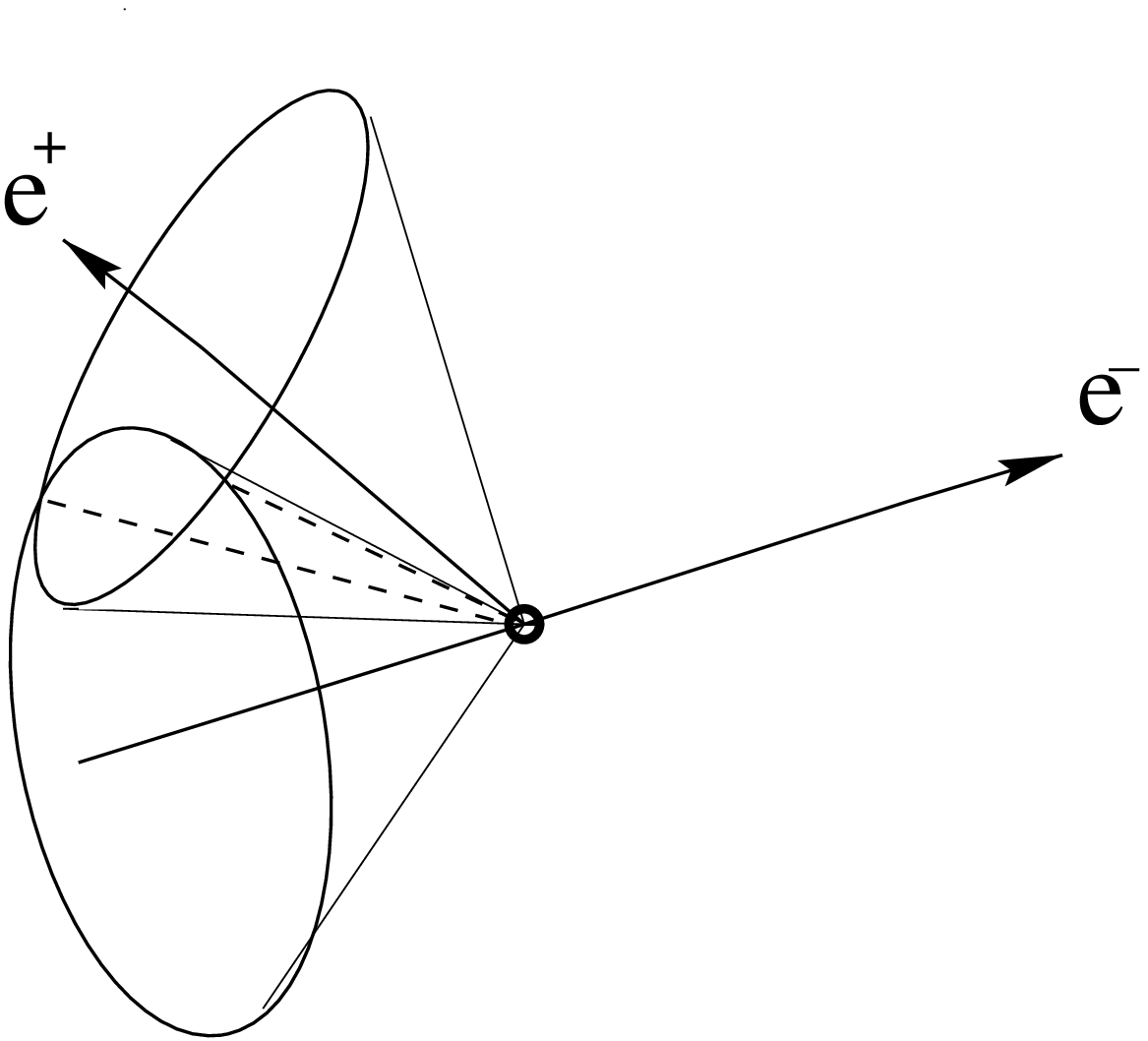}
\includegraphics[angle=90,width=5cm]{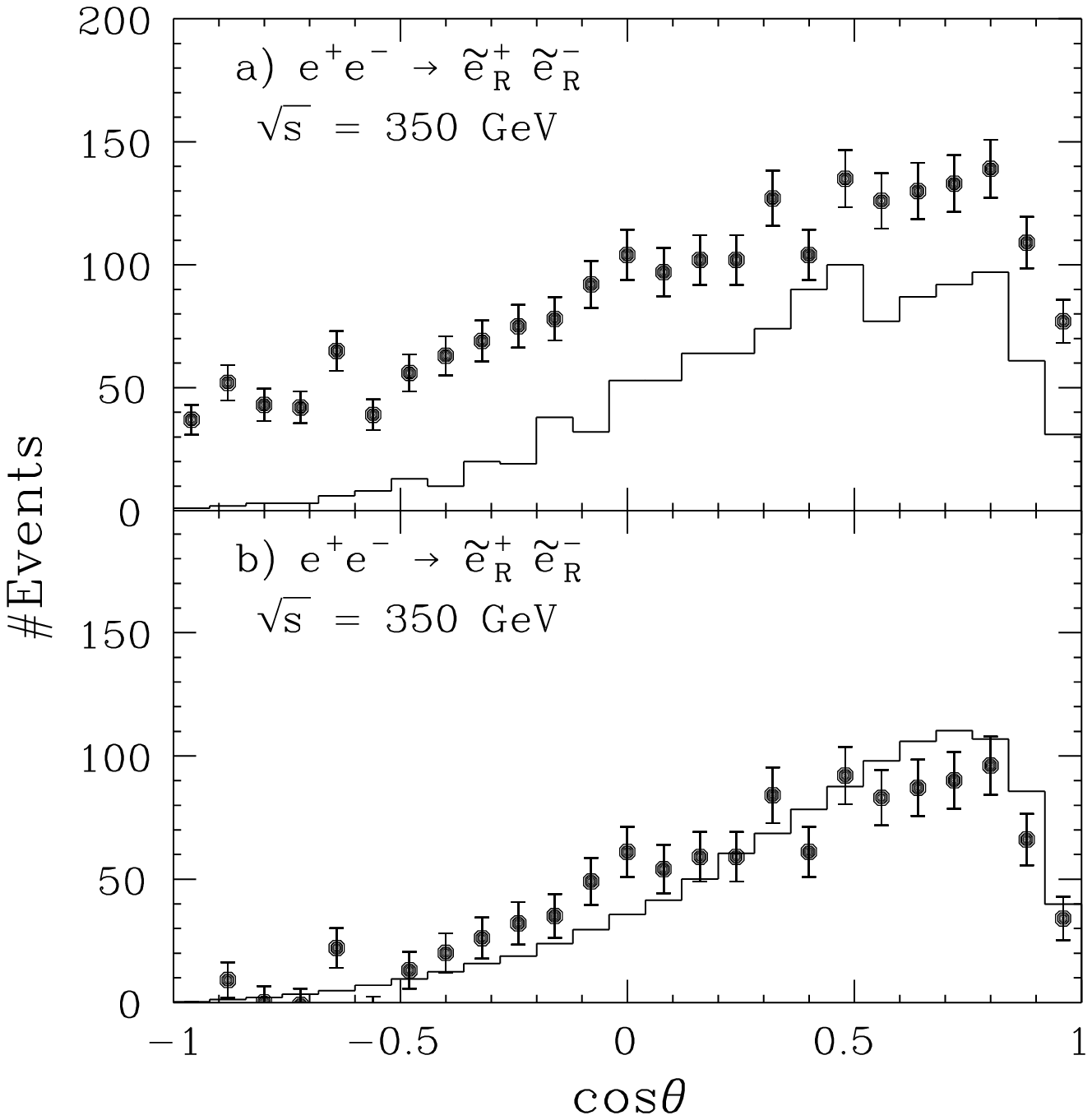}
\end{center}
\caption{\footnotesize Left: Reconstruction of the slepton production
angle~\protect\cite{Tsukamoto:1993gt}.  Each parent slepton's
direction lies on a cone around the observed lepton momentum.  Since
the sleptons are produced back-to-back, the slepton directions are
then determined up to a two-fold ambiguity to lie along the dashed
lines.  The cone for $\tilde{e}^-$ solution has been inverted in the
figure.  The dashed lines represent the solutions for the
$\tilde{e}^+$ direction.  Right: Angular distribution of $e^+e^-
\rightarrow \tilde{e}^+_R \tilde{e}^-_R$ reconstructed from the final
state $e^+e^-$ for underlying parameters $\sqrt{s}=350~\gev$,
$\tilde{e}_R=142~\gev$, and
$\tilde{\chi}^0_1=118~\gev$~\protect\cite{Tsukamoto:1993gt}.  The data
points with error bars show angular distributions with (a) both right
and wrong solutions and (b) with the wrong solution distribution,
assumed flat, subtracted (see text).  The histograms are the
distribution of correct solutions (a) after and (b) before selection
cuts.
\label{Sfig:M11}}
\end{figure}

Given the differential cross section for selectron pair production, as
well as kinematic endpoint information to constrain $m_{\tilde{e}_R}$
and $m_{\tilde{\chi}^0_1}$, one may then constrain $Y_{\tilde B}$.
The results of such an analysis are given in Fig.~\ref{Sfig:M12},
where we see that $Y_{\tilde{B}}$ may be determined at the 2\% level
for integrated luminosity $100~\ifb$~\cite{Nojiri:1996fp}, providing
precise and model-independent evidence that the produced scalar
particle is indeed a selectron.  Constraints on $Y_{\tilde{B}}$ of
even higher precision are possible from $\emem \to \tilde{e}^-_R
\tilde{e}^-_R$ production~\cite{Cheng:1997vy},\index{$\emem$ option,
uses for!tests of supersymmetry} and the equivalent parameters for the
SU(2) and SU(3) couplings may also be constrained by chargino and
squark studies~\cite{Cheng:1997vy} and also through studies of triple
gauge vertices~\cite{Mahanta:1999hx}.  An exhaustive list of
observables with the potential for testing supersymmetry is given in
Ref.~\cite{Cheng:1997sq}.

\begin{figure}[tb]
\begin{center}
\includegraphics[width=8cm]{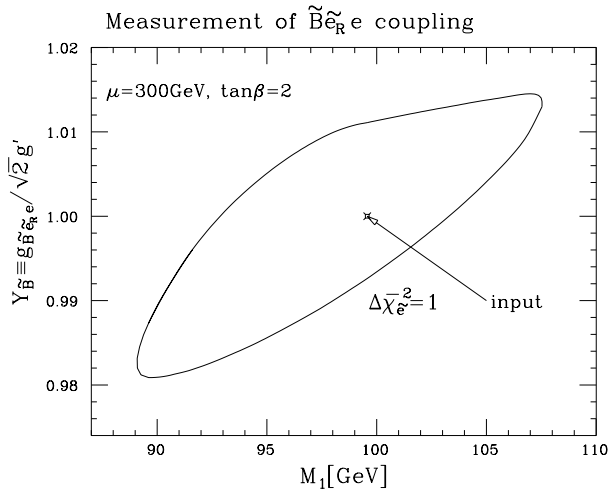}
\end{center}
\caption{\footnotesize $\Delta\bar{\chi}^2_{\se}=1$ contour in the
$M_1$-$Y_{\tilde B}$ plane for integrated luminosity
$100~\ifb$~\protect\cite{Nojiri:1996fp}.  $\Delta
\chi^2_{\tilde{e}}=1$ is defined by constraints from electron energy
and differential cross sections measurements and is roughly equivalent
to a significance of $\Delta{\chi}^2=1$.  Input values are
$m_{\tilde{e}_R} = 200~\gev$, $\mu = 300~\gev$, $M_1 = 99.57~\gev$,
and $\tan\beta = 2$.
\label{Sfig:M12}}
\end{figure}

\subsection{Super-oblique Parameters}
\label{Ssec:super-oblique}

So far we have been discussing the possibility of testing
supersymmetry by verifying dimensionless coupling identities.  These
identities are exact at tree-level.  At the loop-level, however, even
these relations receive corrections~\cite{Chankowski:1990du}.  The
breaking of dimensionless relations is called {\em hard supersymmetry
breaking}~\cite{Hikasa:1996bw}.\index{supersymmetry breaking!hard}

These corrections may be understood by analogy to the oblique
corrections~\cite{Peskin:1990zt} of the standard
model.\index{super-oblique!corrections!analogy to oblique} In the
standard model, SU(2) multiplets with custodial SU(2)-breaking masses,
such as the $(t,b)$ multiplet, induce splittings in the couplings of
the $(W,Z)$ vector multiplet at the quantum level. Similarly, in
supersymmetric models, supermultiplets with soft
supersymmetry-breaking masses, such as the $(\tilde{f},f)$
supermultiplets, induce splittings in the couplings of the $({\rm
gauge\ boson}, {\rm gaugino})$ vector supermultiplet at the quantum
level.  This analogy can be made very
precise~\cite{Cheng:1997sq,Nojiri:1997ma,Katz:1998br}.  Corrections to
hard supersymmetry relations are therefore called {\em super-oblique
corrections},\index{super-oblique!corrections} and the splittings are
typically written in terms of {\em super-oblique
parameters}.\index{super-oblique!parameters}

A enlightening viewpoint is to consider the evolution of dimensionless
couplings from some high scale in these theories. Above the scale of
superpartner masses, the theory is completely supersymmetric, and so
the gaugino and gauge couplings remain identical.  However, at scales
below some sparticle mass, the effective theory is not supersymmetric,
and so the supersymmetry relations get corrected.  If there are still
sparticles in this effective theory, they may be produced at
colliders, and their couplings will show evidence of the heavy
superpartners.  

This viewpoint makes obvious an interesting fact:~the super-oblique
parameters are {\em
non-decoupling}.\index{super-oblique!parameters!non-decoupling of} The
heavier the decoupled superparticles are, the longer the couplings
evolve in the non-supersymmetric theory and the greater their
splitting.  Measurements of super-oblique parameters at the linear
collider therefore provides a method for probing superpartner masses
at arbitrarily high scales.  Super-oblique parameter measurements are
particularly relevant in models with large hierarchies in the
supersymmetry spectrum.  Such hierarchies are certainly possible, and
possibly even favored, given the constraints discussed in
\secref{Ssec:successes}, and may be found in models with heavy colored
superpartners, focus point supersymmetry, or the superheavy
supersymmetry scenarios of \secref{Ssec:superheavy}.\index{upper
bounds!from super-oblique
parameters}\index{super-oblique!parameters!and superheavy
supersymmetry}\index{super-oblique!parameters!and focus point
supersymmetry}\index{models!focus point supersymmetry!and
super-oblique parameters}\index{models!superheavy supersymmetry!and
super-oblique parameters}

For concreteness, let us consider models with
\begin{equation}
M_3, m_{\tilde{q}} \gg m_{\tilde{l}}, M_2, M_1 \ . 
\end{equation}
Characterizing the superparticle mass scale with two parameters,
$M_{\tilde{Q}}$ for all the heavy states, and $M_{\tilde{L}}$ for all
the light states, we find~\cite{Cheng:1997sq}
\begin{eqnarray}
\tilde{U}_1 &\equiv& Y_{\tilde{B}} - 1 = 
\frac{11}{5}\frac{g_1^2}{16\pi^2}
\ln\frac{M_{\tilde{Q}}}{M_{\tilde{L}}} \\
\tilde{U}_2 &\equiv& Y_{\tilde{W}} - 1 = 3\frac{g^2}{16\pi^2} 
\ln \frac{M_{\tilde{Q}}}{M_{\tilde{L}}}\ .
\end{eqnarray}
where $\tilde{U}_1$ and $\tilde{U}_2$ are the super-oblique parameters
for the U(1) and SU(2) gauge groups.  Both of these may be measured at
linear colliders. For $M_{\tilde{Q}}/M_{\tilde{L}} \sim 10$,
$\tilde{U}_2 \approx 2.0\%$ and $\tilde{U}_1 \approx 0.7\%$, leading
to enhancements of $t$-channel contributions of about 8.0\% and 2.8\%,
respectively. Given the results described above, such deviations are
certainly observable at the linear collider.  Non-vanishing
super-oblique parameters are signals of supersymmetry breaking, and
precise measurements could even constrain the mass scale of
superpartners that are far beyond the reach of colliders.

The super-oblique corrections are, of course, just a subset of all
radiative corrections.\index{radiative corrections} They may be
particularly large, given the decoupling picture leading to the $\log
M_Q$ corrections discussed above.  However, additional radiative
corrections appear in all supersymmetry processes. Radiative
corrections to masses have been presented in
Ref.~\cite{Bagger:1995bw}.  For some production and decay processes,
the full one-loop correction is also available~\cite{Diaz:1997kv}.  If
supersymmetry is discovered, the full radiative analysis of all
available processes will become important in precision supersymmetric
physics, just as it was for precision electroweak physics.  These
studies will be important also for global fits and extrapolation to
higher mass scales, which we will discuss below in
\secref{Ssec:extraplanck}.

\section{Determining the Scale of Supersymmetry Breaking}
\label{Ssec:scaleofSUSYbreaking}

If supersymmetry is discovered, the mediation of supersymmetry
breaking will be investigated in detail by measuring the soft
supersymmetry-breaking terms, as we have discussed.  Intimately
connected to the mediation mechanism, however, is the mechanism of
supersymmetry breaking.  As a first step, one would like to know the
scale of supersymmetry breaking $\fdsb$ discussed in
\secref{Ssec:models}.\index{supersymmetry breaking!scale
of!determination of} This scale determines both the gravitino mass and
its couplings.  In frameworks such as supergravity, where the scale of
supersymmetry breaking is high, the gravitino is decoupled from
collider phenomenology, and so colliders cannot provide much
information about the supersymmetry-breaking scale, other than to
provide lower bounds.

In low-scale supersymmetry-breaking scenarios, including the
gauge-mediated models discussed in \secref{Ssec:gmsb}, however,
standard model superpartners may decay to gravitinos on observable
scales.  This decay provides a window on the gravitational sector of
supersymmetry, allowing a quantitative measurement of the scale of
supersymmetry breaking, and possibly providing information about the
gravitino mass $m_{3/2}$, the messenger scale $\mmess$, and so
on.\index{gravitino!mass!measurement}

The robust prediction of gauge-mediated models is the existence of a
very light gravitino LSP with significant couplings to standard model
particles.\index{models!gauge-mediation!signals} In these
scenarios, standard model superpartners may decay into gravitinos in
collider detectors.  The decay length for this decay was given in
\secref{Ssec:gmsb}.  More precisely, for a neutralino NLSP, the decay
length for $\tilde{\chi}^0_1 \to \tilde{G}\gamma$ is
\begin{equation}
L \simeq 0.10~\text{mm} \frac{1}{\kappa_{\gamma}^2}
\Biggl[\frac{100~\gev}{m_{\tilde{\chi}_0^1}}\Biggr]^5
\Biggl[\frac{\sqrt{\fdsb}}{10^5~\gev}\Biggr]^4
\Biggl[\frac{E_{\tilde{\chi}_0^1}^2 - m_{\tilde{\chi}_0^1}^2}
{m_{\tilde{\chi}_0^1}^2}\Biggr]^{\frac{1}{2}} \ ,
\end{equation}
where $\kappa_{\gamma}= | N_{11} \cos\theta_W + N_{12}\sin\theta_W|$.
For typical parameters, this is a macroscopic distance.  The NLSP's
decay products therefore typically do not point back to the
interaction point.  Depending on what superpartner is the NLSP, we
would then see the following unusual signals at the linear
collider:\index{displaced vertices}
\begin{eqnarray}
e^+e^- &\to& \tilde{l}^+\tilde{l}^-\to l l 
\tilde{G}\tilde{G} + {\rm displaced \ vertices}
\nonumber\\
e^+e^-&\to& \chi^0_1\chi^0_1\to \gamma\gamma
\tilde{G}\tilde{G}
+{\rm displaced\  vertices} \ .
\label{SNLSPdecay}
\end{eqnarray}
These signals are spectacular, and may be used to differentiate
supersymmetry from backgrounds.\footnote{Note that displaced vertices
are particularly helpful at the LHC.  For example, given displaced
vertices, one can reconstruct the masses of various sparticles, such
as $\tilde{\chi}^0_3$ and $\tilde{\chi}^0_4$, which might not be
accessible in supergravity scenarios~\cite{AtlasTDR:1999fq}.}

The signals of \eqref{SNLSPdecay} have been studied carefully in
Ref.~\cite{Ambrosanio:1999iu}. For slepton NLSPs with decay lengths
longer than $\sim 10\ \mu \text{m}$, the $\tilde{l}$ momentum can be
measured by the inner tracking detector.  The NLSP lifetime can then
be reconstructed with relative ease.  In conjunction with NLSP mass
measurements, $\fdsb$ will determined to within a few percent.

The case of neutralino NLSPs is more difficult, but still promising.
The linear collider calorimeter has some angular resolution, and will
measure decay lengths between 5 cm and 2 m with a statistical
precision of a few percent.  For decay lengths above $\sim 10\
\mu\text{m}$ but below 10 cm, three-body decays $\tilde{\chi}^0_1\to
f\bar{f}\tilde{G}$ may be used to reconstruct the vertex. The
branching fraction for such decays is typically a few
percent. Finally, for very long decay lengths of the order of 10 m,
one can compare the number of 2 photon events to the number of 1
photon events, where, in the latter case, one of the two NLSPs decays
outside the detector.  This may be sensitive to supersymmetry-breaking
scales $\sqrt{\fdsb}$ as large as $2000~\tev$. In this last case, high
luminosity is of great importance.

\section{Extrapolation to the Planck Scale}
\label{Ssec:extraplanck}

In \secsref{Ssec:slstudy}{Ssec:chaneustudy}, we saw numerous examples
in which the weak-scale parameters of the MSSM can be determined in a
model-independent manner.\index{Planck scale!extrapolation to} As
described in \secref{Ssec:models}, these soft supersymmetry-breaking
parameters are expected to derive from soft masses at higher scales,
and the weak-scale parameters are determined by the more fundamental
high energy parameters by renormalization group
evolution.\index{renormalization group!evolution}

In practice, the renormalization group flow will be inverted.
Measurements of superpartner properties at the linear collider and the
LHC will provide determinations of weak-scale supersymmetry
parameters. These can then be evolved to high scales to determine
their fundamental, microscopic values.

An example of the power of this approach is given in
Fig.~\ref{Sfig:M13}~\cite{Blair:2000gy}.  The assumed physical
framework is a minimal supergravity model.\index{Planck
scale!extrapolation to!in minimal supergravity} The precision with
which the weak-scale parameters will be measured depends, of course,
on the collider and detector parameters.  The authors assumed mass
precisions of
\begin{eqnarray}
\Delta m_{\tilde{\chi}^{\pm,0}}&\approx& 0.3~\gev \cr
\Delta m_{\tilde{l}} \approx 
\Delta m_{\tilde{\nu}}&\approx& 0.1~\gev\cr
\Delta m_{\tilde{\tau}}&\approx&0.6~\gev \cr
\Delta m_{\tilde{t},\tilde{b}}&\approx& 1~\gev \ .
\end{eqnarray}
Along with analyses of cross sections and other quantities in slepton
and chargino/neutralino events, these measurements would provide
determinations of the gaugino parameters $M_1$ and $M_2$ at the
percent level.  The gluino mass $M_3$ is assumed to be measured by a
combination of LHC and linear collider data.  Note that the linear
collider information improves the $M_3$ precision by over an order of
magnitude through the complementarity described in
\secref{Ssec:slmasses}.\index{complementarity!of LC and LHC}

Extrapolations of the supersymmetry parameters from the weak scale to
the GUT scale are shown in Fig.~\ref{Sfig:M13}.  The unification of
gaugino masses is striking, providing clear evidence for gauge group
unification.\index{tests!of gaugino mass unification} The sfermion
masses also unify at the same scale, with great precision for the
sleptons.\index{tests!of scalar mass unification} The quark and Higgs
parameters are less well-determined, and the uncertainty in weak-scale
values is noticeably magnified as one evolves to higher energies.
Nevertheless, taken as a whole, the sfermion mass trajectories also
show clear evidence for unification at the GUT scale.

\begin{figure}[tb]
\begin{center}
\includegraphics[width=5cm]{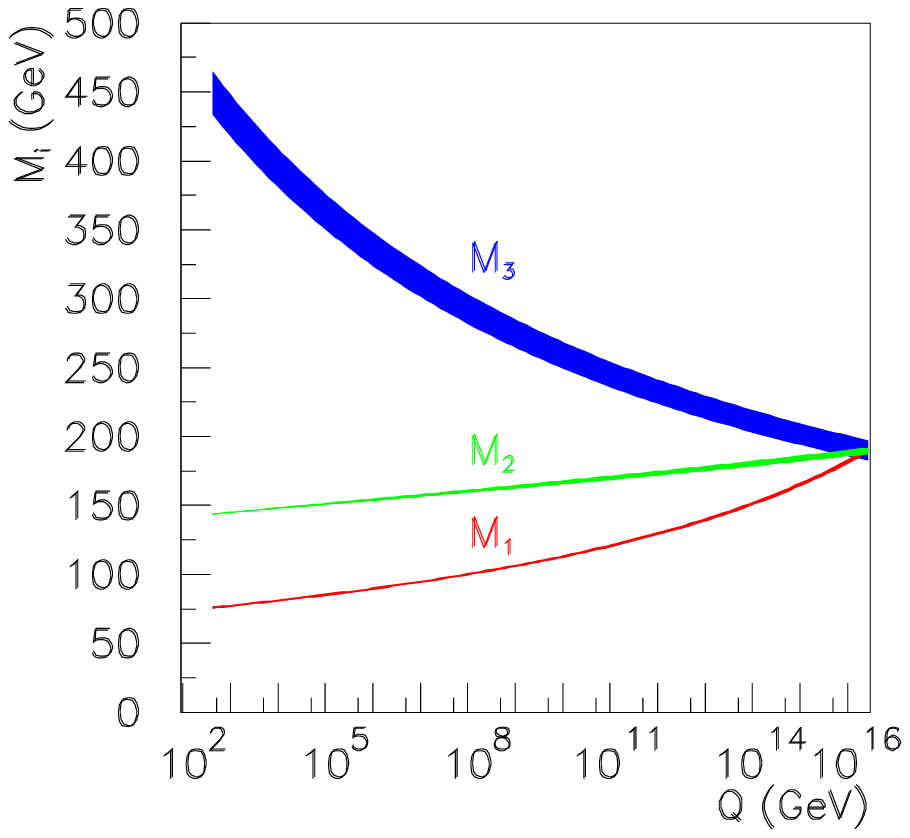}
\includegraphics[width=5cm]{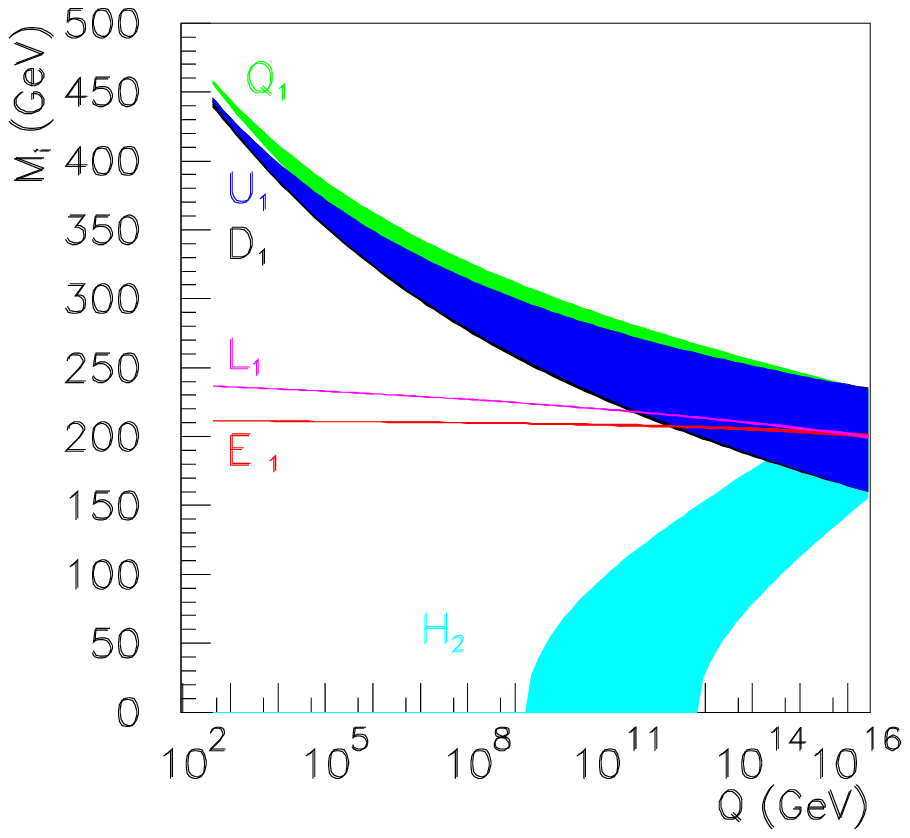}
\end{center}
\caption{\footnotesize Renormalization group evolution of gaugino
(left) and sfermion (right) mass parameters from the weak scale to the
GUT scale in a minimal supergravity model with $m_0=200~\gev$,
$\mgaugino = 190~\gev$, $A_0=500~\gev$, $\tan\beta=30$, and
$\mu<0$~\protect\cite{Blair:2000gy}. $H_u$ denotes the $m_{H_u}$
renormalization group trajectory.  The weak-scale experimental inputs
are discussed in the text, and the bands indicate 95\% CL contours.
\label{Sfig:M13}}
\end{figure}

The identical analysis has been performed for a gauge-mediated
supersymmetry breaking model.\index{Planck scale!extrapolation to!in
gauge-mediation} The results are show in Fig.~\ref{Sfig:M14}.  Recall
that the simplest gauge-mediated models predict gaugino mass relations
identical to those in unified scenarios, but the scalar mass
predictions of \eqref{Sgmsbmasses} are very different.  The results of
Fig.~\ref{Sfig:M14} clearly disfavor scalar mass unification at any
scale, and would thereby exclude minimal supergravity and many
GUTs.\index{tests!of scalar mass unification} In addition, the
renormalization group trajectories show unification of the the
left-handed slepton $L_1$ and $H_u$ masses at $10^8~\gev$.  In
gauge-mediated scenarios this identifies the messenger
scale.\index{messenger!scale!measurement}

\begin{figure}[tb]
\begin{center}
\includegraphics[width=5cm]{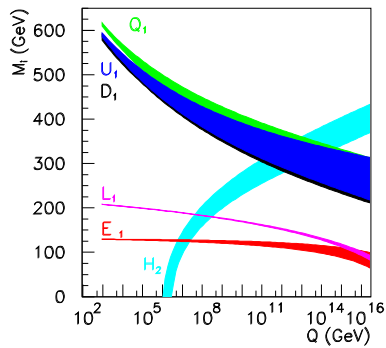}
\end{center}
\caption{\footnotesize Evolution of sfermion mass parameters in a
gauge-mediated supersymmetry breaking model with $\mmess=2 \times
10^8~\gev$, $\Lambda = \langle F \rangle/\mmess = 28~\tev$, $N=3$,
$\tan\beta=30$ and $\mu<0$~\protect\cite{Blair:2000gy}.  $H_u$ denotes
the $m_{H_u}$ renormalization group trajectory.
\label{Sfig:M14}}
\end{figure}

\section{Connections to Cosmology} 
\label{Ssec:cosmology}

The discovery of supersymmetry with a potentially stable LSP implies
the discovery of dark matter candidates. These discoveries will open
many avenues for dark matter studies, with a wealth of connections
between particle physics and cosmology.\index{cosmology} A schematic
picture of the resulting investigation of supersymmetric dark matter
is given in Fig.~\ref{Sfig:darksusy}.\index{complementarity!of
particle and astroparticle physics}

In the MSSM, suitable dark matter candidates are the neutralino and
the gravitino.\index{dark matter} In the case of neutralino dark
matter, the thermal relic density is determined by the neutralino pair
annihilation cross section as described in \secref{Ssec:dm}, and dark
matter may be detected either directly through its interactions with
ordinary matter or indirectly through its annihilation decay
products.\index{dark matter!thermal relic density} At linear
colliders, supersymmetry parameters may be determined at the percent
level, and the thermal relic density and neutralino-nucleon scattering
cross sections may be determined with similar precision.  Such
progress corresponds to the upper-half of Fig.~\ref{Sfig:darksusy},
and the completion of this program will provide a great deal of
information about the suitability of neutralinos as dark matter
candidates.

\begin{figure}[tb]
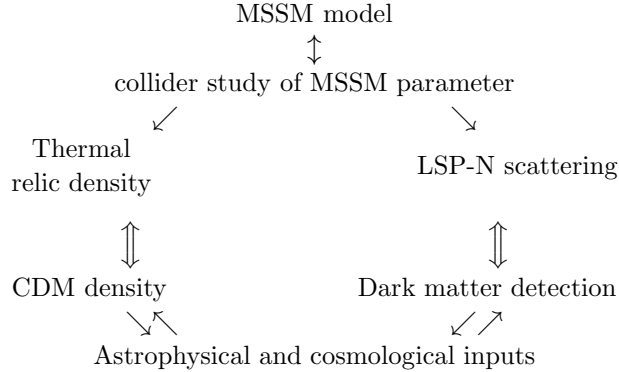

\begin{center}
MSSM model 
\\
{\large
$\updownarrow$}
\\
collider study  of MSSM parameter
\\
$\swarrow$ \hskip 3.5cm $\searrow$ 
\\
\begin{tabular}{c}
Thermal \cr
relic density
\end{tabular}
\hskip 3cm    
\begin{tabular}{c}
LSP-N scattering
\end{tabular}
\end{center}
\begin{center}
$\Big\Updownarrow$ \hskip 4.5cm $\Big\Updownarrow$ \\
\begin{tabular}{c}
CDM density
\end{tabular}
\hskip 2cm 
\begin{tabular}{l}
Dark matter detection 
\end{tabular}
\\
$\searrow$$\nwarrow$ \hskip 3.5cm 
$\swarrow$$\nearrow$ 
\\
Astrophysical\  and\  cosmological\  inputs
\end{center}
\caption{The road to understanding supersymmetric dark matter.
\label{Sfig:darksusy}}
\end{figure}

As examples, recall that the thermal relic density and detection rates
are sensitive to gaugino-Higgsino mixing, as explained in
\secsref{Ssec:dm}{Ssec:focuspoint}.  The mixing depends on the ratios
$M_i/|\mu|$ and may be measured either by studying both chargino
states, if available, or by studying a single chargino state with
polarized beams, as discussed in \secref{Ssec:chmass}. The relic
density is also modified if there are other superpartners with masses
within roughly 5\% of the LSP mass~\cite{Griest:1990kh}.\index{dark
matter!co-annihilation} In the case of minimal supergravity, for
example, the lighter $\tilde{\tau}$ is lighter than all other charged
superpartners and may be nearly degenerate with the neutralino LSP.
In this case, the thermal relic density is reduced by coanihilation
effects and is highly sensitive to the
$\tilde{\tau}$-$\tilde{\chi}^0_1$ mass splitting.  Precise mass
measurements are therefore needed to pin down the relic density. The
mass determination of highly degenerate sparticles is very challenging
in general; in this case, it is complicated by the $ee\tau\tau$
background discussed in \secref{Ssec:slsgbg} and requires careful
detector design to retain sensitivity to low momentum jets and
leptons.

Even if the supersymmetry parameters are all precisely measured, this
is not the whole story, however.  The identification of the thermal
relic density with the present day cold dark matter density is subject
to cosmological assumptions.  For example, the calculation of the
thermal relic density assumes that the dominant source of dark matter
is from dark matter particles falling out of thermal equilibrium.  It
is possible, however, that the bulk of the dark matter is created not
through thermal equilibrium and freeze-out, but through the decay of a
supermassive particle after freeze-out, but before $t=1~\s$. For
example, in the anomaly-mediated supersymmetry breaking models of
\secref{Ssec:amsb}, the LSP thermal relic density is negligible, as
Winos annihilate extremely efficiently in the early
universe.\index{dark matter!in anomaly-mediation}\index{dark
matter!Wino-like} However, anomaly-mediated scenarios also have
supermassive gravitinos, with mass $\sim 10~\tev$.  These may decay to
LSPs after the usual freeze-out temperatures, producing Wino LSPs with
a cosmologically interesting mass
density~\cite{Moroi:1999zb}.\index{dark matter!non-thermal production}

The thermal relic density calculation also assumes that nothing
unusual happens once the dark matter is produced at temperatures of
$T\sim {\cal O}(10)~\gev$. At present, the thermal history of the
universe is on sure footing only for times after Big Bang
nucleosynthesis, that is, times $t \agt 1~\s$ and temperatures $T \alt
1~\mev$.  The thermal relic density calculation therefore requires an
extrapolation of four orders of magnitude in temperature.  Large
entropy production by late-decaying particles may drastically alter
calculated relic densities, reducing a seemingly too-large relic
density to the ideal range, for example.  The bottom line is that the
cold dark matter density obtained following the path from the bottom
of Fig.~\ref{Sfig:darksusy} need not coincide with the thermal relic
density obtained by following the path from the top. Instead,
discrepancies might provide new insights into the history of our
universe.

In a similar vein, the neutralino-nucleon cross sections are not
necessarily in one-to-one correspondence with dark matter detection
rates.\index{dark matter!detection} This correspondence requires
information about the local density and velocities of dark
matter.\index{dark matter!velocity distribution} The uncertainties and
problems associated with these issues have been discussed
extensively~\cite{Sikivie:1996nn,Gelmini:2000dm,Calcaneo-Roldan:2000yt}.
If neutralinos are identified as the dark matter, future colliders
will determine their properties and the neutralino-nucleon scattering
cross sections.  To the extent that these may be checked against
actual detection rates, future colliders will also provide important
information about dark matter halo densities and velocity
distributions.

The second possible supersymmetric dark matter candidate is the
gravitino.\index{gravitino!as dark matter}\index{dark
matter!gravitino} Gravitino dark matter is somewhat beyond the scope
of this paper; for an excellent review, see
Ref.~\cite{Giudice:1998bp}.  Recall, however, that the NLSP lifetime
is related to the gravitino mass, as both are determined by $\fdsb$ as
in \eqsref{Smgravitino}{Sgravitinointeractions}.  The gravitino's
relic density is proportional to the gravitino's mass, and so the
observed dark matter density provides an upper limit on the
gravitino's mass of $m_{3/2} \alt 0.2~\kev$ typically.  In this range,
the gravitino is hot dark matter.  Gravitinos with masses around 1~keV
may be interesting warm dark matter candidates.  In this range, the
displaced vertices of NLSP decays to gravitinos may be accessible at
linear colliders through the studies discussed in
\secref{Ssec:scaleofSUSYbreaking}.

\section{Conclusions}
\label{Ssec:conclusions}

Supersymmetry and the linear collider are a near perfect fit --- it is
difficult to envision a richer linear collider program than that
provided by the superpartner spectrum, and it is hard to imagine a
more incisive tool for studying supersymmetry than the linear collider.
Of course, the program of supersymmetry studies at the linear collider
depends on what superpartners are kinematically accessible. However,
as evident from the many examples discussed above, even if only one or
a few superpartners are kinematically accessible, the linear collider
will be able to provide model-independent measurements of a host of
soft supersymmetry-breaking parameters.  In conjunction with the
information provided by hadron colliders, this information will likely
provide precise information about sparticles beyond the linear
collider's reach.

In this overview, we have highlighted many of the fundamental
questions that will be addressed by such supersymmetry measurements.
These range from issues specific to supersymmetry, such as the testing
of supersymmetric identities, the resolution of the supersymmetric
flavor and CP puzzles, and the determination of the scale of
supersymmetry breaking, to grand universal questions, such as the
nature of dark matter, the unification of forces, and the geometry of
spacetime.  As discussed in this review, if supersymmetry is within
reach, the linear collider may shed light on all of these issues, and
will, in some cases, provide definitive answers.

At the same time, there are many outstanding
problems.\index{problems!outstanding} To name but a few, the study of
radiative corrections\index{radiative corrections} in supersymmetry
and its impact on precision measurements is still in its early days.
In addition, the importance of experimental uncertainties, such as in
the luminosity spectrum, beam energy, and polarimetry have been
considered in a few studies, but have not been systematically
investigated.\index{collider parameters} And of course, new ideas for
physics beyond the standard model continually arise, and the potential
of linear colliders to probe such new ideas and to differentiate these
ideas from other new physics possibilities will continue to be of
interest.  To fulfill the potential of linear colliders to study new
physics, dedicated experimental and theoretical efforts are still
needed. We hope this chapter will be of use to people who are
interested in continuing the exploration of supersymmetry at new
colliders.

\section*{Acknowledgments}
\addcontentsline{toc}{section}{Acknowledgements}

We thank our collaborators for their many insights through the years,
Keisuke Fujii, David Miller, and Amarjit Soni for their support of
this review, and Michael Peskin for sparking our initial interest in
this subject.

\clearpage

\addcontentsline{toc}{section}{Index}

\printindex		%to generate and print out for index text  

\end{document}